\documentclass[aps,amsmath,amssymb,twocolumn]{revtex4}
\usepackage[dvips]{graphicx,color}
\usepackage{times}
\usepackage{mathrsfs}
\usepackage{amsmath}
\usepackage{graphicx}
\usepackage{epsfig}
\usepackage{dcolumn}
\usepackage{bm}
\definecolor{r}{rgb}{1.0,0,0}

\begin{document}
\title{Lateral density and arrival time distributions of Cherenkov 
photons in extensive air showers: a simulation study}
\author{P. Hazarika$^1$, U. D. Goswami$^1$, V. R. Chitnis$^2$, 
B. S. Acharya$^2$, G. S. Das$^1$, B. B. Singh$^2$, R. J. Britto$^3$}
\affiliation{$1.$ Department of Physics, Dibrugarh University,
Dibrugarh 786 004, Assam, India,}
 \affiliation{$2.$ Department of High Energy Physics, Tata Institute of 
Fundamental Research, Homi Bhabha Road, Mumbai 400 005, India,}
\affiliation{$3.$ Department of Physics, University of Johannesburg, 
Auckland Park 2006, Johannesburg, South Africa.} 
\begin{abstract}
We have investigated some features of the density and arrival time 
distributions of Cherenkov photons in extensive air showers using  
the CORSIKA simulation package. The main thrust of this study is to see the 
effect of hadronic interaction models on the  
production pattern of Cherenkov photons with respect to distance from the 
shower core. Such studies
are very important in ground based $\gamma$-ray astronomy for an effective
rejection of huge cosmic ray background, where the atmospheric Cherenkov 
technique is being used extensively within the energy range of some hundred
GeV to few TeV. We have found that for all primary particles, the density 
distribution patterns of Cherenkov photons follow the negative exponential
function with different coefficients and slopes depending on the type of 
primary particle, its energy and the type of interaction model combinations. 
Whereas the arrival time distribution patterns of Cherenkov photons follow the 
function of the form $t (r) = 
t_{0}e^{\Gamma/r^{\lambda}}$, with different values of the function parameters.
There is no significant effect of hadronic interaction model combinations on 
the density and arrival time distributions for the $\gamma$-ray primaries. 
However, for the hadronic showers, the effects of the model combinations are 
significant under different conditions. 
\end{abstract}

\maketitle

\section{Introduction}
The Atmospheric Cherenkov Technique (ACT) is being used extensively to detect
the $\gamma$-rays emitted by celestial sources using the ground-based 
telescopes within the energy range of some hundred GeV to few TeV.
This technique is based on the effective detection of Cherenkov photons 
emitted by the relativistic charged particles present in the Extensive Air 
Showers (EASs) initiated by the  primary $\gamma$-rays in the atmosphere 
\cite{Hoffman, Weekes, Acharya}. It is worthwhile to mention that, 
the celestial sources which emit $\gamma$-rays also emit Cosmic Rays 
(CRs). But CRs being charged particles, are deflected in the 
intergalactic magnetic fields and hence they reach us isotropically  loosing 
the direction(s) of their source(s). However, as the $\gamma$-rays are  
neutral,  by detecting them we can pinpoint  the locations of such 
astrophysical sources. 

As the ACT is an indirect method, detailed Monte Carlo 
simulation studies of atmospheric Cherenkov photons have to be carried out to 
estimate the energy of incident $\gamma$-ray. Also it is necessary to reject huge CR 
background as CRs also produce the EAS in the atmosphere like 
$\gamma$-rays but with a slight difference. The EAS originated from primary 
$\gamma$-rays are of pure electromagnetic in nature whereas those due to CRs
are a mixture of electromagnetic and hadronic cascades. Many extensive studies 
have already been carried out on the arrival time as well as on the density 
distributions of atmospheric Cherenkov photons in EASs at the observation 
levels in the high as well as low energy regimes using available detailed 
simulation techniques \cite{Badran, Versha}. As the simulation study is 
interaction model dependent, it is also necessary to carry out the model 
dependent study 
for a reliable result. Although such type of studies have been also carried 
out extensively in the past, there are not many studies applicable particularly 
to high altitude observation levels. In keeping this point in mind, in this 
work we have studied the density and the arrival time distributions of 
Cherenkov photons in EASs of $\gamma$ and CR primaries, at different energies 
and at high altitude observation level, using different low  and high energy 
hadronic interaction models available in the present version of the CORSIKA 
simulation package \cite{Heck}.

CORSIKA is a detailed Monte Carlo simulation package to study the 
evolution and properties of extensive air showers in the atmosphere. This
allows to simulate interactions and decays of nuclei, hadrons, 
muons, electrons and photons in the atmosphere up to energies of some 
10$^{20}$eV. For the simulation of hadronic interactions, presently 
CORSIKA has seven options for high energy hadronic interaction models
and three low energy hadronic interaction models. 
It uses EGS4 code \cite{Nelson} for the simulation of electromagnetic 
component of the air shower \cite{Heck}.

This paper is organized as follows. In the next section, we discuss in 
detail our simulation process. The section III is devoted for the 
details about the analysis of the simulated data and the results of the 
analysis. We summarize our work with conclusions and future outlook in the 
section IV.
              
\section{Simulation of atmospheric Cherenkov photons}
We used CORSIKA 6.990 simulation package to simulate the Cherenkov photons 
emitted in the earth's atmosphere by the relativistic charged particles of 
the EAS generated by $\gamma$ and CR primaries. We have considered 
two high energy hadronic interaction models, viz., QGSJET 01C and VENUS 4.12 
together with all three low energy hadronic interaction models. These two sets 
of models are combined together in all six possible combinations, viz., 
QGSJET-GHEISHA, VENUS-GHEISHA, VENUS-UrQMD etc. to generate EASs for the 
vertically incident monoenergetic $\gamma$, proton and iron primaries. The 
motivation behind the choice of QGSJET01 and VENUS high energy hadronic 
interaction model is that, they are being used extensively in the simulation 
works of CR and $\gamma$-ray experiments \cite{Versha, Oshima}. Moreover, 
these two models are relatively old and both are based on Gribov-Regge theory 
\cite{Heck, Goswami}. However, it is important to have this type of study by 
using other high energy hadronic interaction models also.
 
Using all these six combinations of hadronic interaction models, few thousand 
showers of different energies are generated for $\gamma$-ray, proton 
and iron primaries. Details of the simulated sample are given in Table 
\ref{tab1a}. The energies of these primaries selected here belong to the 
typical ACT energy range of respective primaries in terms of the equivalent 
number of Cherenkov photons produced by them. The altitude of HAGAR experiment 
at Hanle (longitude: 78$^o$ 57$^\prime$ 51$^{\prime\prime}$ E, 
latitude: 32$^o$ 46$^\prime$ 46$^{\prime\prime}$ N, altitude: 4270 m) 
\cite{Versha1} is used as the observational level in the generation of all 
these showers. However, to demonstrate the altitude 
effect in our study, we have also generated the showers  
using QGSJET-GHEISHA combination only over the Pachmarhi observation level 
(longitude: 78$^o$ 26$^\prime$ E, 
latitude: 22$^o$ 28$^\prime$ N,  altitude: 1075 m), the site of 
PACT experiment \cite{Majumdar}. Apart from all these showers, for  specific 
purposes mentioned in the concerned sections, some more showers are generated 
for the Hanle observation level by using the VENUS-GHEISHA combination also.
\begin{center}
\begin{table}[ht]
\caption{\label{tab1a} Number of generated showers for different primaries at
different energies.}
\begin{tabular}{ccc}\hline
Primary particle & ~~~~Energy & ~~~~Number of Showers \\\hline
$\gamma$-ray  &  100 GeV  & 10000 \\
              &  500 GeV  &  5000 \\
              &  1 TeV    &  2000 \\
  && \\
 Proton       &  250 GeV  & 10000 \\
              &  1 TeV  &  5000 \\
              &  2 TeV    &  2000 \\
  && \\
 Iron        &  5 TeV     & 5000 \\
             & 10 TeV     & 2000 \\
\hline  
\end{tabular}
\end{table}
\end{center}  

As we have used the QGSJET01 high energy hadronic interaction model in our 
study, it is important to compare the output of QGSJET01 with that of 
QGSJETII. For this purpose, we have also generated 2000 showers each for 1 TeV 
$\gamma$, 2 TeV proton and 10 TeV iron primary using the QGSJETII-GHEISHA model 
combination. 
  
We have taken the detector geometry as a horizontal flat detector array, where 
there are 17 telescopes in the E -- W direction with a separation of 25 m and 
21 telescopes in the N -- S direction with a separation of 20 m and the mirror
area of each telescope as 9 m$^2$. So we have an array of 17$\times$21 i.e.
357 telescopes covering an area of 400 m$\times$400 m, which is equivalent to
several 25 telescope sectors identical to PACT experiment \cite{Majumdar}. Thus we 
have selected this detector geometry keeping in view of our earlier simulation 
works \cite{Versha}. It doesn't really matter as our idea is to study
core distance dependence of parameters of vertical showers and this geometry is 
quite adequate to simulate vertical showers within 
the ACT energy range. However, it may be noted that, for inclined showers of 
ACT energy range, this geometry may not be sufficient as such showers cover 
more area depending on the angle of inclination for a given energy. The EASs 
produced by above mentioned primaries are considered to be incident vertically 
with their core at the centre of the array. For the longitudinal 
distribution of Cherenkov photons, photons are counted only in 
the step where they are emitted. Also the  Cherenkov light emission angle is 
chosen as wavelength independent. The  Cherenkov radiation produced within the 
specified bandwidth 200 -- 650 nm by the charged secondaries is propagated to 
the ground. The position and time 
(with respect to the first interaction) of each photon hitting the detector on 
the observation level are recorded. The variable bunch size option of 
Cherenkov photon is set to "5", so that the size of the data file can be 
reduced. Multiple scattering length for e$^-$ and e$^+$ is decided by the 
parameter STEPFC in EGS code \cite{Nelson} which has been set to 0.1. The 
low energy cutoff's for the particle kinetic energy is chosen for 
hadrons, muons, electrons and photons as 3.0, 3.0, 0.003 and 0.003 GeV 
respectively. The US standard atmosphere parametrized by Linsley has been used 
\cite{US}. 

\section{Analysis and Results}
For each shower, we have calculated the Cherenkov photon density at each 
detector. The arrival time of Cherenkov photons at each detector is 
calculated with respect to the first photon of the shower hitting the array. 
Since there are several photons hitting each detector, average of these
arrival times is calculated for each detector. As there are shower to 
shower fluctuations in Cherenkov photon density and arrival time at each 
detector, the variation of Cherenkov photon density and arrival time with 
core distance is obtained by calculating average values over the 
specified number of showers. To demonstrate the fluctuations of photon 
density and arrival time as a function of core distance (or 
for each detector), the ratio of their r.m.s to mean values has  been 
calculated. Moreover, to see the model dependent variations of these parameters
clearly, we have calculated their percentage relative 
deviations with respect to the corresponding parameters for a reference model  
using the following formula:
\begin{equation}
\Delta_{\xi} = \frac{\xi_{mp} - \xi_{rp}}{\xi_{rp}}\times 100\%,
\label{eq1}
\end{equation}                 
where $\Delta_{\xi}$ is the percentage relative deviation of the parameter, 
$\xi_{rp}$ is the reference model parameter and $\xi_{mp}$ is the given model 
parameter. Different features of Cherenkov photon density and arrival time 
distributions, and their model dependent behaviours are discussed in 
the following subsections:       
\begin{figure*}[hbt]
\centerline
\centerline{\includegraphics[width=5.4cm, height=4cm]{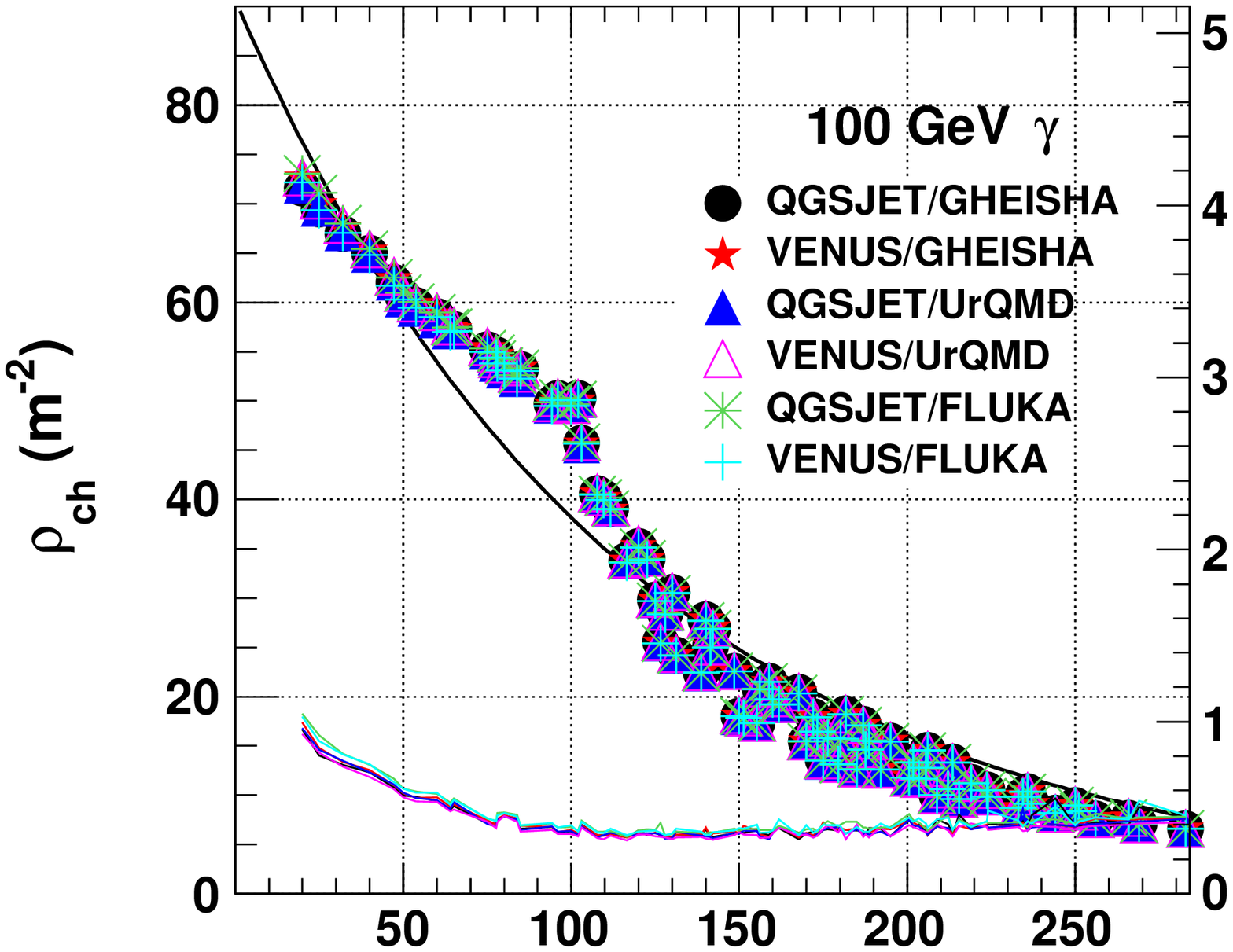}
\includegraphics[width=4.8cm, height=4cm]{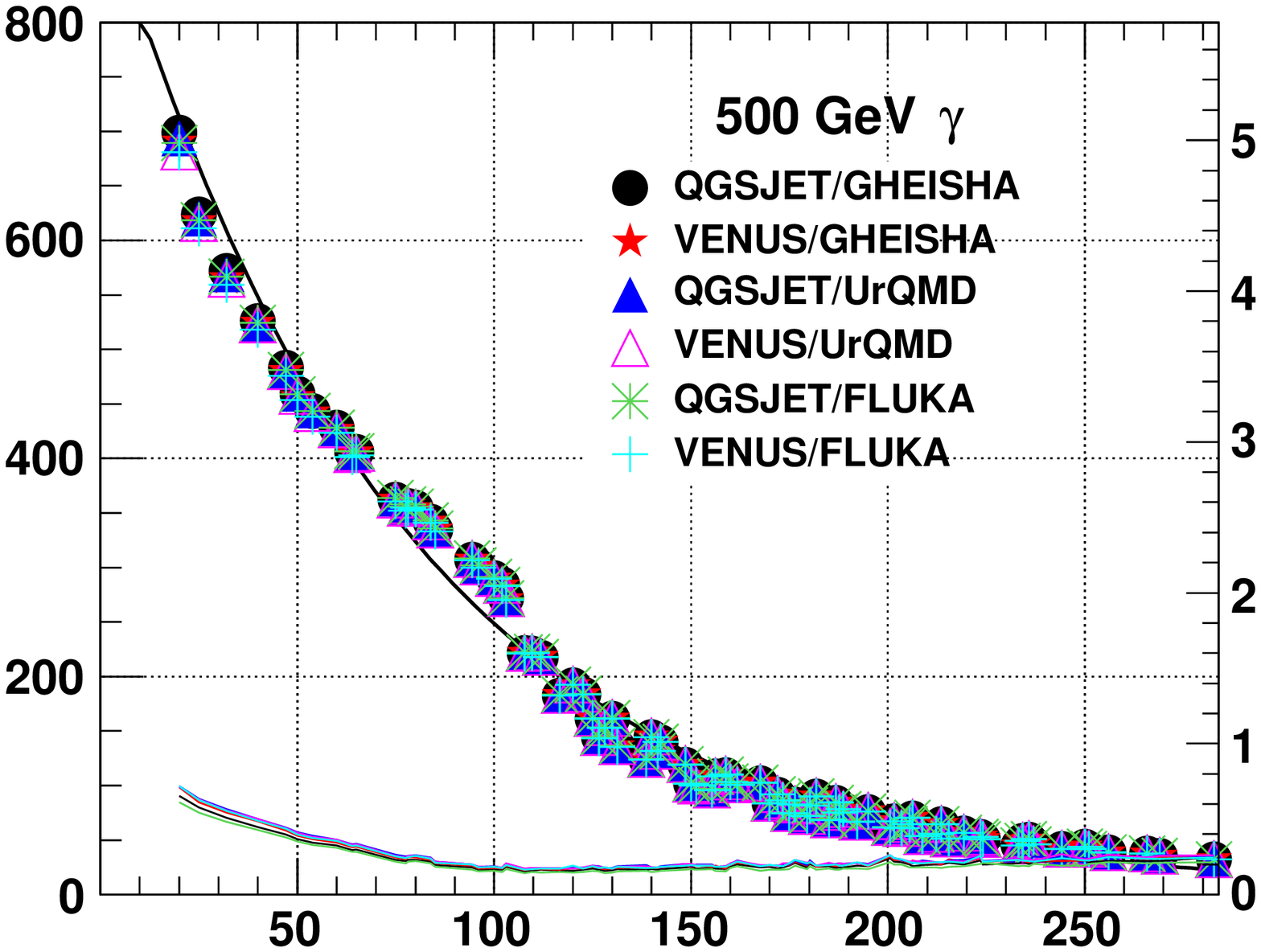}
\includegraphics[width=5.4cm, height=4cm]{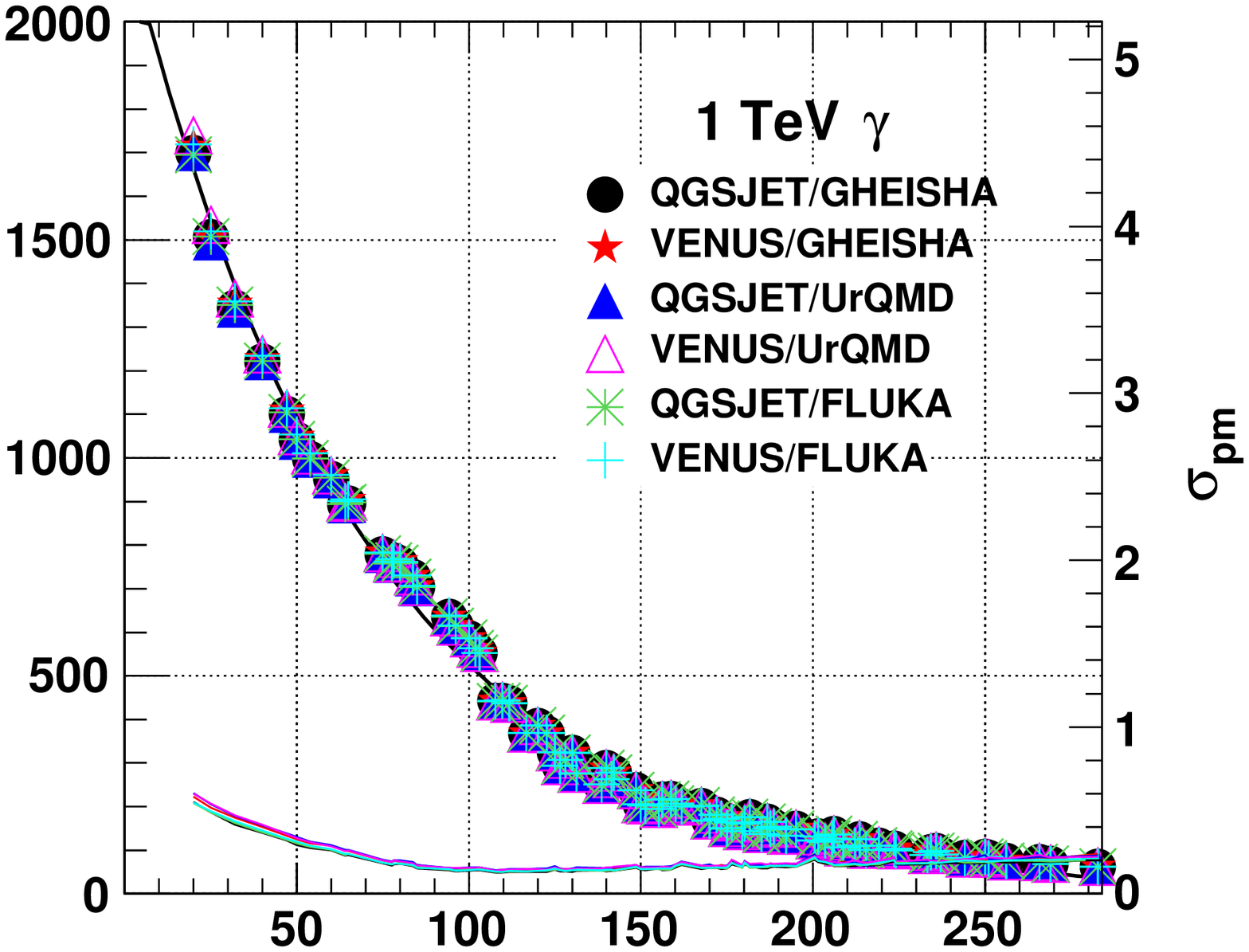}}
\centerline{\includegraphics[width=5.4cm, height=4cm]{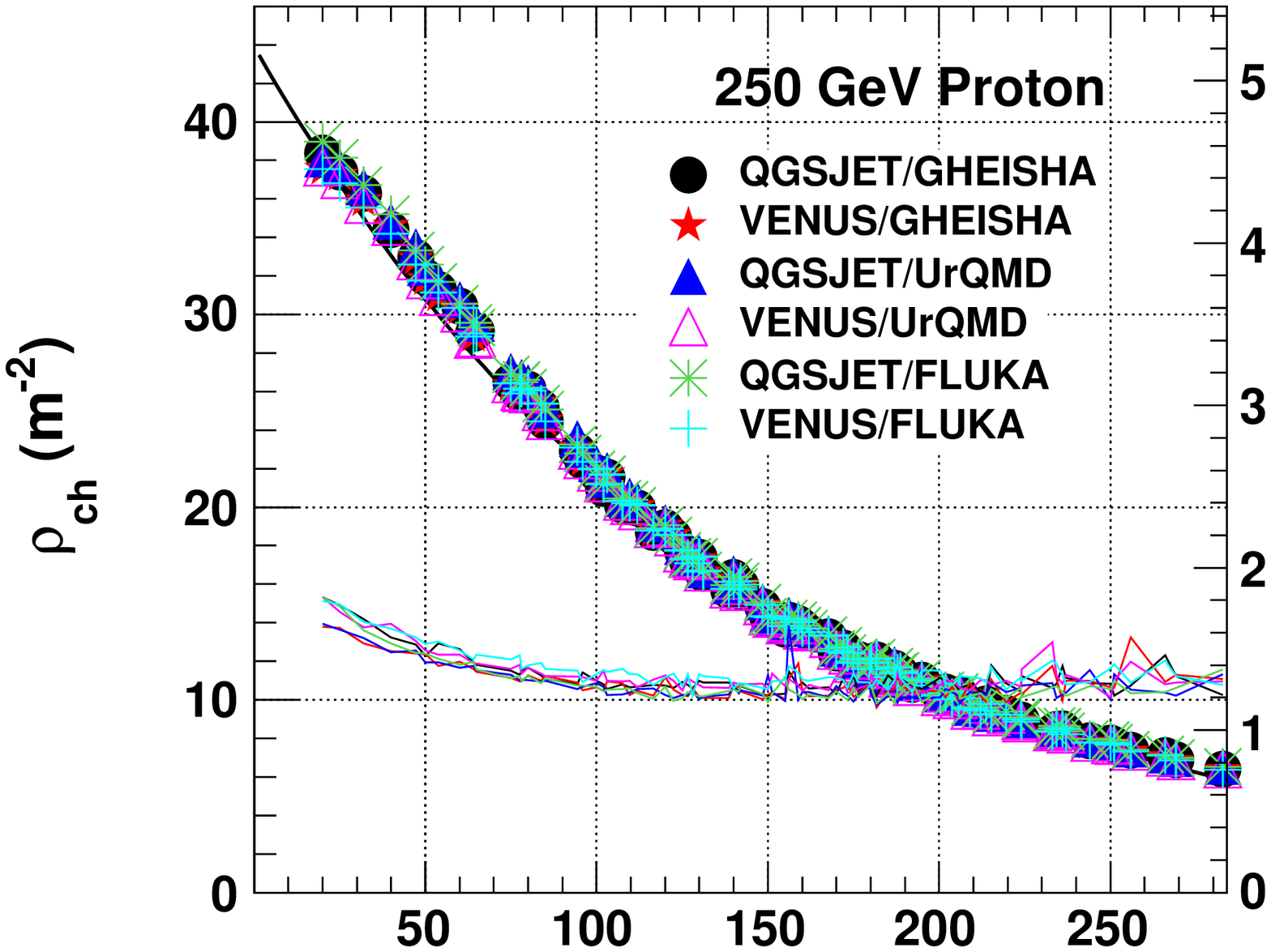}
\includegraphics[width=5cm, height=4cm]{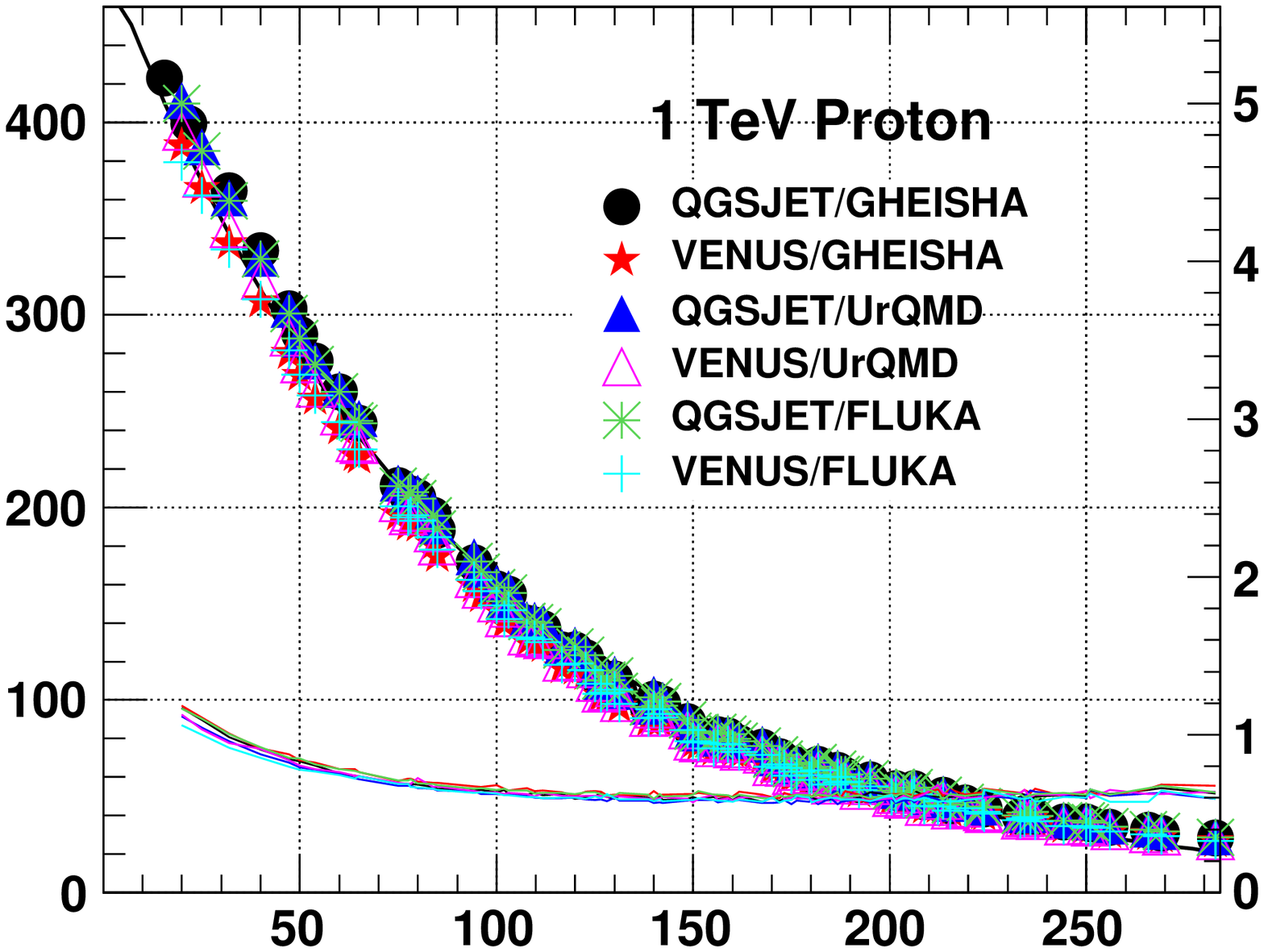}
\includegraphics[width=5.4cm, height=4cm]{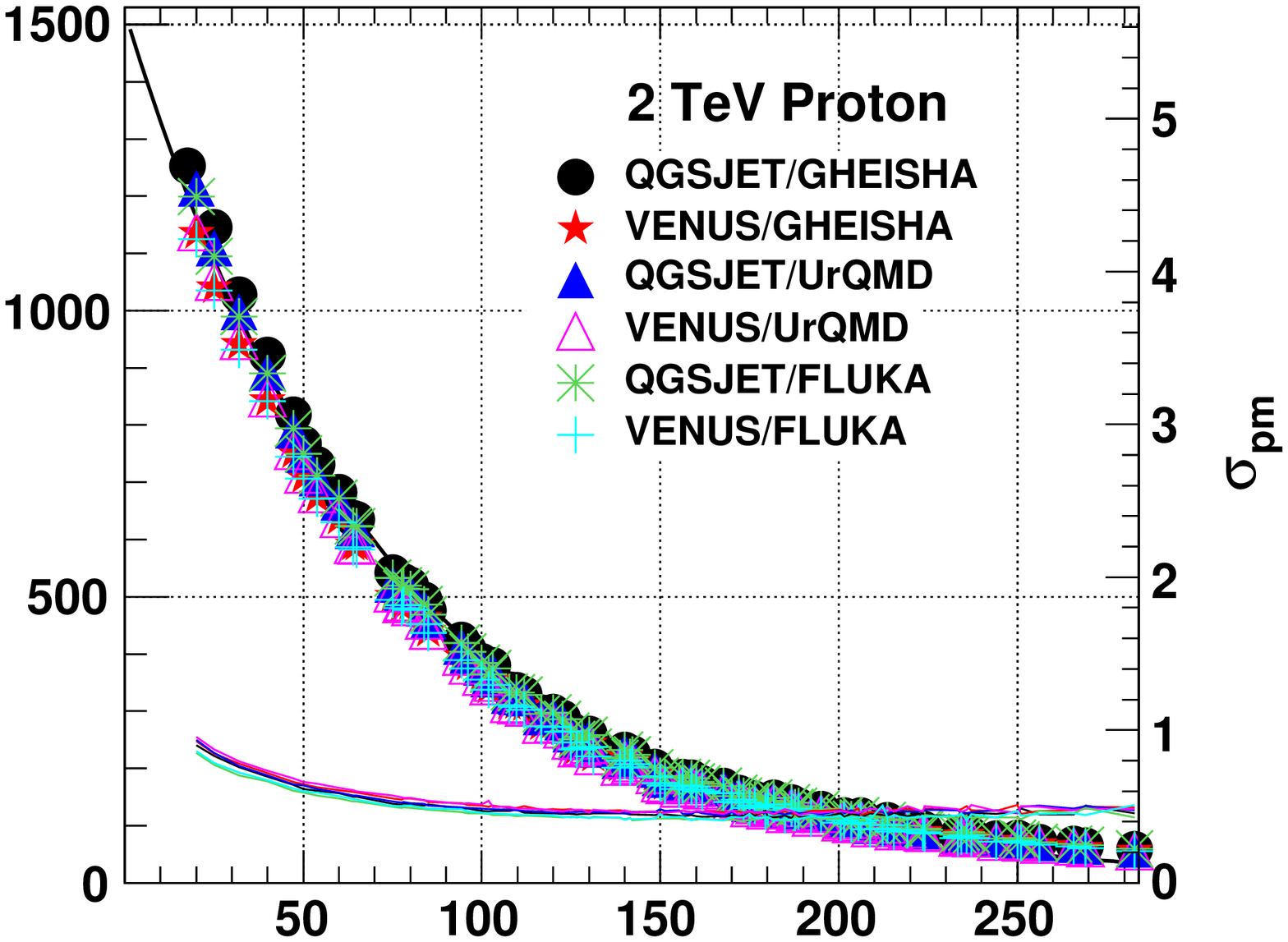}}
\centerline{\includegraphics[width=5.6cm, height=4.5cm]{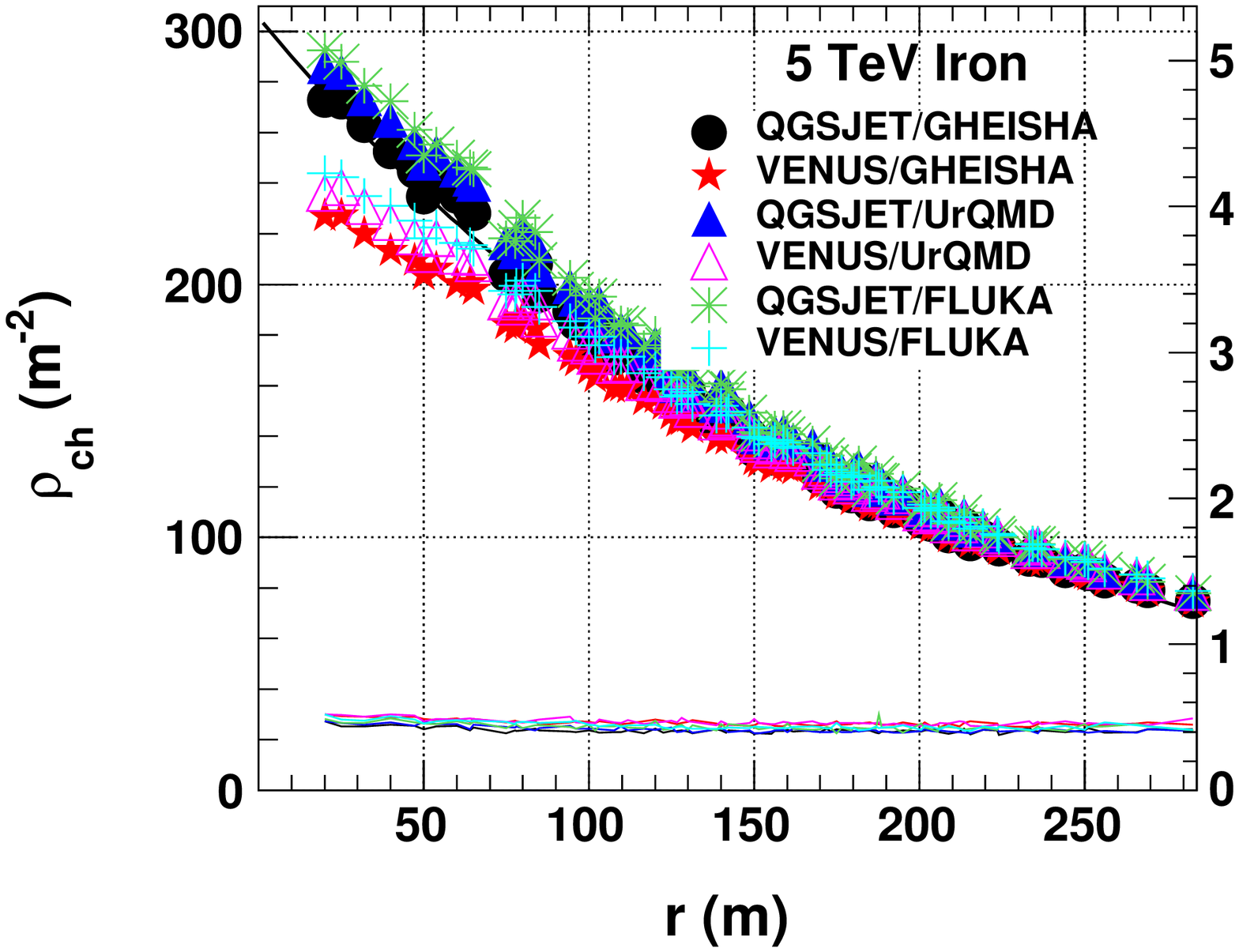}
\includegraphics[width=5.6cm, height=4.5cm]{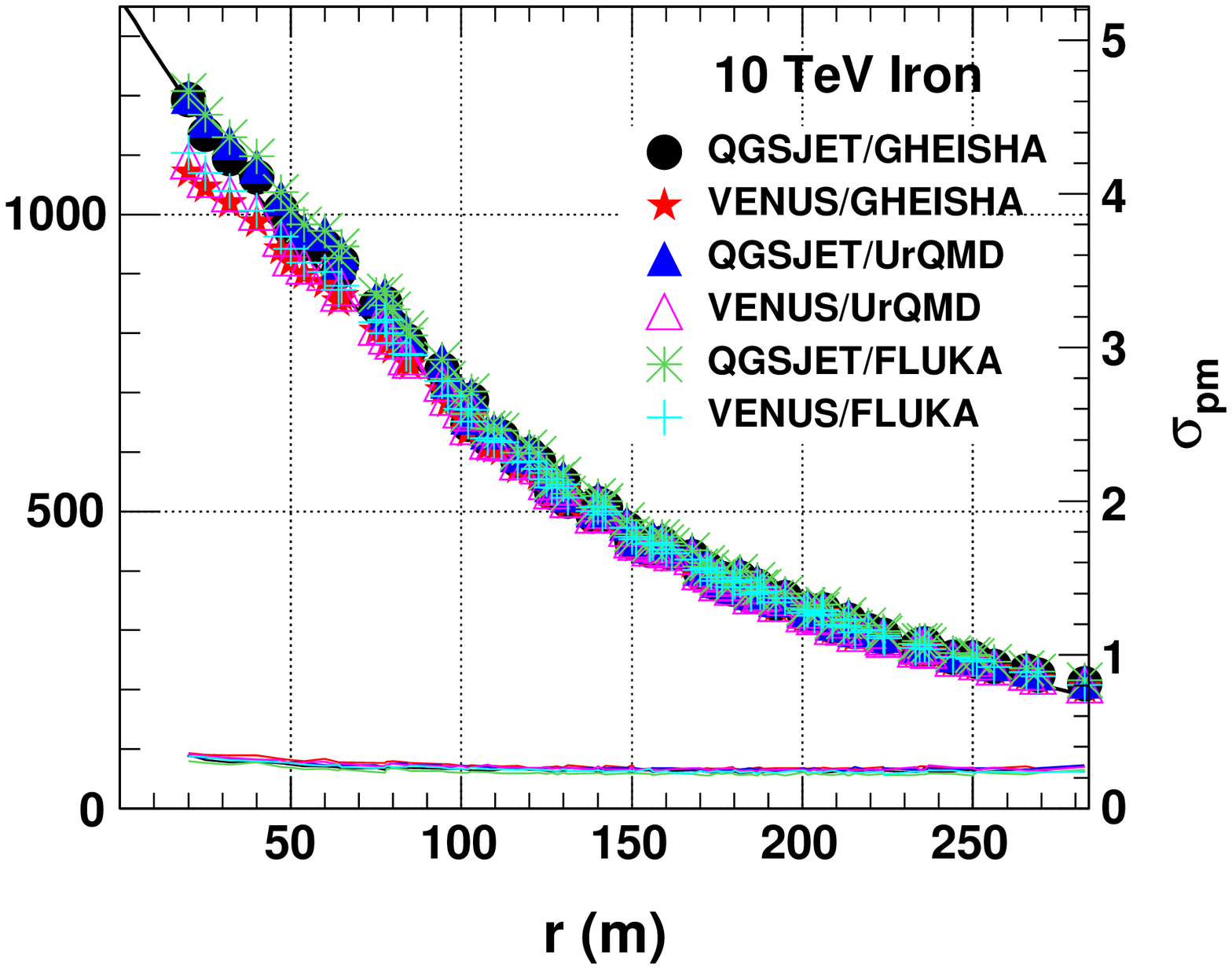}}
\caption{Variations in the average Cherenkov photon densities ($\rho_{ch}$)  
and their r.m.s. values per mean ($\sigma_{pm}$) with respect to distance from 
the shower core of different primaries for different high and low energy 
hadronic interaction models. The solid lines in respective plots indicate the 
results of the best fit function (\ref{eq2}) for the QGSJET-GHEISHA combination.
Fits are done by using the $\chi^2$-minimization method in the ROOT software 
\cite{Root} platform.}
\label{fig1}
\end{figure*}

\subsection{Cherenkov photon density}
\subsubsection{General feature}
Fig.\ref{fig1} shows the variations of average Cherenkov photon densities 
($\rho_{ch}$s) and their r.m.s. values per mean ($\sigma_{pm}$) as a function 
of core distance for different combinations of low and high energy hadronic 
interaction models and for different shower primaries. It is seen that, the 
distribution of $\rho_{ch}$s 
falls off exponentially with core distance for all hadronic interaction model 
combinations, primary particles and their energies, as given by the equation
\begin{equation}
\rho_{ch}(r) = \rho_{0}\;e^{-\beta r},
\label{eq2}
\end{equation}
where $\rho_{ch}(r)$ is the position dependent density function of Cherenkov 
photons, $r$ the distance from the shower core, $\rho_{0}$ the coefficient and 
$\beta$  the slope of the function. Values of $\rho_{0}$ and $\beta$ are 
different for different primaries. As an example, the best fit negative 
exponential functions for QGSJET-GHEISHA hadronic interaction model 
combination are indicated by solid lines in the plot. These fits are done by 
using the $\chi^2$-minimization method in the ROOT software \cite{Root} 
platform. It needs to be mentioned that for 100 GeV $\gamma$-ray primary, 
the fit is not good due to the presence of hump at a 
core distance of about 100 m. Although the distributions follow the same 
mathematical function almost for all the cases with different coefficients and
slopes, the geometry of the distributions is  different for different primaries
at a particular energy. It is seen that at a given energy, the distribution 
shows larger curvature for the $\gamma$-ray primary (except for 100 GeV) with 
higher values of coefficient and slope of the 
exponential function with a parameter $\beta$ smaller for iron primaries than 
for protons and increasing with energy.
\begin{figure*}[hbt]
\centerline
\centerline{\includegraphics[width=5.5cm, height=4cm]{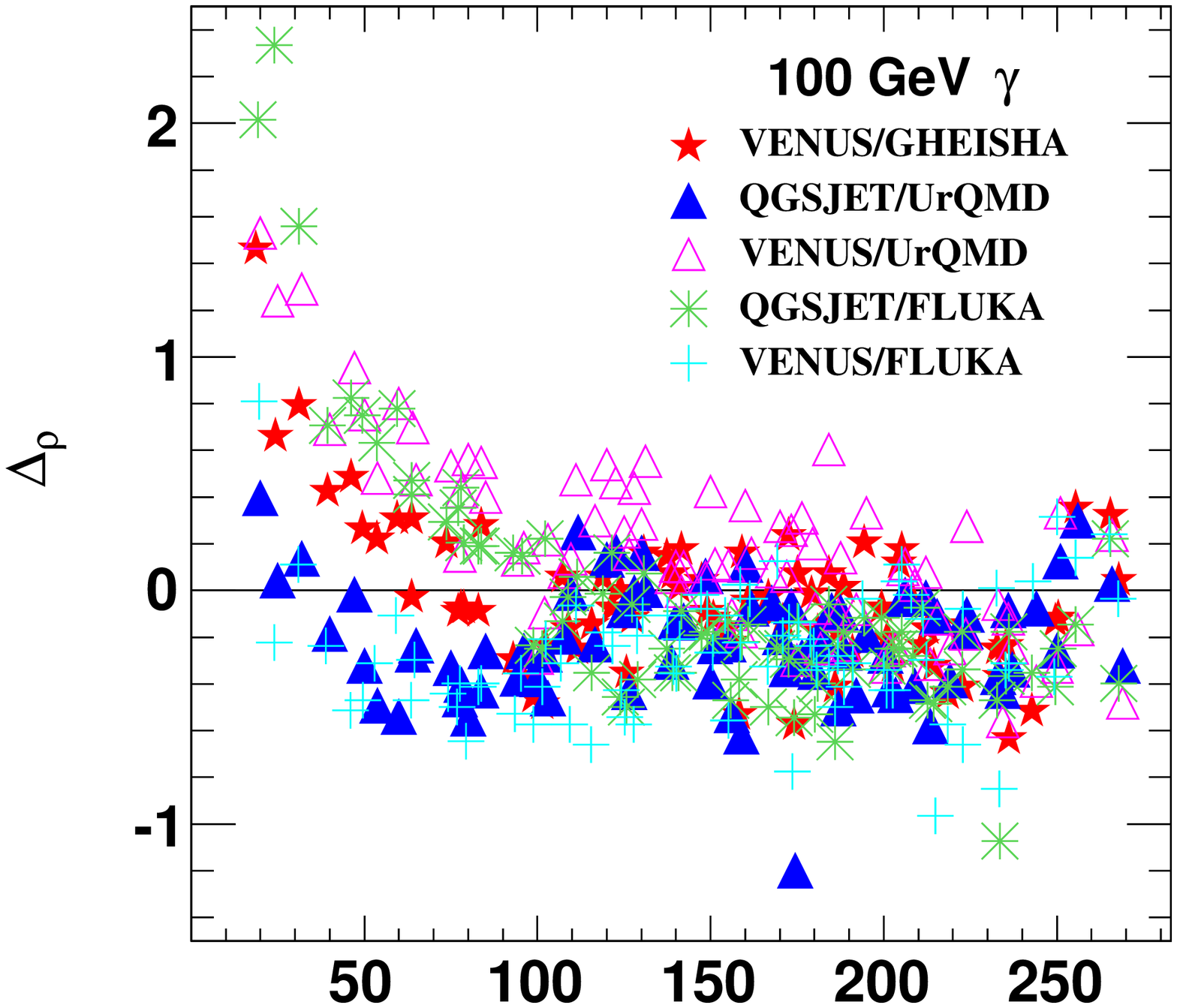}
\includegraphics[width=5cm, height=4cm]{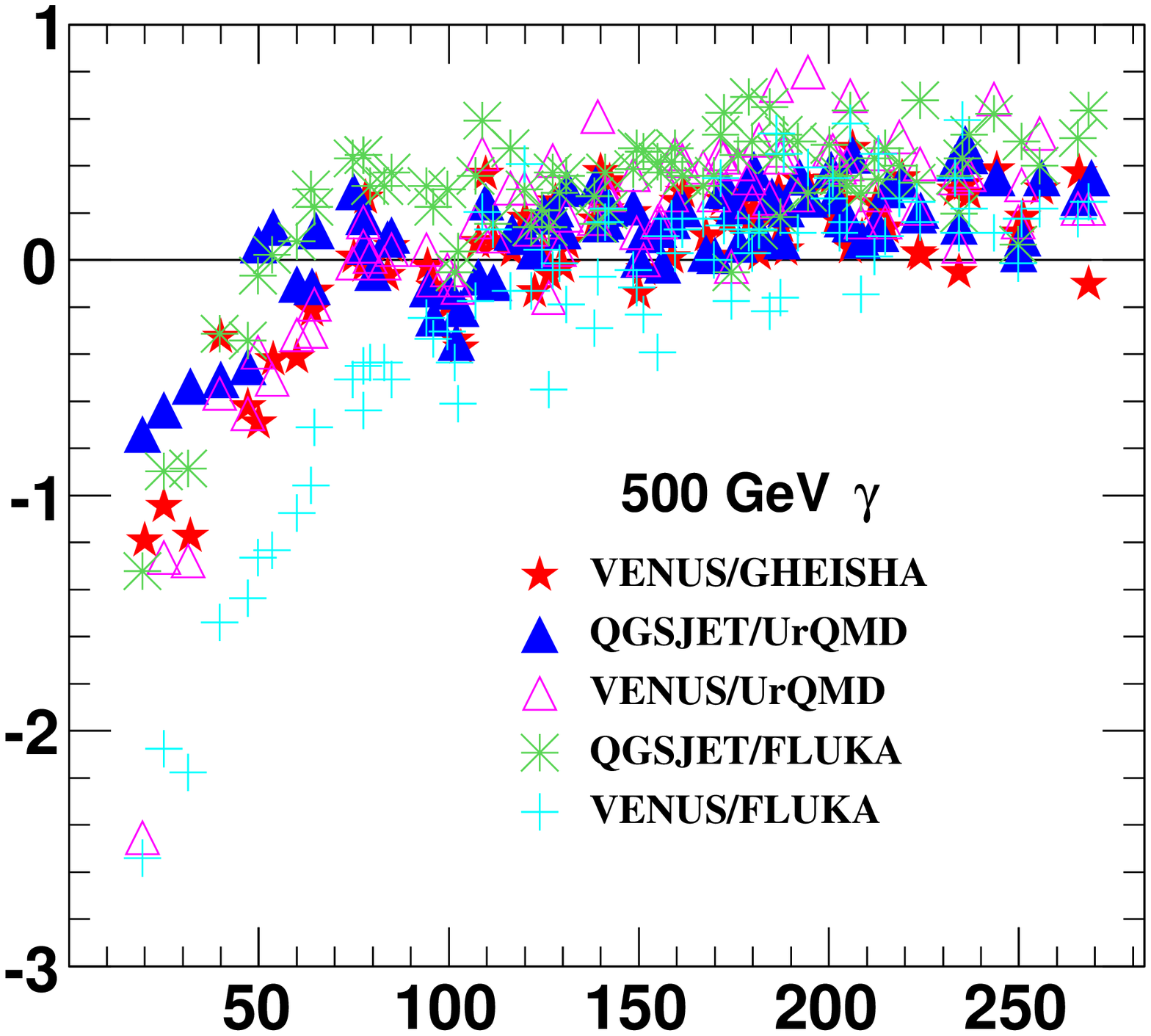}
\includegraphics[width=5cm, height=4cm]{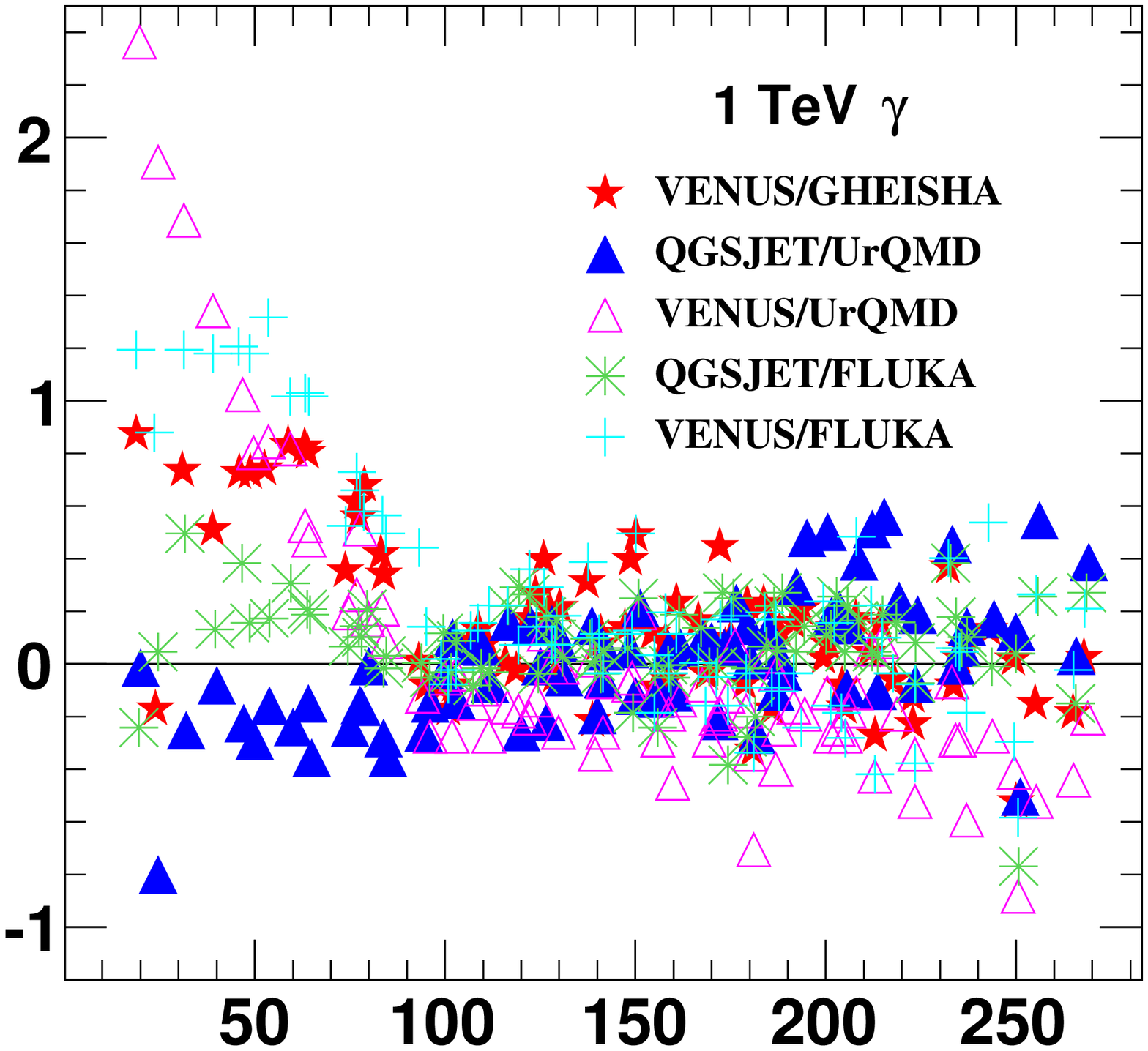}}
\centerline{\includegraphics[width=5.5cm, height=4cm]{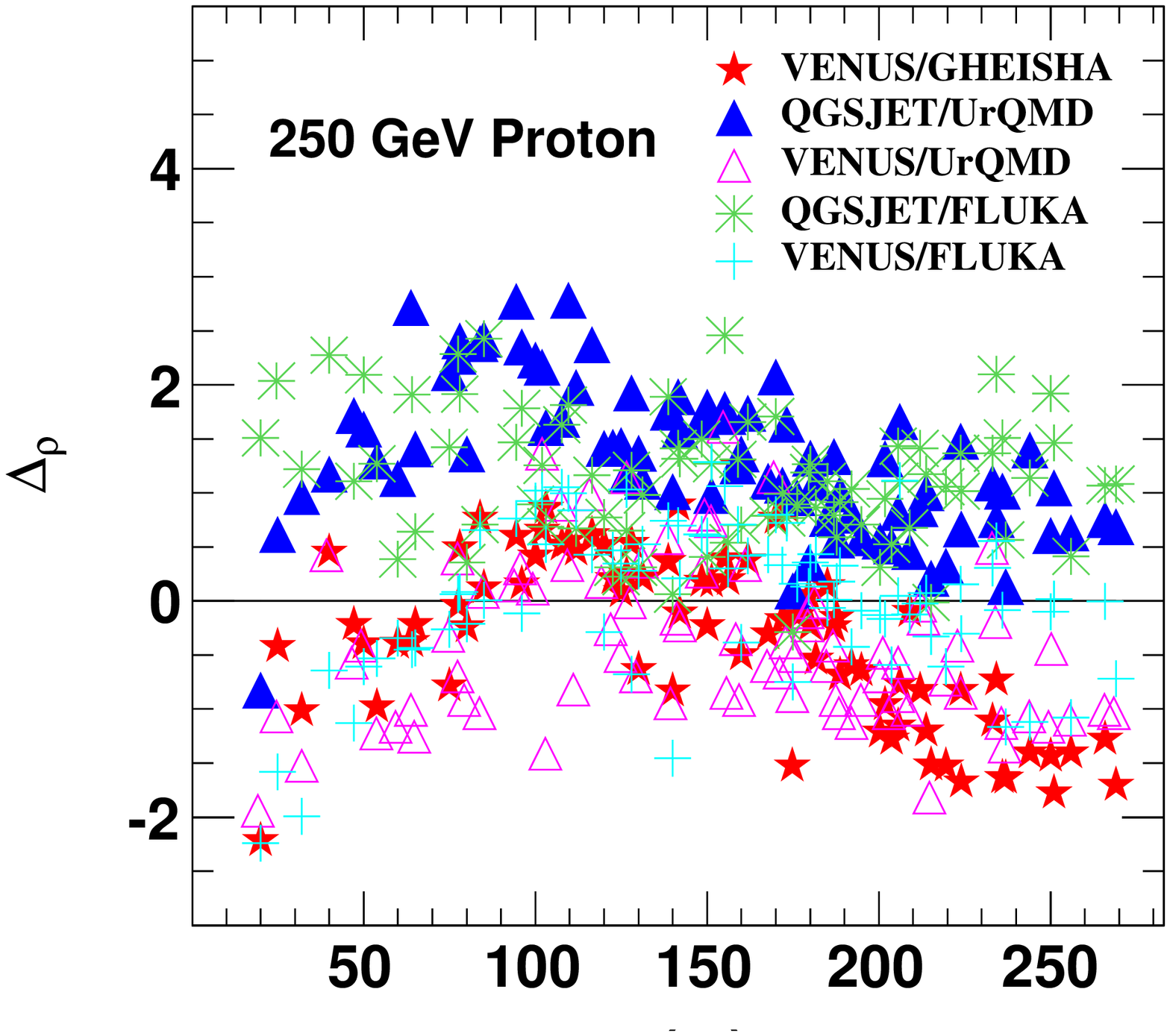}
\includegraphics[width=5cm, height=4cm]{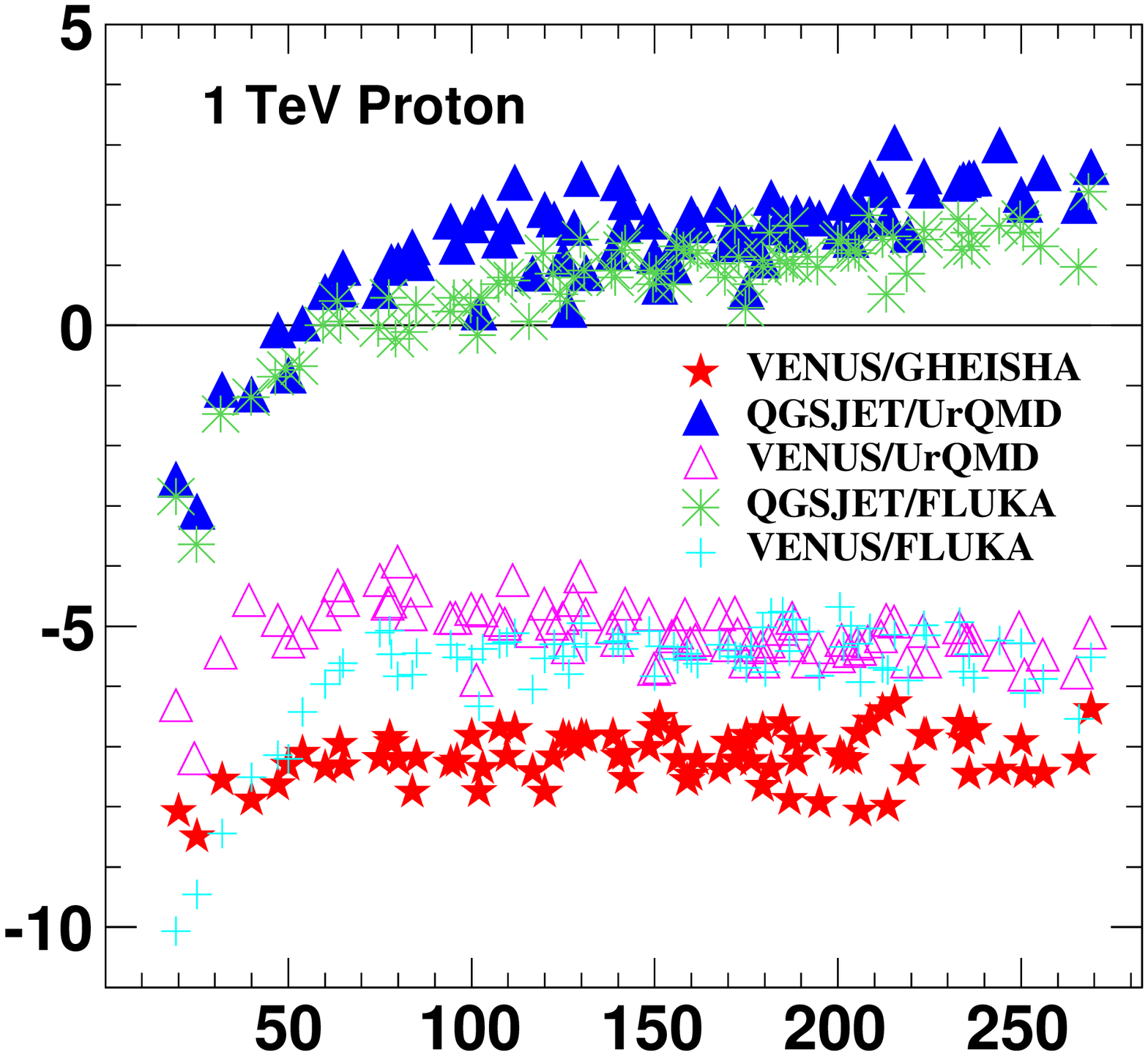}
\includegraphics[width=5cm, height=4cm]{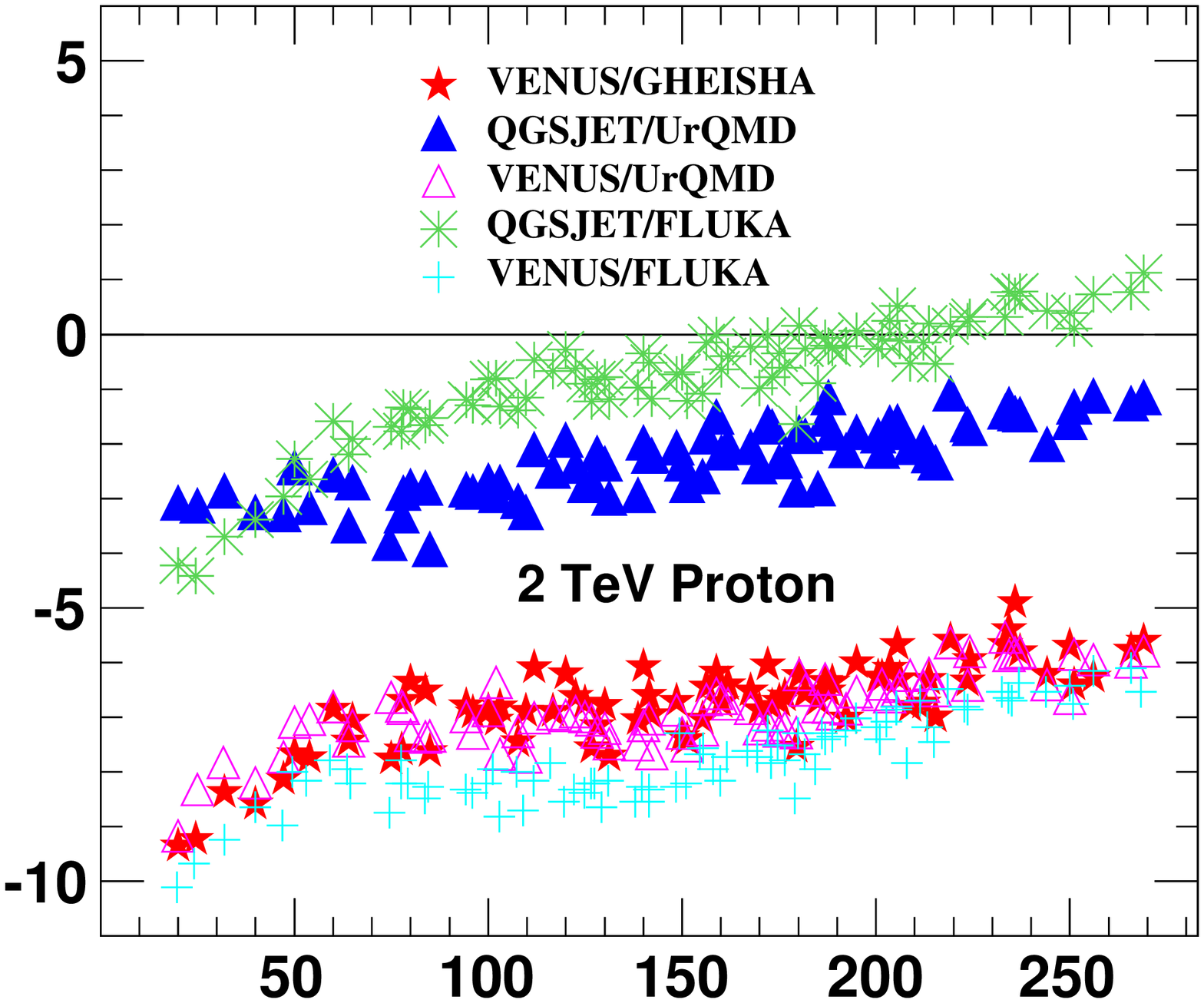}}
\centerline{\includegraphics[width=5.5cm, height=4.5cm]{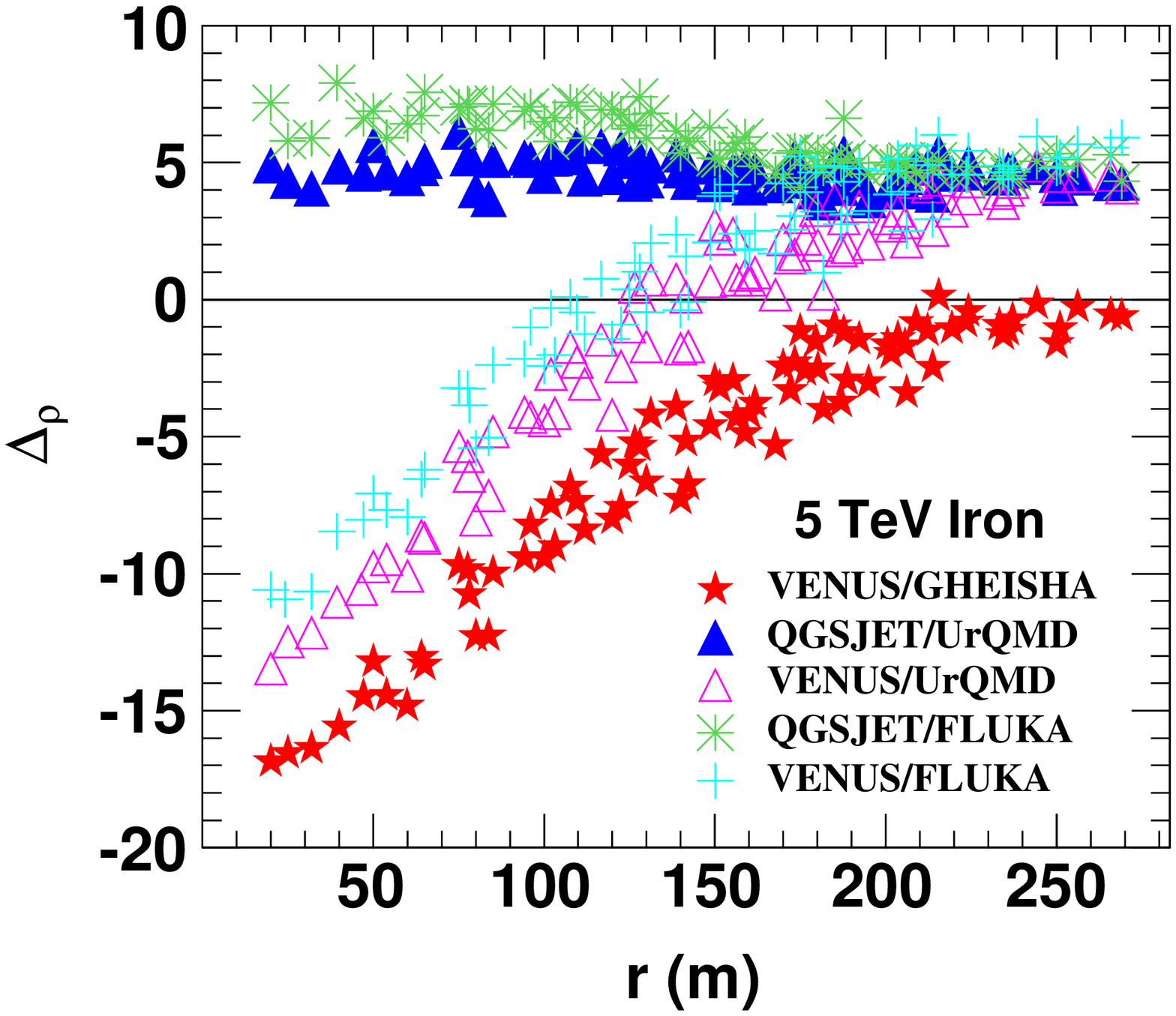}
\includegraphics[width=5cm, height=4.5cm]{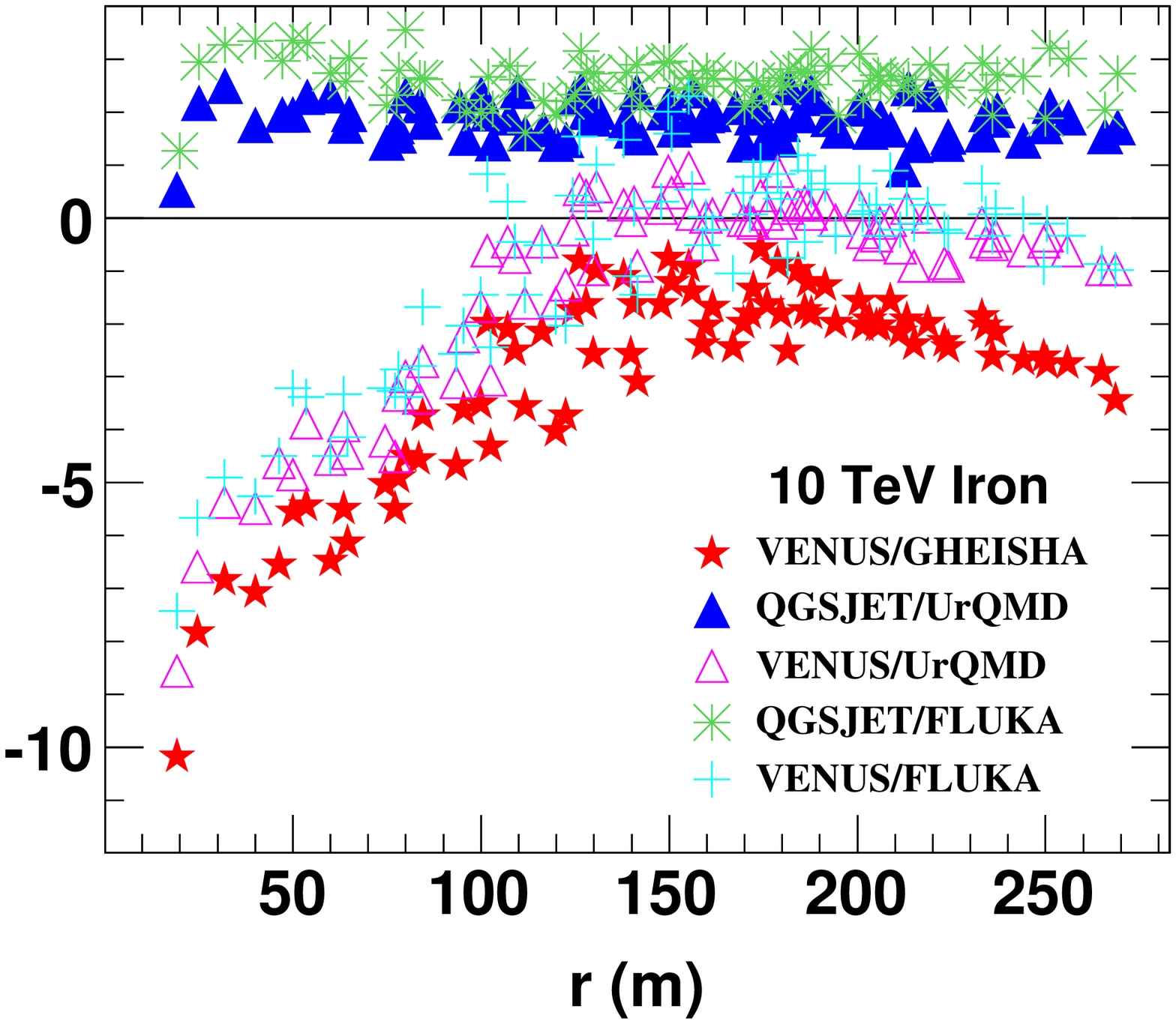}}
\caption{Variations of \% relative Cherenkov photon densities
($\Delta_{\rho}$s) with respect to
distance from the shower core of different primaries for different high and
low energy hadronic interaction models. The horizontal solid lines in all
plots indicate the QGSJET-GHEISHA model combination, which is considered as
the reference for the calculation.}
\label{fig2}
\end{figure*}
\subsubsection{Dependence on hadronic interaction model}
It is seen from the Fig.\ref{fig1} that, there is almost no visible effect of 
hadronic interaction models on the $\rho_{ch}$ distributions for the 
$\gamma$-ray primaries of energies upto 1 TeV. For proton primaries of 
energies upto 2 TeV, some slight discrepancies in densities are observed near 
the shower core upto the distance of $\le$100 m, for different model 
combinations. In the case of iron primaries, the clear effect of hadronic 
interaction models is seen depending on the energy of the primary particle. In 
this case, QGSJET lead group of models (i.e. QGSJET-GHEISHA, QGSJET-UrQMD and 
QGSJET-FLUKA) have generated higher $\rho_{ch}$s than those 
generated by the VENUS lead group of models (i.e. VENUS-GHEISHA, VENUS-UrQMD 
and VENUS-FLUKA), in the region near shower core, depending on the energy of 
the primary particle. The magnitude of this difference and the effective core 
distance over which differences are seen is higher for the low energy primary 
particle. 

To quantify the effect of different interaction models, we have
calculated the \% relative deviation of $\rho_{ch}$ ($\Delta_{\rho}$) 
for different model combinations taking QGSJET-GHEISHA as reference. Results 
are shown in the Fig.\ref{fig2}. This choice of reference is arbitrary and it
doest not effect the results as our aim is to see the effect of one model 
combination over the other.    
From this figure it is clear that, for the $\gamma$-ray primaries at all 
energies of our interest, there are no significant differences due to 
interaction models. Very near to the shower core ($<$50 m) some model 
combinations show deviations of $\sim$3\% and beyond this distance the density 
deviations of all model combinations are within $\sim$$\pm$1\%. 

For the case of proton primary, the deviations from the reference 
combination are significant in comparison to $\gamma$-ray primary. The QGSJET 
lead group of models show deviations in densities within $\sim$$\pm$5\%. At 
higher primary energies these deviations become increasingly negative. Within 
this group QGSJET-UrQMD 
combination gives maximum deviation in densities almost at all distances 
from the shower core for 1 TeV and 2 TeV primaries. Whereas for the VENUS 
lead group of models the density deviations are very large in comparison to 
deviations for QGSJET lead group of models. However, for the 250 GeV proton 
these deviations are insignificant as they are only within $\sim$$\pm$2\%. On 
the other hand, for 1 TeV and 2 TeV 
proton primaries these deviations are negative: $\sim$-4\% to -10\% for 1 
TeV and $\sim$-6\% to -10\% for 2 TeV. Within this group the maximum deviation
is shown by different model combinations depending on the energy and the 
distance from the shower core.  
 
Finally, in the case of iron primaries, the deviations are considerable 
for both types of model combinations. 
Unlike in the case proton primaries, the deviation is higher at lower 
primary energy.  However, in this case the deviations for the QGSJET lead 
group of models are always positive and remain almost steady over all 
core distances. These deviations are $\sim$4 -- 8\% for the 5 TeV primary 
and $\sim$1 -- 5\% for the 10 TeV primary. On an average, here the maximum 
deviation is given by the QGSJET-FLUKA combination within QGSJET led group. 
The density deviations for VENUS lead group of models in this case vary from 
negative to positive values with core distance. For higher energies,
deviations again become negative for higher core distances. This tendency
is highest for VENUS-GHEISHA combination. For the 5 TeV primary deviations 
vary from $\sim$-17\% to $\sim$6\% with core distance whereas it is 
$\sim$-10\% to $\sim$2\% for the 10 TeV primary. On an average the 
VENUS-GHEISHA combination shows highest density deviations from the 
QGSJET-GHEISHA combination. 

\subsubsection{Behaviour of fluctuations} 
The values of $\sigma_{pm}$ of $\rho_{ch}$ for the $\gamma$-ray primaries 
decrease with increasing core distance upto a distance of $\sim$100 m and 
remain constant at larger core distances. This trend is seen for all energies 
and for all model combinations. This ratio decreases also with increasing 
energy (see Fig.\ref{fig1}). 
There are no considerable differences due to different 
combinations of models. Although similar trend of variation of $\sigma_{pm}$
is seen in proton primaries, the values are higher compared to $\gamma$-ray
primary for all model combinations. In case of iron primaries, $\sigma_{pm}$
is very small compared to proton primaries. For iron primaries also,
values decrease with increasing energy.  Moreover, it should be 
noted that, in this case the values of $\sigma_{pm}$ remain almost constant 
over all core distances. As for the case of $\gamma$-ray primaries, there are 
no considerable differences in $\sigma_{pm}$ due to different combinations of 
models for iron primaries.

\subsubsection{Primary energy dependence}
\begin{figure}[hbt]
\centerline
\centerline{\includegraphics[width=6cm, height=5cm]{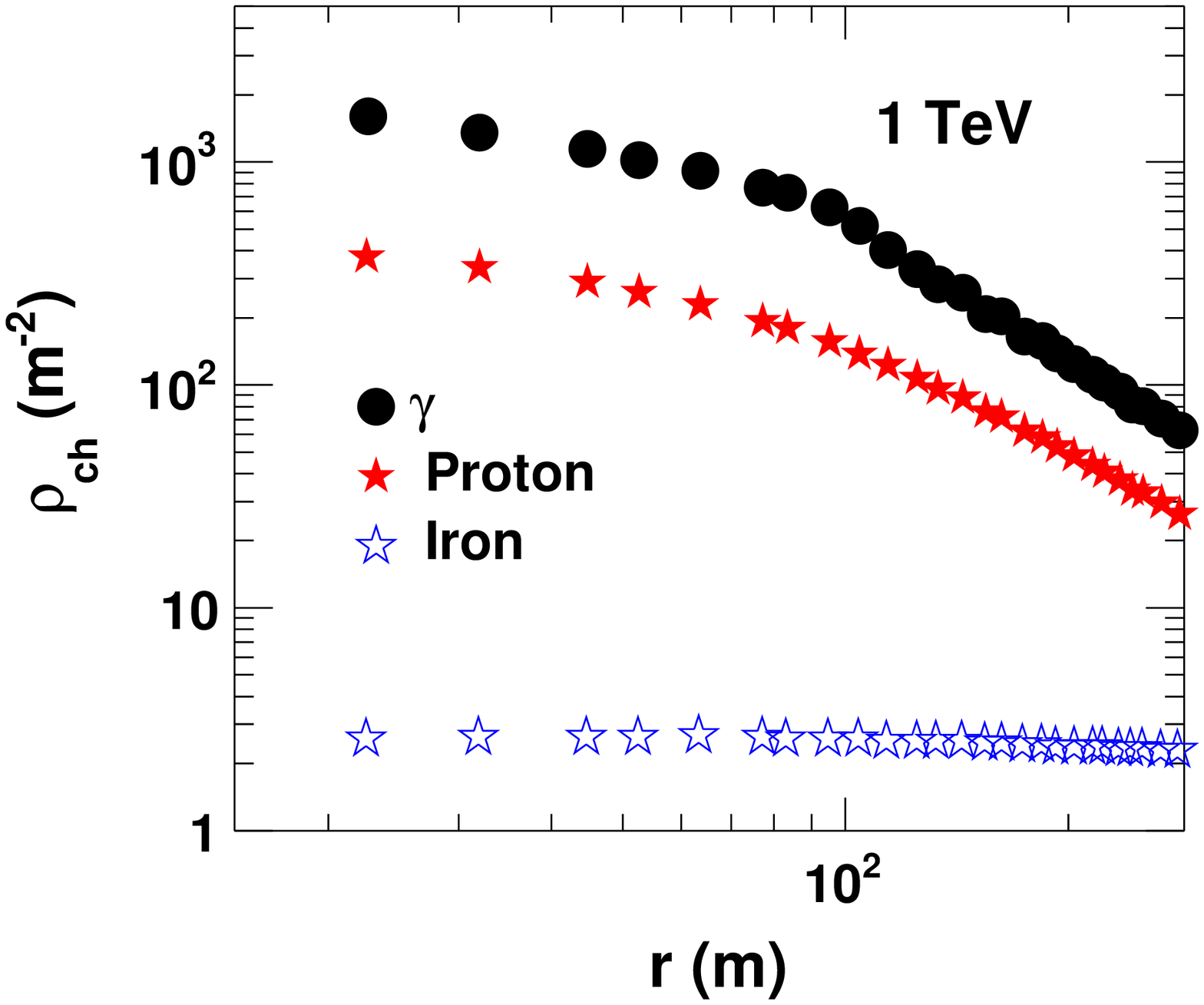}
\vspace{5mm}\\
\includegraphics[width=6cm, height=5cm]{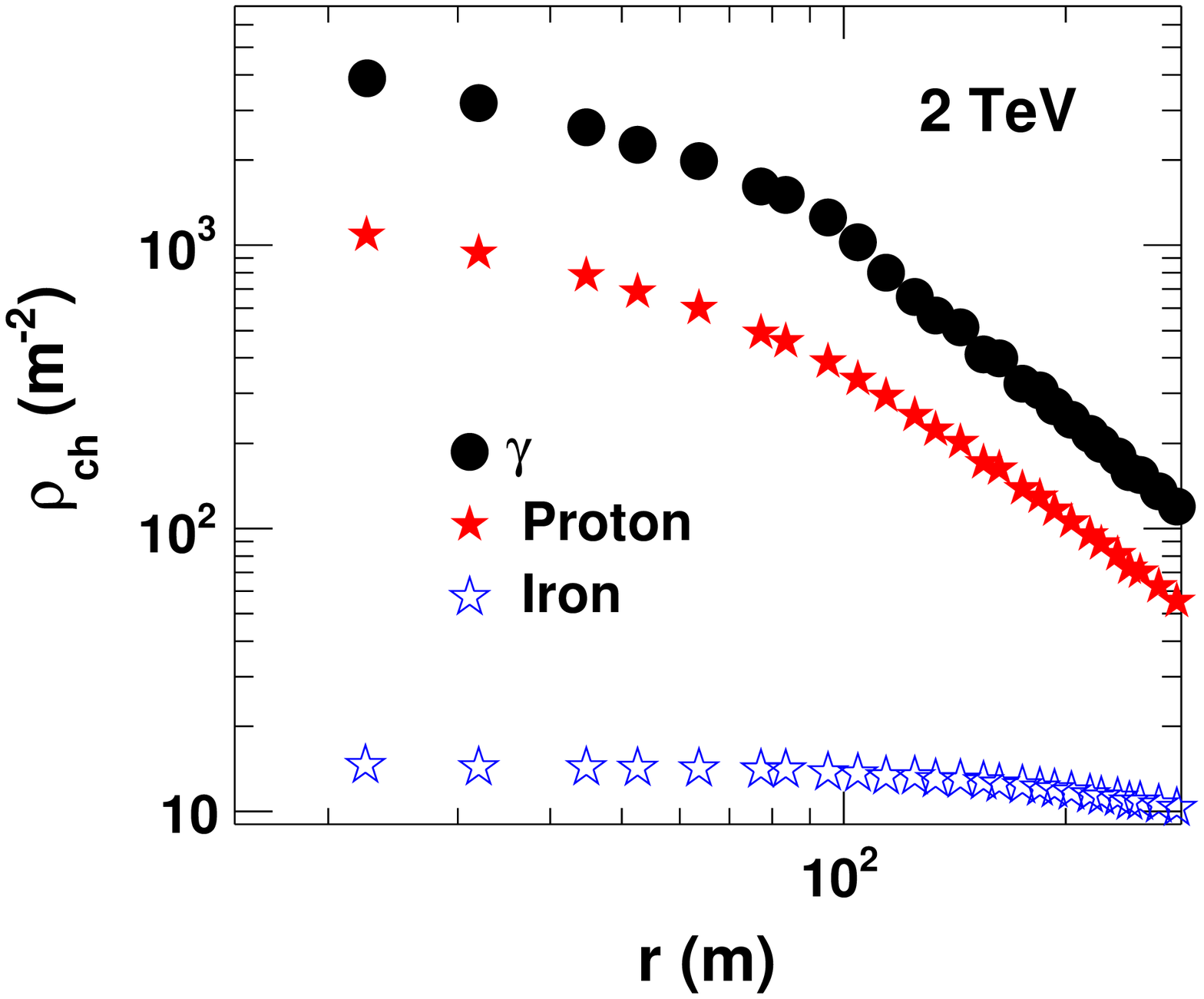}
\vspace{5mm}\\
\includegraphics[width=6.2cm, height=5cm]{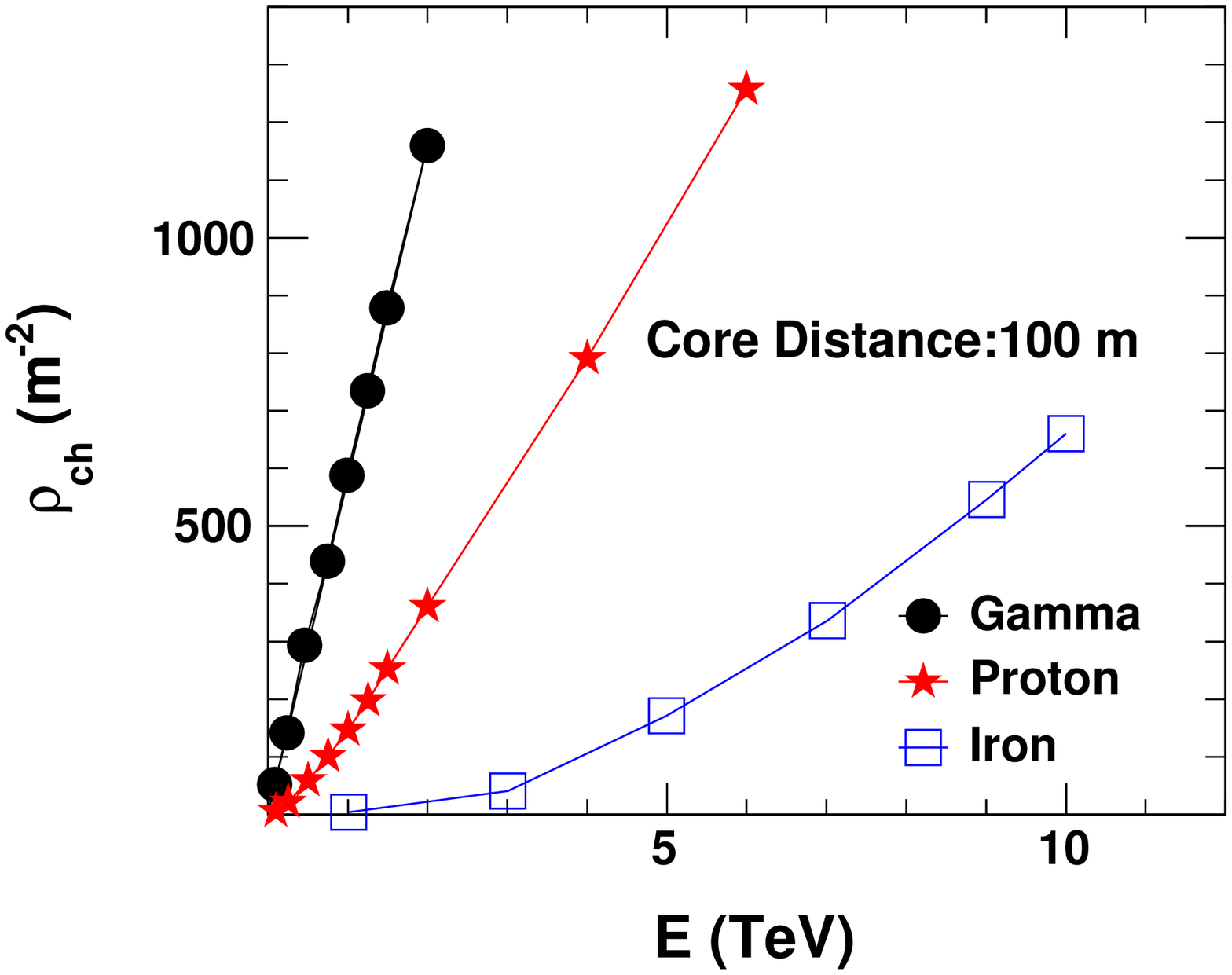}
}
\caption{Top and middle panels: Distributions of $\rho_{ch}$s with respect to 
distance from the shower core of $\gamma$, proton 
and iron primaries at 1 TeV and 2 TeV energies given by the VENUS-GHEISHA 
model combination. Bottom panel: Variations of $\rho_{ch}$s with respect to 
primary energy at 100 m from the shower core of $\gamma$, proton and iron 
primaries for the same  model combination as for the other two panels.}
\label{fig3}
\end{figure}

At a given energy, for all low and high energy hadronic interaction 
model combinations, the $\gamma$-ray primary produces maximum number of 
Cherenkov photons and the iron primary produces least number of it. To  
demonstrate this, we have plotted in the top and middle panels of the 
Fig.\ref{fig3}, the $\rho_{ch}$ distributions with respect to 
the core distance produced by the $\gamma$, proton and iron primaries of 1 TeV 
and 2 TeV energies respectively. These are obtained by using the VENUS-GHEISHA 
model combination. Moreover, in the bottom panel of this figure, we have shown 
the variation of $\rho_{ch}$ at core distance of 100 m with 
respect to energy for the all primary particles. It is seen that, for the 
$\gamma$-ray primary, the $\rho_{ch}$ increases very rapidly and linearly with 
the primary energy. Whereas in the cases of proton and iron primaries, it 
increases slowly and non-linearly with the 
primary energy. The slowness and non-linearity effect is more prominent in the 
case of iron primary. These are due to the fact that, almost all the secondary 
EAS particles produced by $\gamma$-ray primary are electrons and positrons. 
Whereas the secondary particles produced by proton and iron primaries consist 
of not only electrons and positrons but muons and other hadronic particles 
as well. Moreover, in the case of iron nuclei primary, energy per nucleon is 
56 times smaller than that for proton at a given energy. As electrons and 
positrons are the lightest particles, at a given energy more number of 
secondary particles of $\gamma$-ray primary cross threshold to produce 
Cherenkov photons compared to those produced by proton and iron primaries, 
with iron primary having the least number of such particles.               

\subsubsection{Altitude effect}
\begin{figure}[hbt]
\centerline
\centerline{\includegraphics[width=6cm, height=5cm]{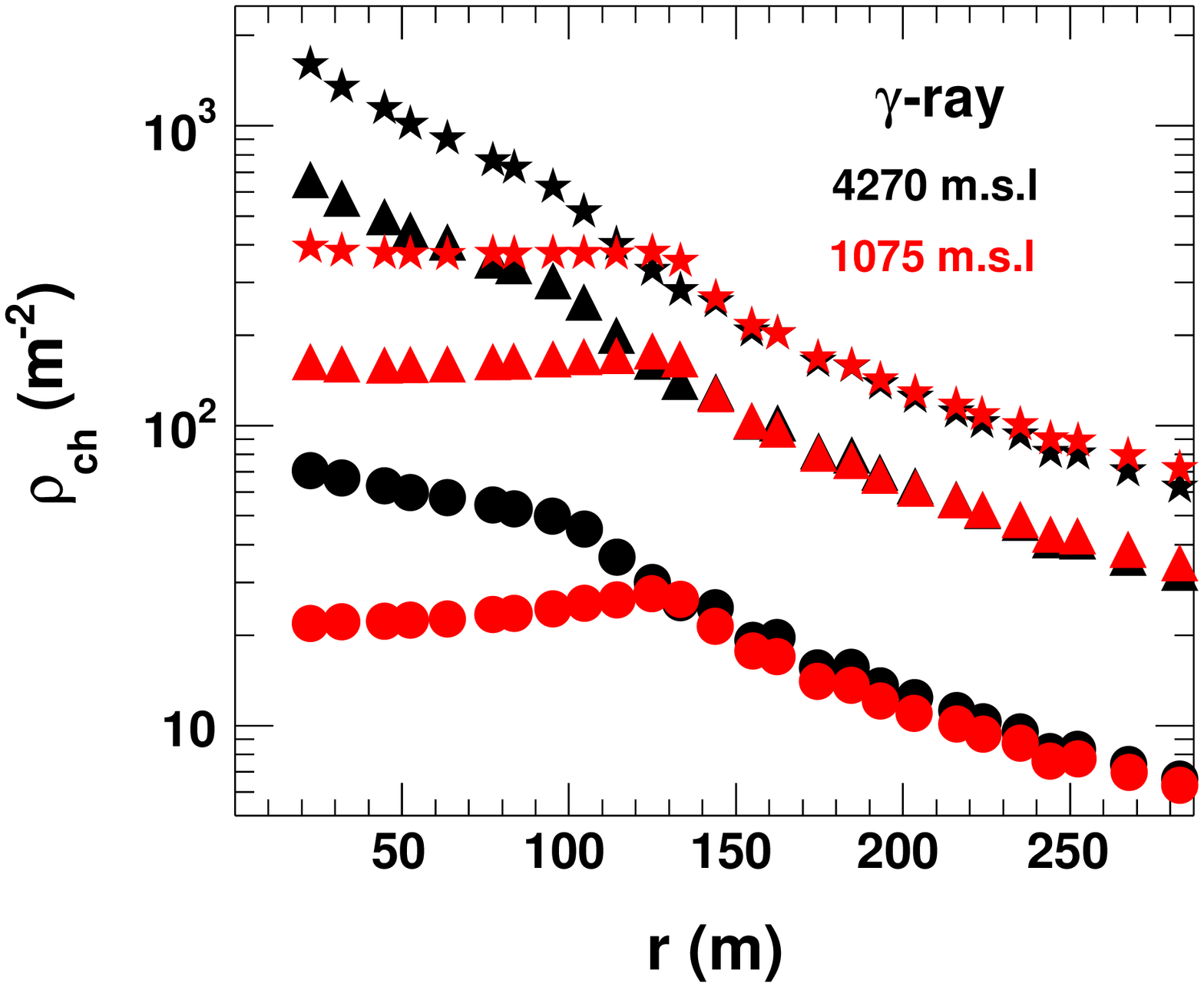}
\vspace{5mm}\\
\includegraphics[width=6cm, height=5cm]{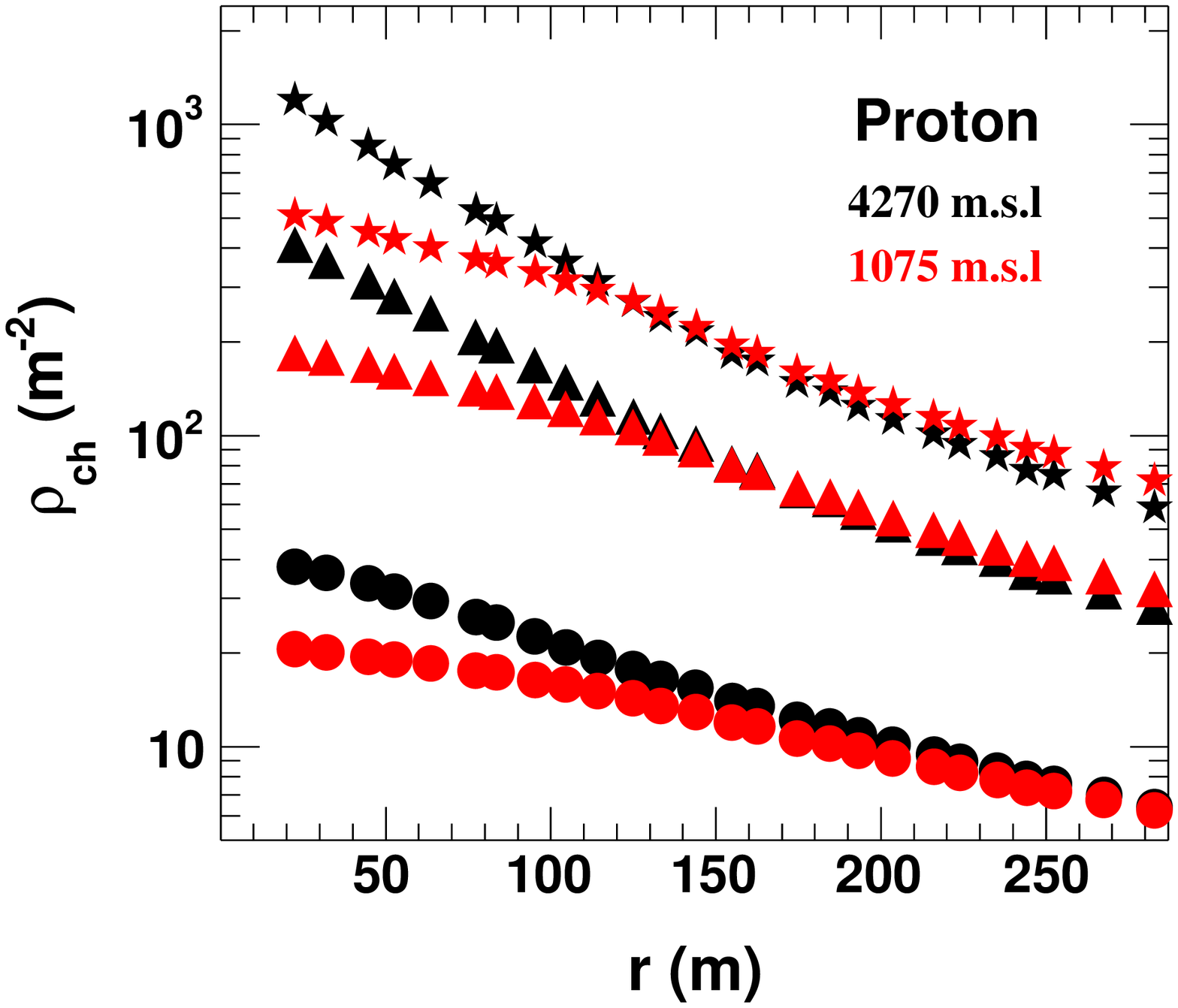}
\vspace{5mm}\\
\includegraphics[width=6cm, height=5cm]{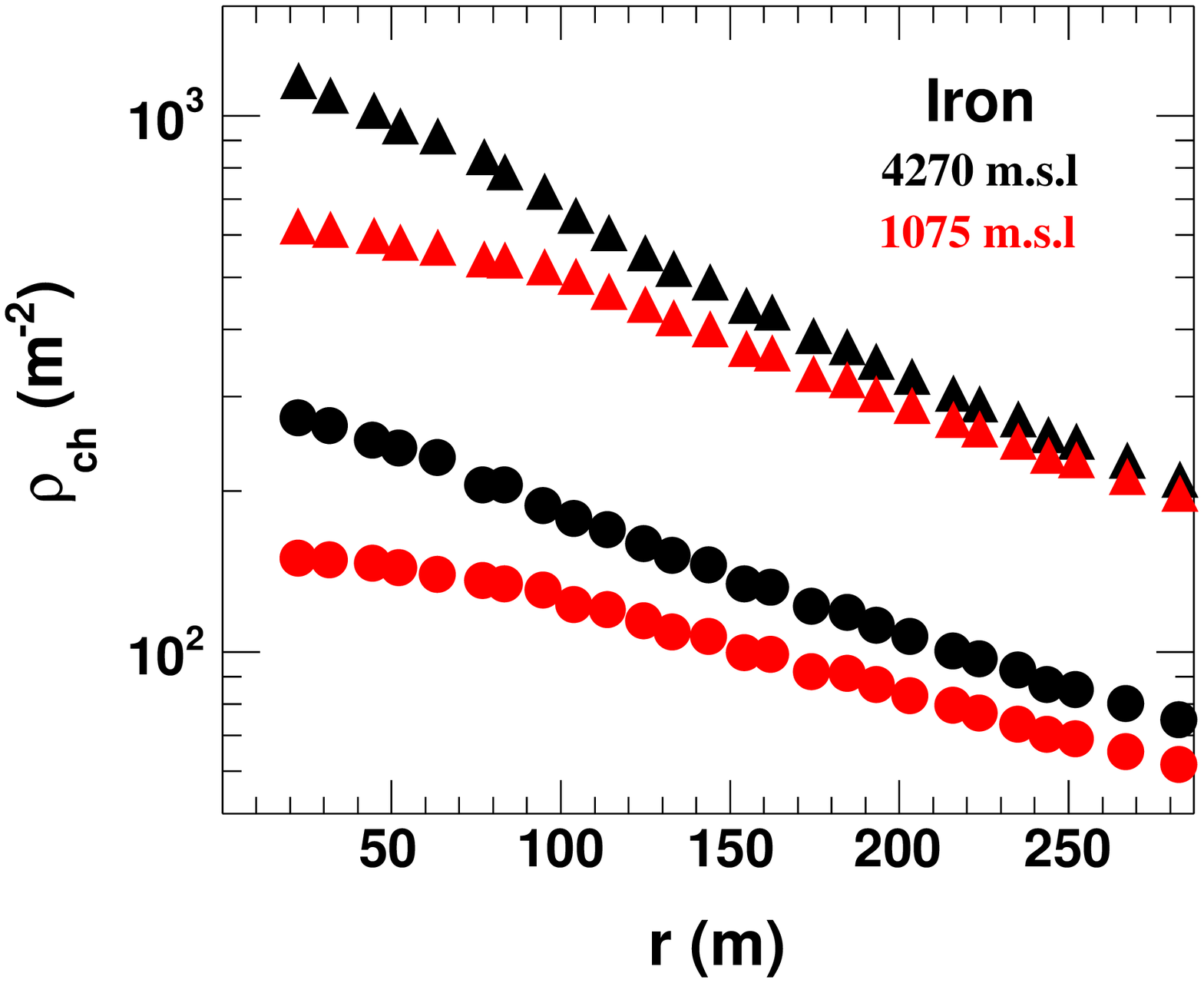}
}
\caption{Distributions of $\rho_{ch}$s
with respect to distance from the shower core of $\gamma$, proton
and iron primaries at different energies given by the QGSJET-GHEISHA model 
combination over the observation levels of Hanle (black symbols) and 
Pachmarhi (red symbols). In the
respective plots, {\large $\bullet$/\textcolor{r}{$\bullet$}} indicates for 
100 GeV $\gamma$, 250 GeV proton and 5 TeV iron primaries;  
$\blacktriangle$/\textcolor{r}{$\blacktriangle$} indicates for 500 GeV  
$\gamma$, 1 TeV proton and 10 TeV iron primaries; and 
$\bigstar$/\textcolor{r}{$\bigstar$} indicates for 1 TeV $\gamma$ and 2 TeV 
proton primaries.}     
\label{fig4}
\end{figure}

To study the effect of height of the observation level on the 
lateral distribution of $\rho_{ch}$, we have 
selected $\gamma$-ray, proton and iron nuclei initiated showers from our 
CORSIKA database for two observational levels, one at the height of 4270 
m from mean sea level (i.e. at the altitude of Hanle) and the other at the
height of 1075 m (i.e. at the altitude of Pachmarhi) (see 
Sec.I). We considered different energies for primaries for QGSJET-GHEISHA 
model combination. The 
results of this analysis are shown in the Fig.\ref{fig4}. This figure shows 
that, at higher observation level near the shower core, all primaries at all 
energies, produced considerably higher number of Cherenkov photons than that
produced at the lower observation level. This difference decreases with 
increasing core distance and then at a particular core distance depending 
on the primary particle and its energy, the density 
become almost equal for both the observational levels. For 
the $\gamma$-ray primary, this distance is $\sim$130 m at all energies, which 
is different for proton and iron primaries. 

It is seen
that at higher observation level, for the 100 GeV $\gamma$-ray primary, 
the rate of decrease of the $\rho_{ch}$ with core 
distance is slower upto a core distance of $\sim$100 m and faster at 
larger core distances. Because of this behaviour a clear hump is seen 
for the 100 GeV $\gamma$-ray primary at a core distance of $\sim$100 m.         

\begin{figure*}[hbt]
\centerline
\centerline{\includegraphics[width=5.5cm, height=4cm]{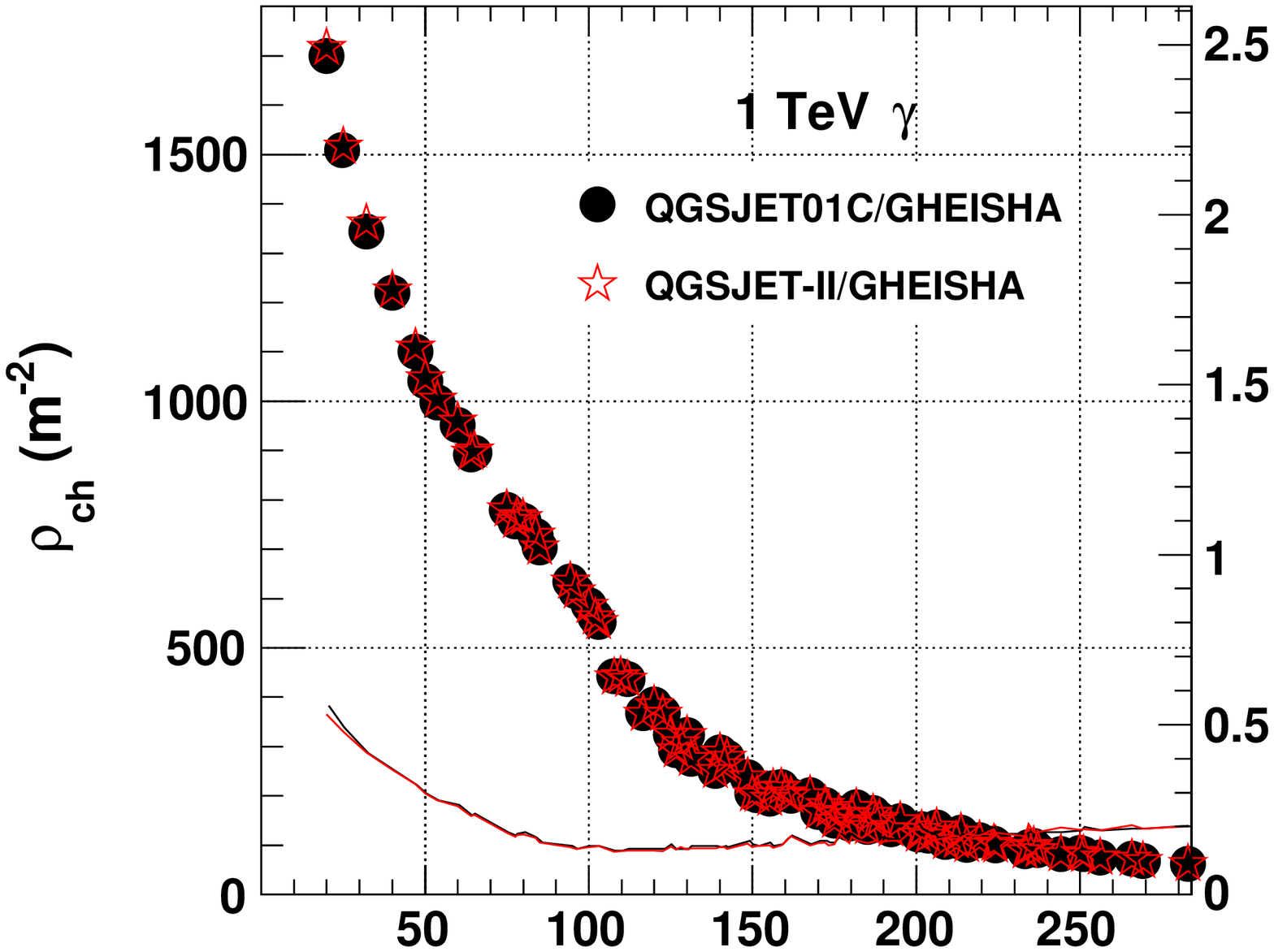}
\includegraphics[width=4.9cm, height=4cm]{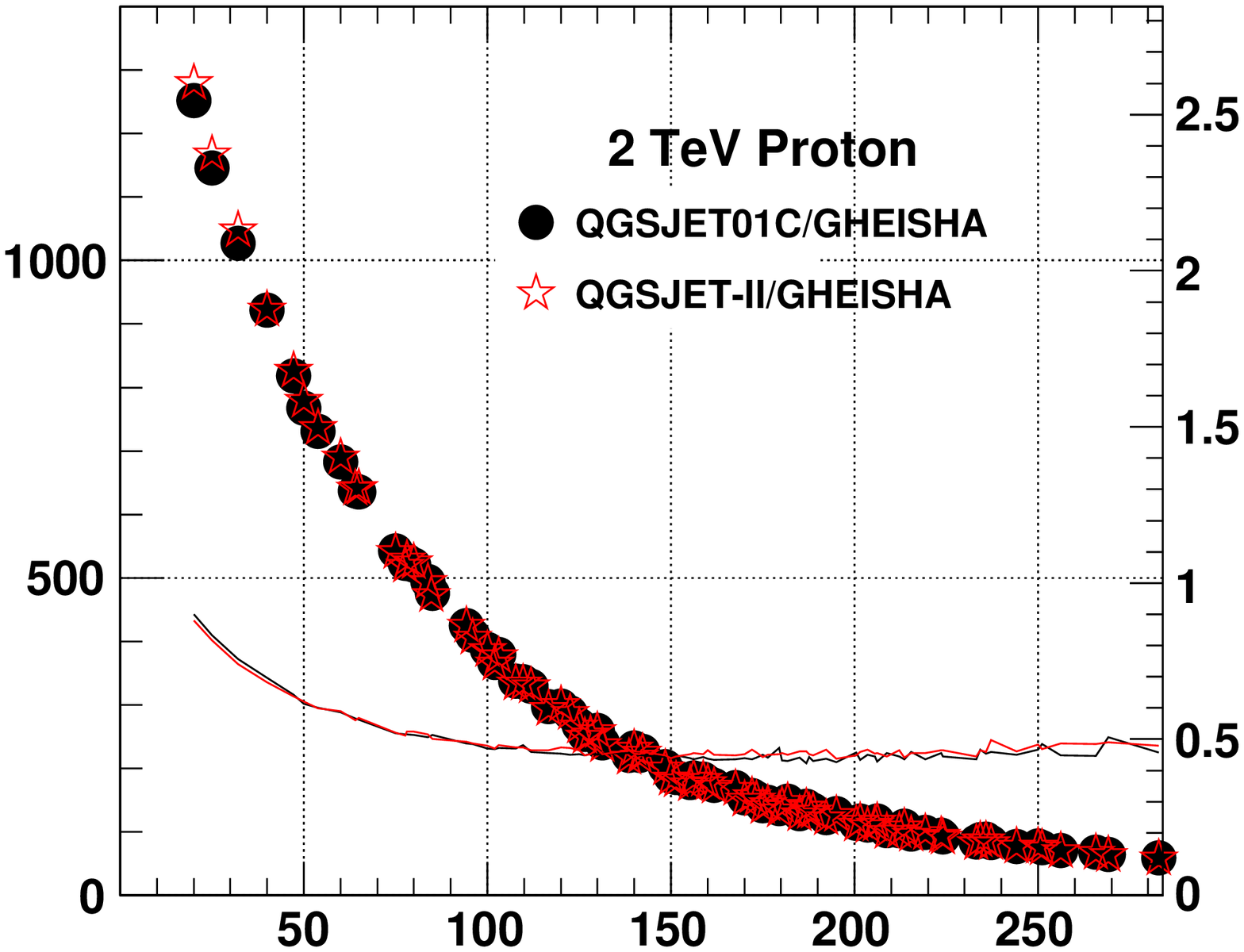}
\includegraphics[width=5.3cm, height=4.1cm]{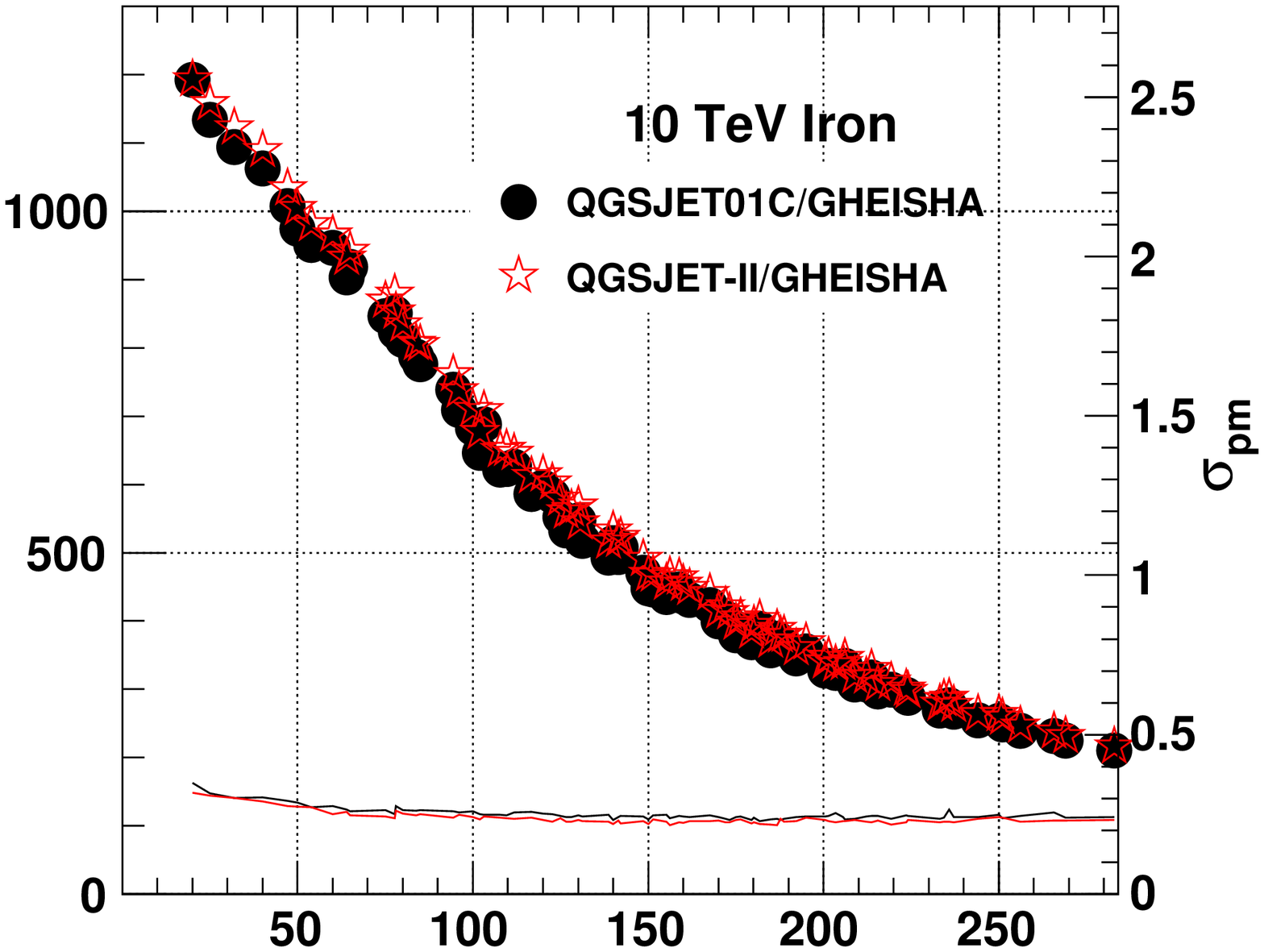}}
\centerline{\hspace{-0.5cm}
\includegraphics[width=5.2cm, height=4.5cm]{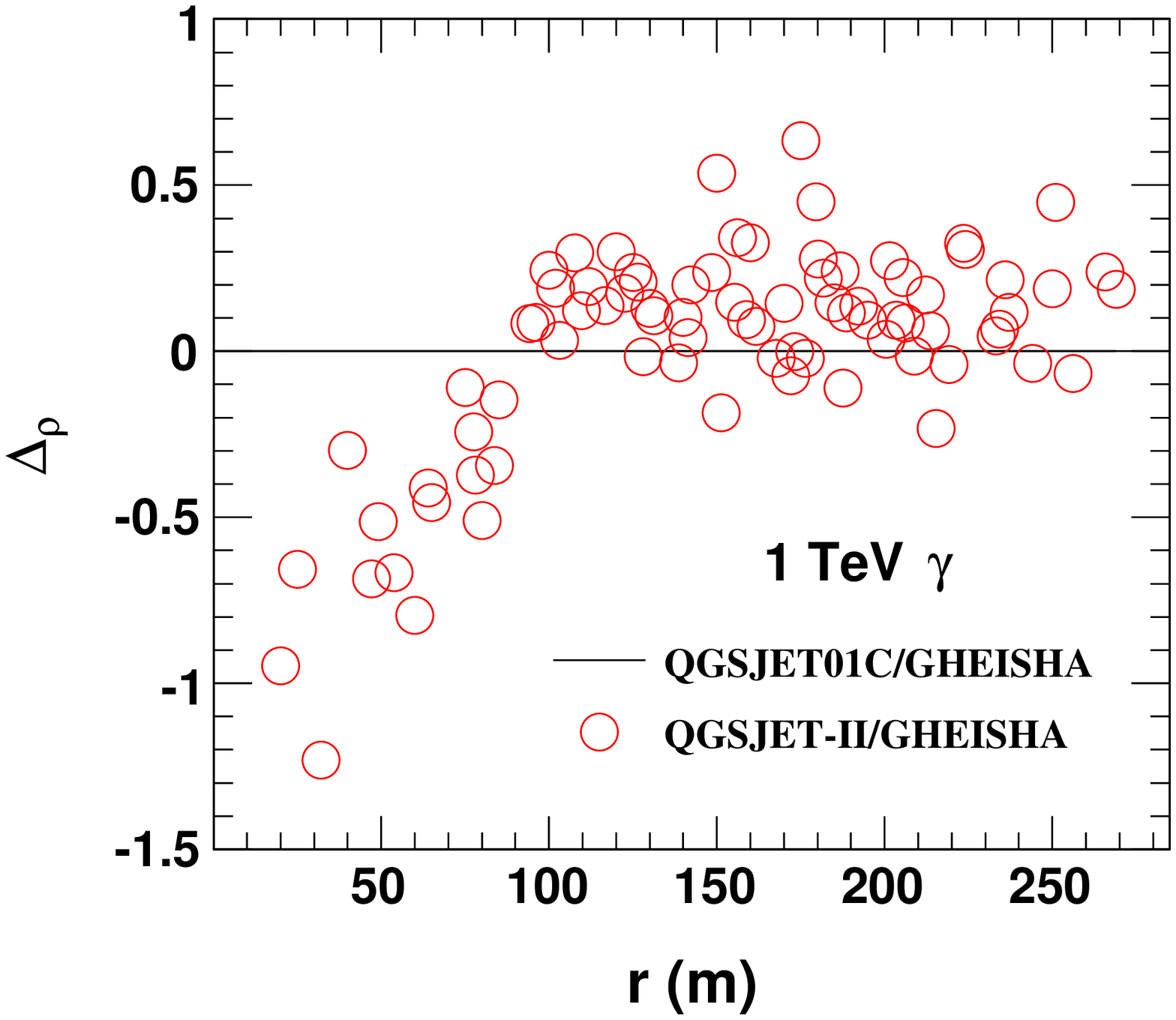}
\hspace{0.2cm}
\includegraphics[width=4.7cm, height=4.5cm]{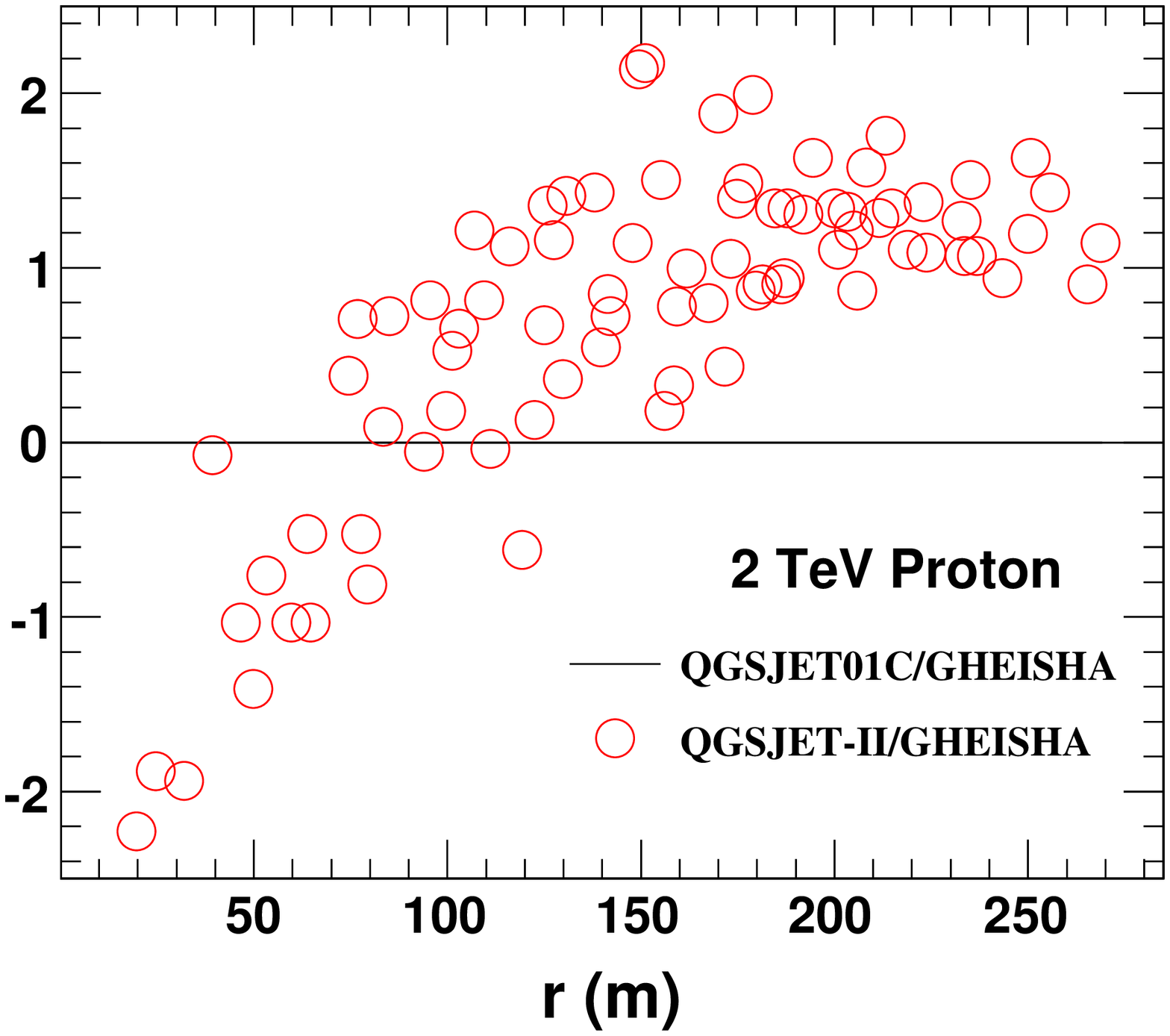}
\hspace{0.2cm}
\includegraphics[width=4.7cm, height=4.5cm]{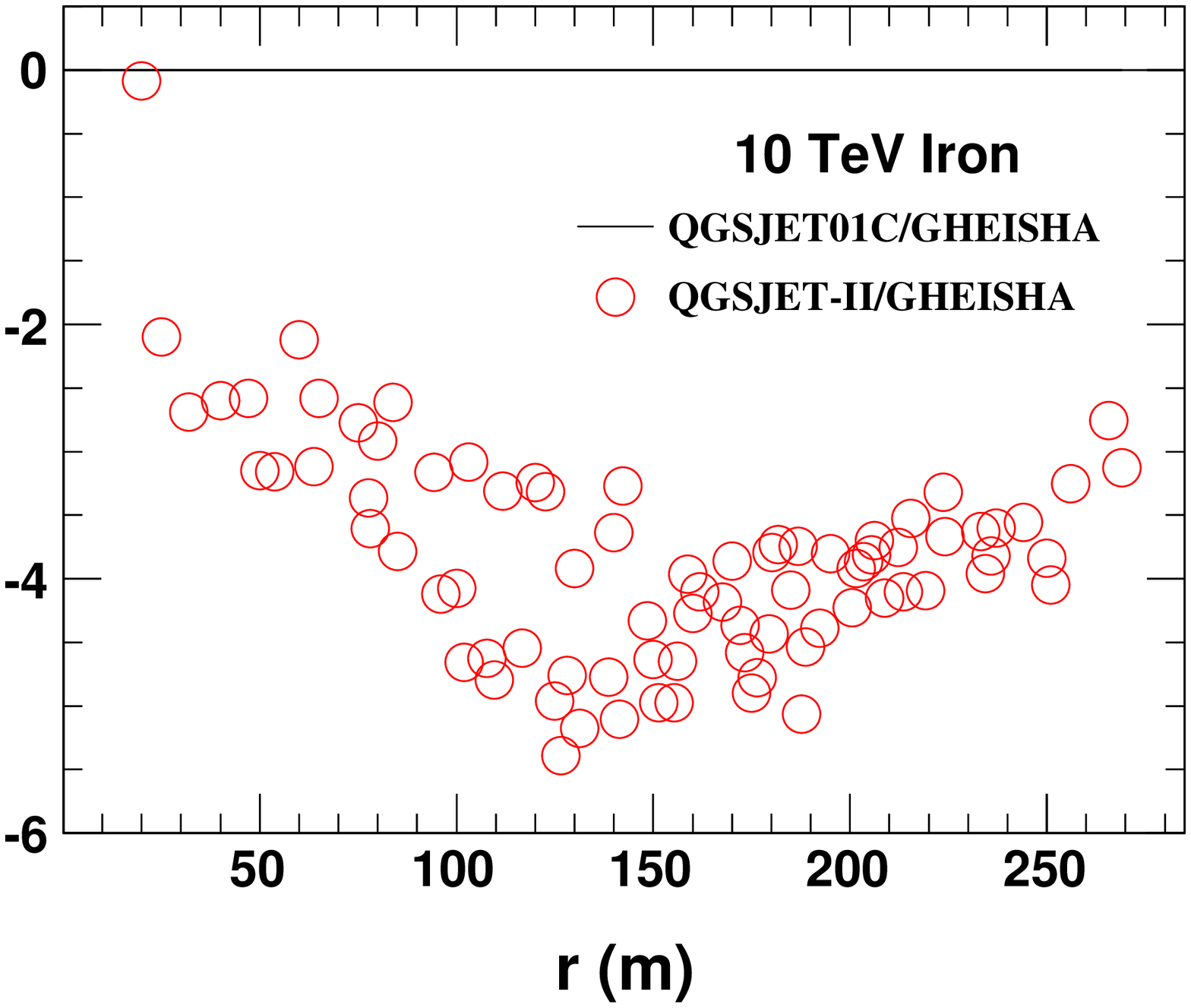}}

\caption{Top panels: Distributions of $\rho_{ch}$ with respect
to distance from the shower core for 1 TeV $\gamma$, 2 TeV proton and 10 TeV
iron primary particles obtained for the QGSJET01-GHEISHA and QGSJETII-GHEISHA
model combinations. Bottom panels: $\Delta_{\rho}$ of densities with
respect to distance from the shower core of primaries for QGSJETII-GHEISHA
model combination from the densities of QGSJET01-GHEISHA combination.}
\label{fig4a}
\end{figure*}
At the lower observational level, the $\rho_{ch}$ 
remains constant with increasing core distance upto $\sim$130 m. At this
core distance, hump is seen and $\rho_{ch}$ falls rapidly at
larger core distances. 
In the case of proton and iron primaries, the $\rho_{ch}$ falls 
slowly with increasing core distance upto a distance of $\sim$100 m 
followed by faster decrease for larger core distances beyond this point,
without forming any clearly visible hump. This behaviour of $\rho_{ch}$ at 
lower observational  level for different primary particles is 
due to the fact that, most of the low energy soft component (electrons and 
positrons) of a shower gets absorbed as the shower travels a long distance 
(depending on the altitude of observational level) through the atmosphere 
to reach the lower observational level. The low energetic soft component 
of a shower is concentrated near to the shower core. As mentioned above, 
the $\gamma$-ray primary produces only the soft component shower, whereas 
the proton and iron primaries produce both the soft and hard components of 
the shower as well as other heavy particles depending on the energy of the 
concerned particles. Again, the iron primary has 56 times less energy per 
nucleon than proton primary of the same energy.         

\subsubsection{Comparison of QGSJET01 and QGSJETII}
In this work, for convenience and as a convention we have used QGSJET01 as one 
of our high energy hadronic interaction models. Since the improved version of 
the QGSJET model is QGSJETII,  it is important to compare results of the 
QGSJET01 and QGSJETII to verify the reliability of results from QGSJET01 model
discussed so far. With this motivation, in the Fig.\ref{fig4a} we have 
compared $\rho_{ch}$ distributions for 1 TeV $\gamma$, 2 TeV
proton and 10 TeV iron primaries as obtained by using QGSJET01-GHEISHA 
and QGSJETII-GHEISHA model combinations. These three primaries are used as the 
representatives of their categories for this comparison. 
It is clear from the figure that the difference of the QGSJET01 and QGSJETII 
models is very small as far as the $\rho_{ch}$ is concerned. 
In fact, for the $\gamma$-ray primary there is no real difference between 
these two models as the $\rho_{ch}$ agrees within 
$\sim$1\% at all core distances. For the proton primary the agreement is 
within $\sim$$\pm$2\%. On the other hand for the iron primary the deviations 
range from 0 to $\sim$-5\%, which are also negligible in comparison to 
deviation of other model combinations discussed  above for this primary.     

\subsection{Arrival time of Cherenkov photons}
\subsubsection{General feature}
\begin{figure*}[hbt]
\centerline
\centerline{\includegraphics[width=5.4cm, height=4cm]{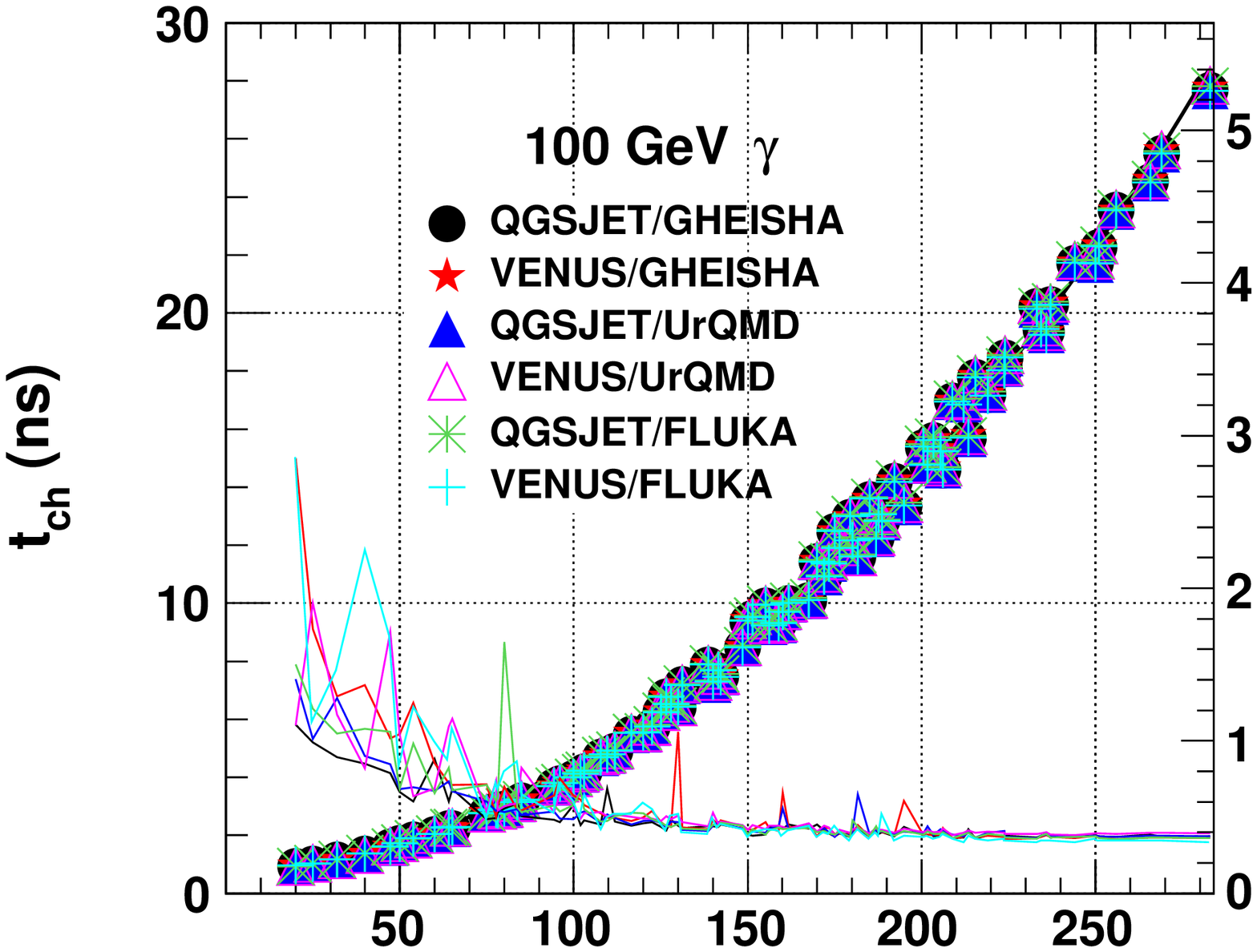}
\includegraphics[width=4.9cm, height=4cm]{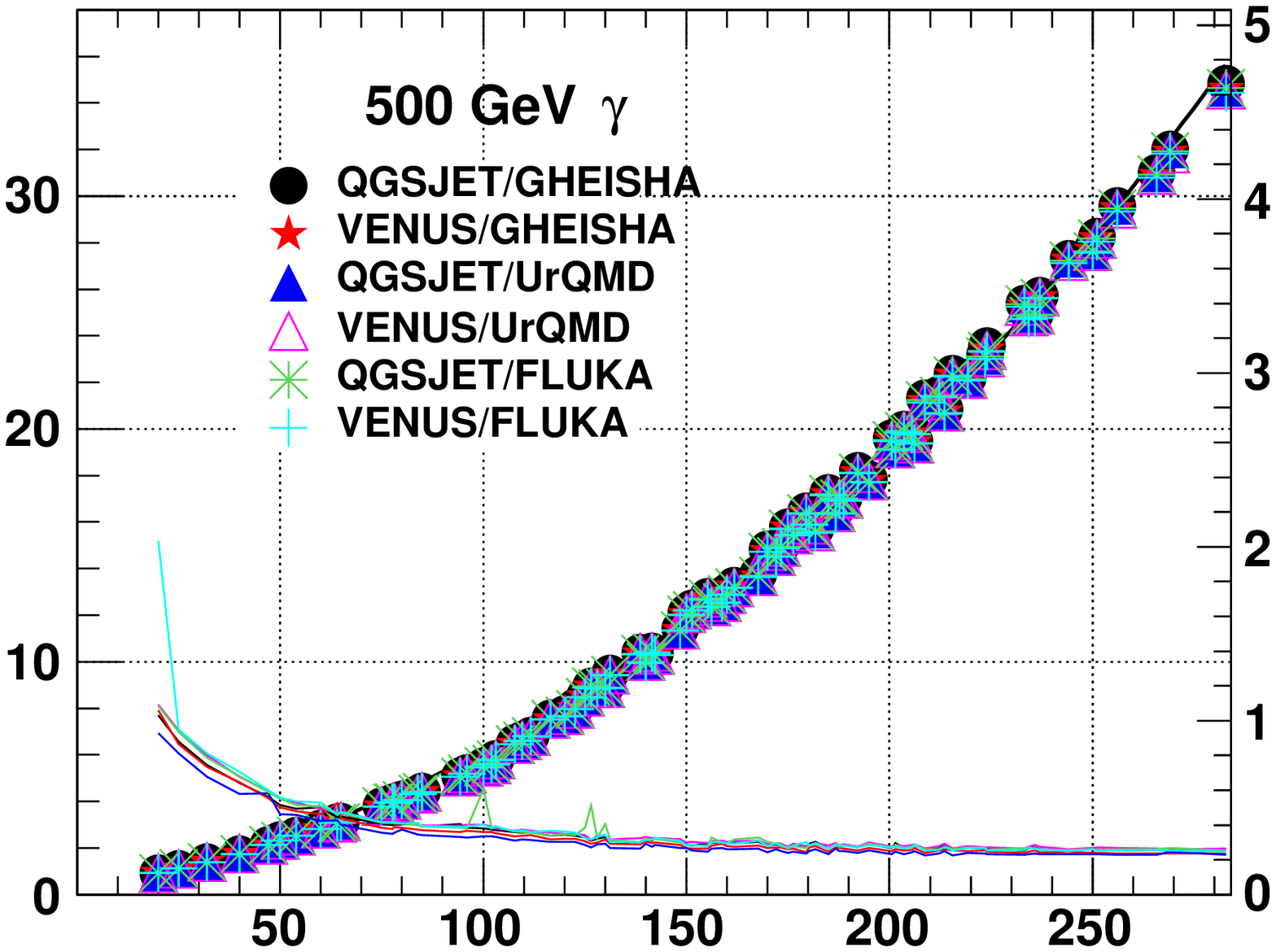}
\includegraphics[width=5.3cm, height=4cm]{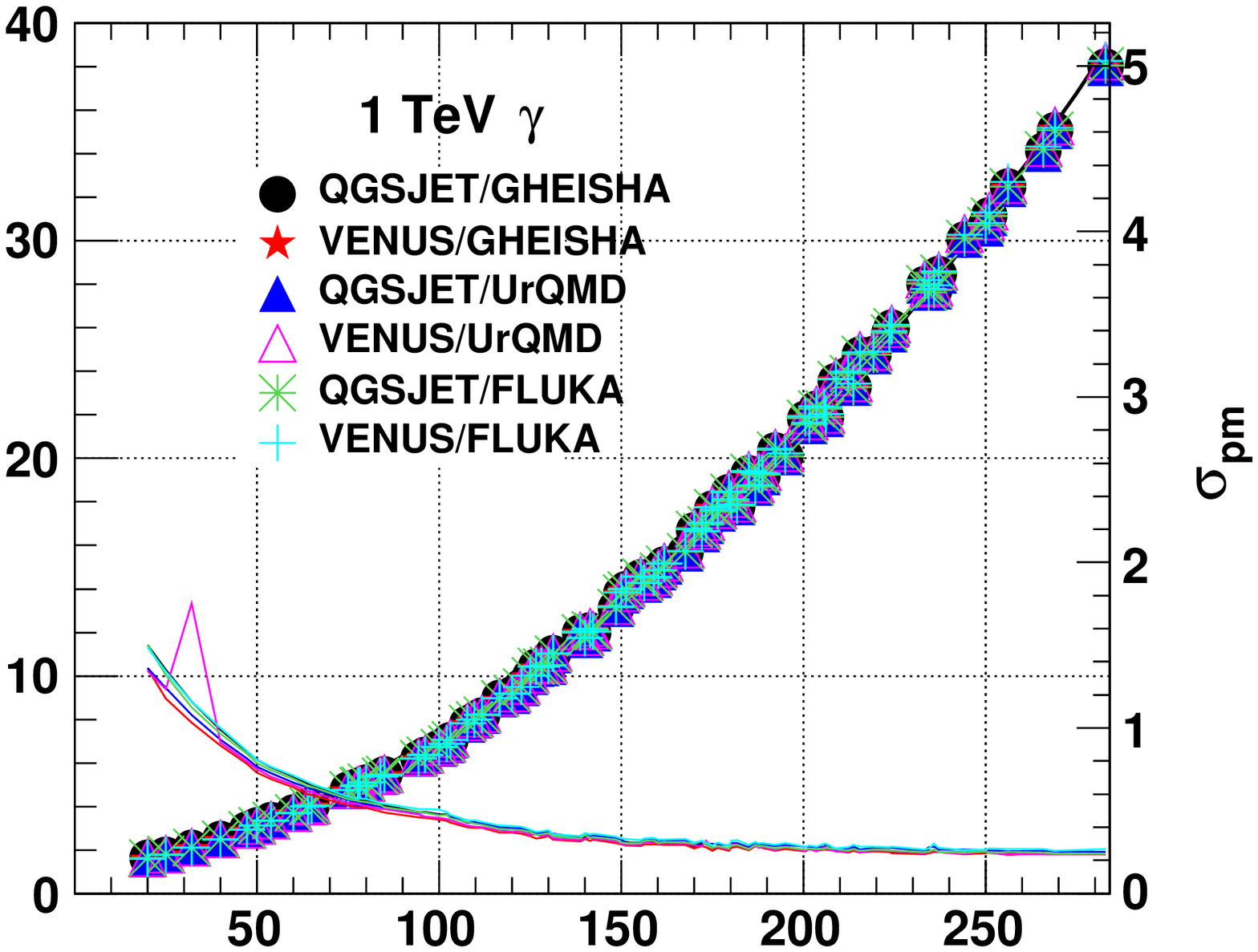}}
\centerline{\includegraphics[width=5.4cm, height=4cm]{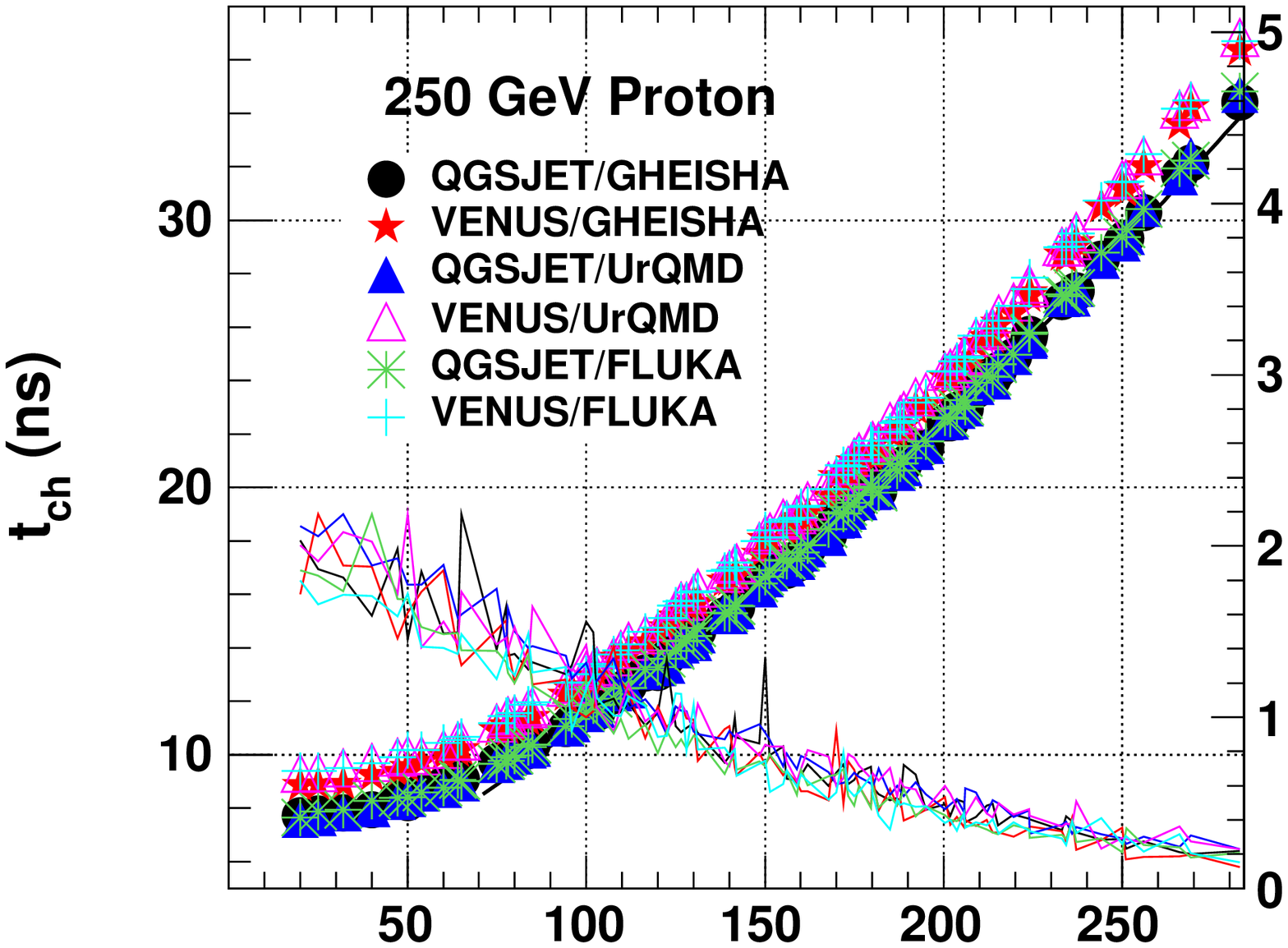}
\includegraphics[width=5cm, height=4cm]{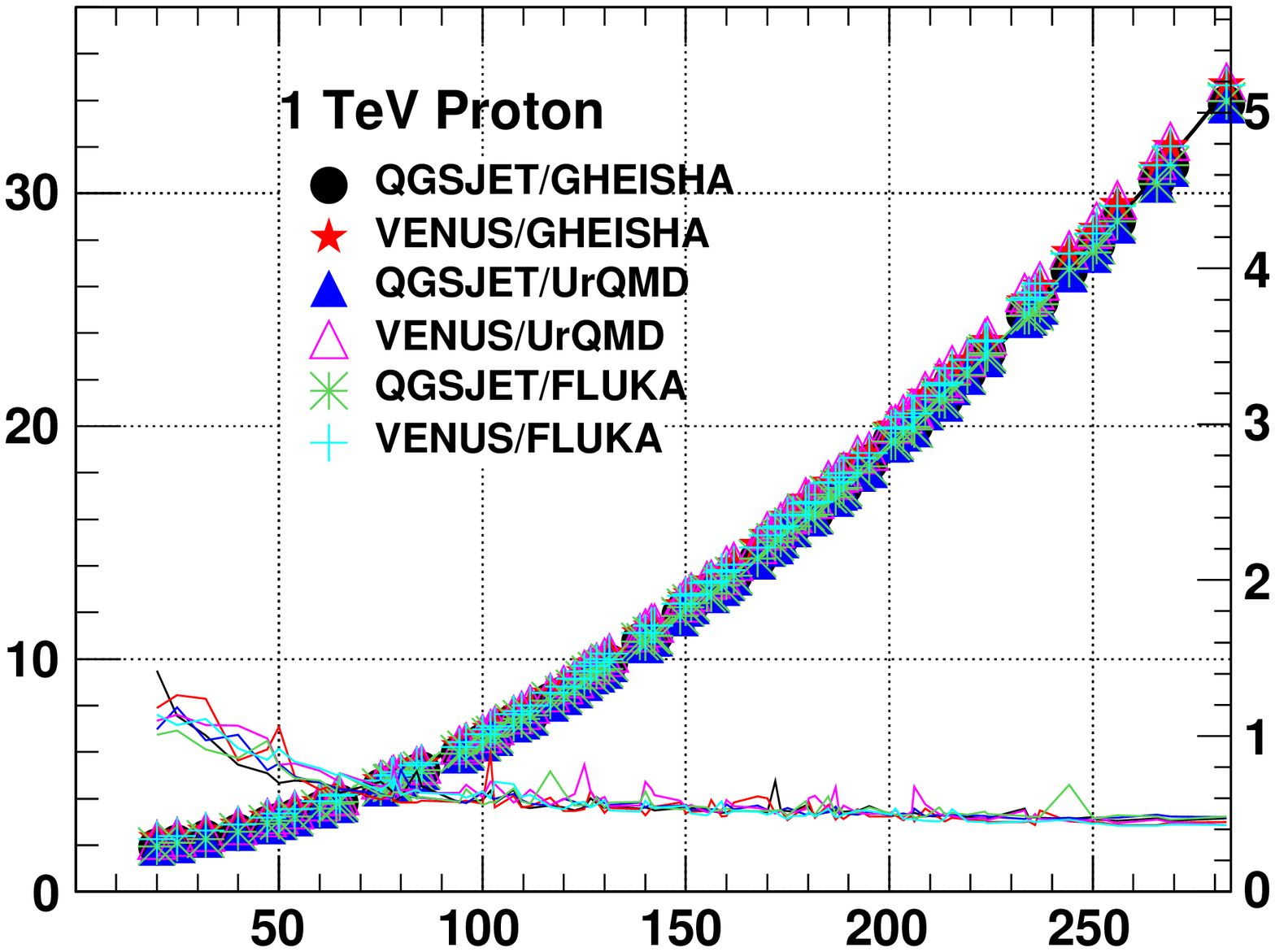}
\includegraphics[width=5.3cm, height=4cm]{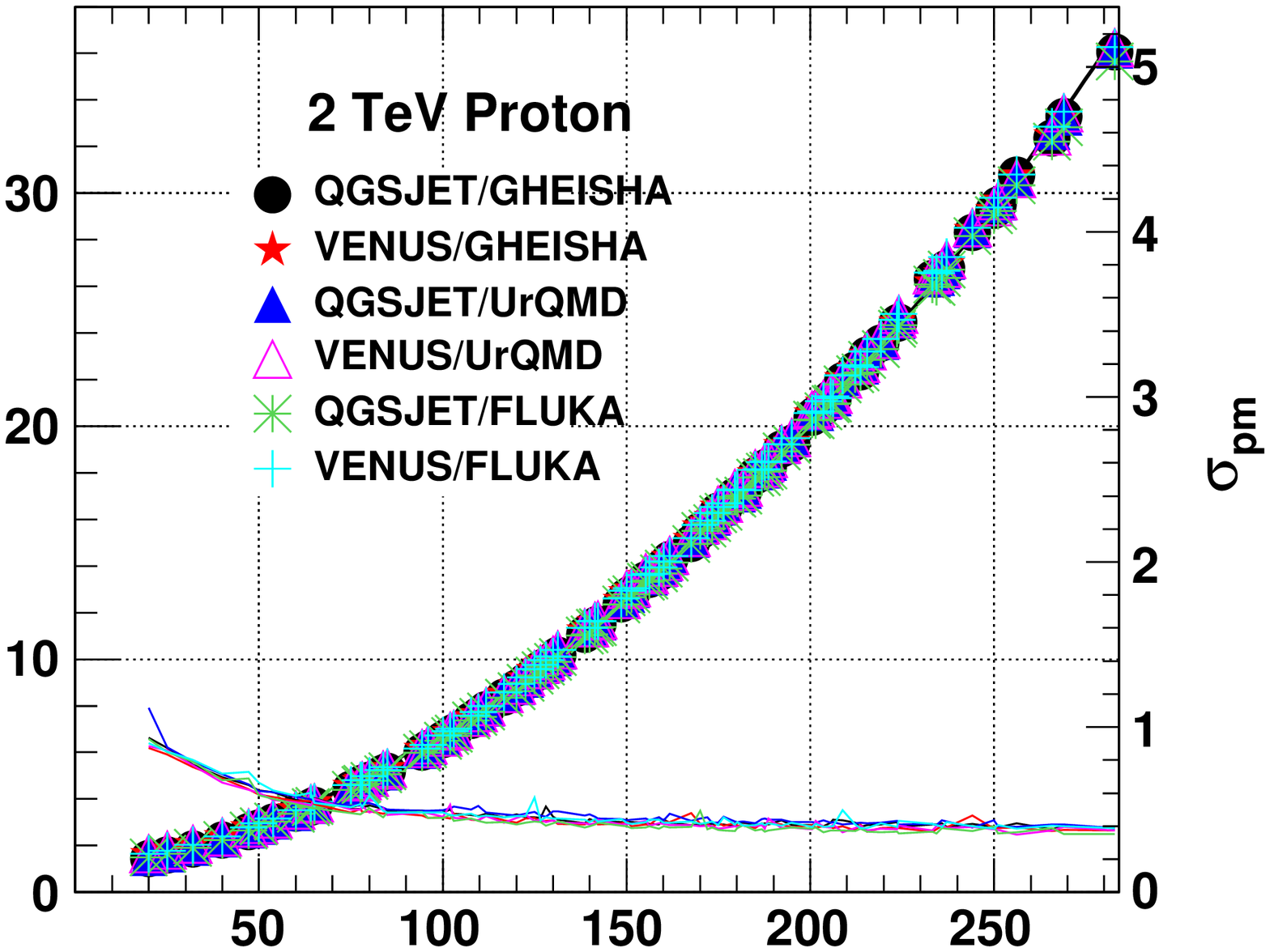}}
\centerline{\includegraphics[width=5.3cm, height=4.5cm]{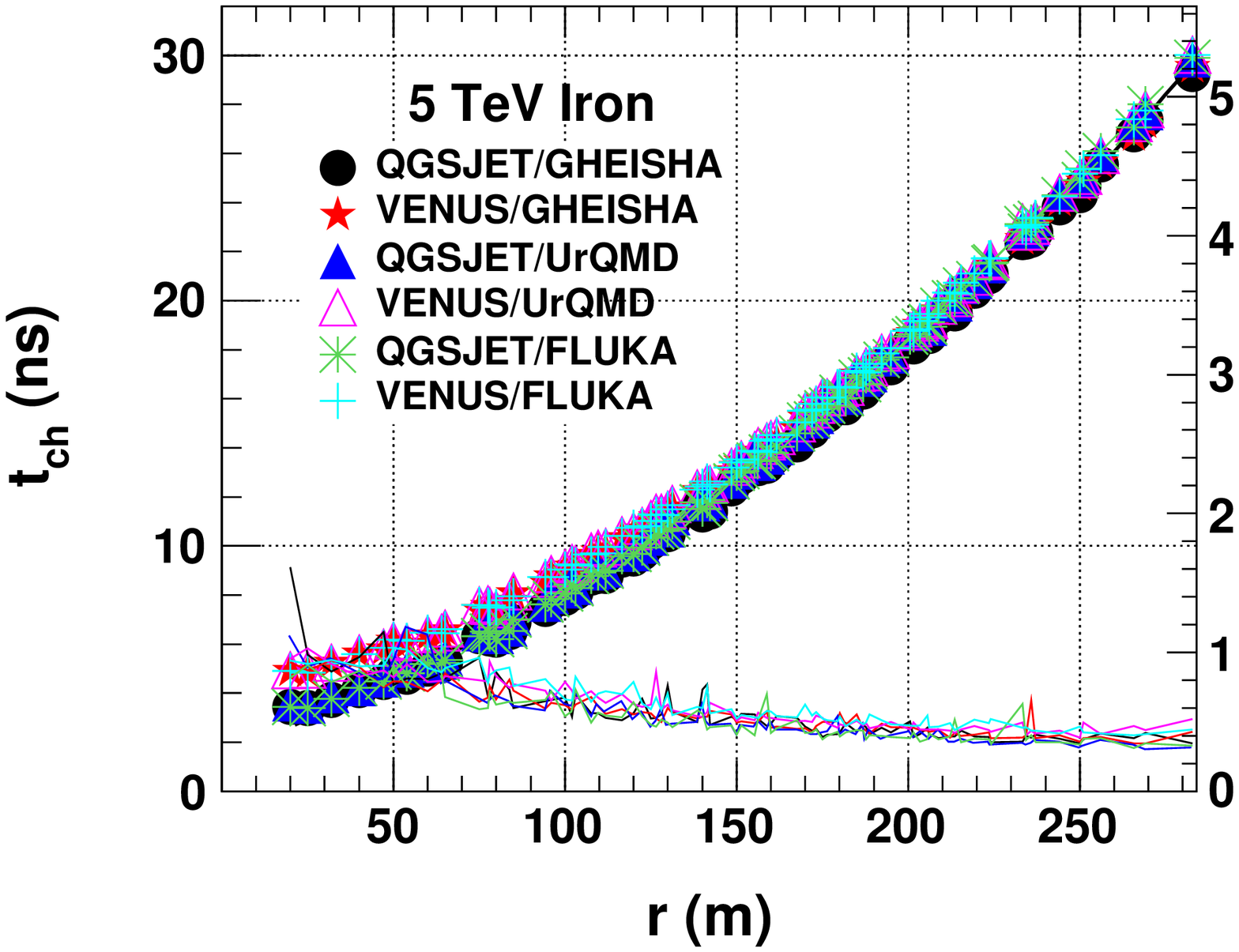}
\includegraphics[width=5.4cm, height=4.5cm]{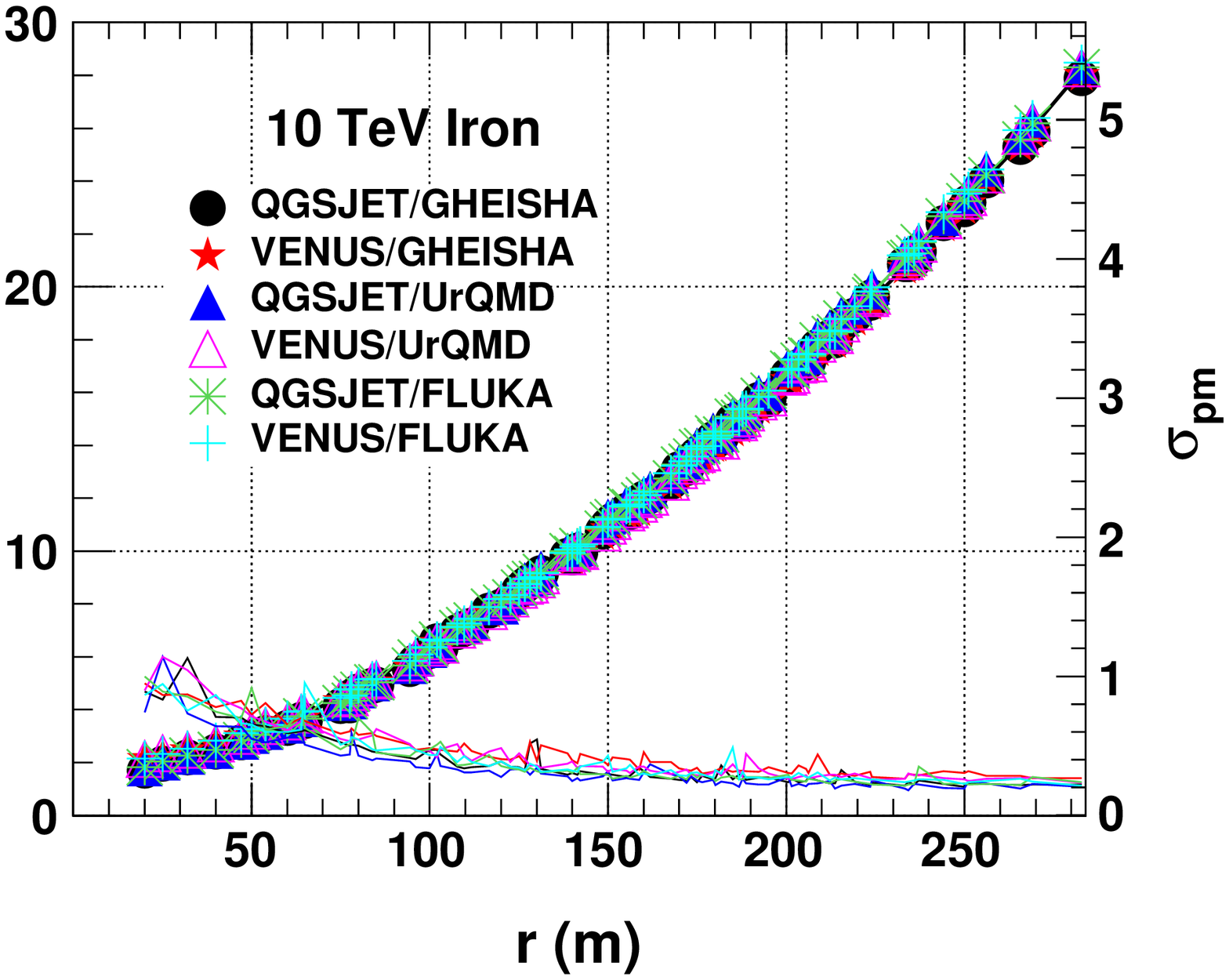}}
\caption{Variations of average arrival times of Cherenkov photons ($t_{ch}$) 
and their $\sigma_{pm}$ with respect to distance from the shower core 
of different primaries at different energies, given by different combinations 
of high and low energy hadronic interaction models. The solids line indicate 
the results of the best fitted function (\ref{eq3}) to the respective plots
for QGSJET-GHEISHA combination. Fits are done by using the 
$\chi^2$-minimization method in the ROOT software \cite{Root} platform.}
\label{fig5}
\end{figure*}

The variation of mean arrival time of Cherenkov photons ($t_{ch}$) with 
respect to core distance is studied for all six combinations of low and 
high energy hadronic interaction models. These are shown in the Fig.\ref{fig5} 
for different primaries and their energies. It is observed that, the shower 
front is nearly spherical in shape for all primary particles, energies and 
model combinations. However, near shower core (core distance $\le$50 m), 
the increase of the $t_{ch}$ with respect to core distance is 
comparatively slow, as a result the shape of the distribution deviates slightly 
from the spherical symmetry. This deviation increases with decreasing 
energy and increasing mass of the primary particle. Also, at a given energy, 
the deviation is more noticeable for the hadronic primaries. We have also found 
that, all these $t_{ch}$ distributions, irrespective of primary particles, 
their energies and the combinations of hadronic interaction models, follow 
the pattern as given by the function,
\begin{equation}
t_{ch}(r) = t_{0}e^{\Gamma/r^{\lambda}},
\label{eq3}
\end{equation} 
where $t_{ch}(r)$ is the position dependent arrival time, $r$ is the distance 
from the shower core, $t_{0}$, $\Gamma$ and $\lambda$ are constant parameters 
of the function. The values of these constant parameters depend on the type and 
energy of the primary particle for a given combination of  
low and high energy hadronic interaction models. In this figure, as an 
example, we have shown the best fit functions to the data from 
the QGSJET-GHEISHA combination as solid lines. Fits are done by using same
method as mentioned in the density subsection. 

\subsubsection{Dependence on hadronic interaction model} 
There are no visible differences in $t_{ch}$,
produced by different combinations of low and high energy 
hadronic interaction models, in the case of $\gamma$-ray primary at all
energies, as shown in the Fig.\ref{fig5}. But, in the cases of 250 GeV proton 
and 5 TeV iron primaries, the noticeable differences due to model combinations 
are seen. For the 250 GeV proton primary the 
VENUS lead group of models give consistently higher $t_{ch}$ than the 
QGSJET lead group for all core distances. The difference is more 
prominent near shower core. Whereas in the case of 5 TeV iron primary,
the difference is clearly noticeable for near core distance only, where the 
VENUS lead group of models give higher arrival time than the QGSJET lead group. 
 
\begin{figure*}[hbt]
\centerline
\centerline{\includegraphics[width=5.4cm, height=4cm]{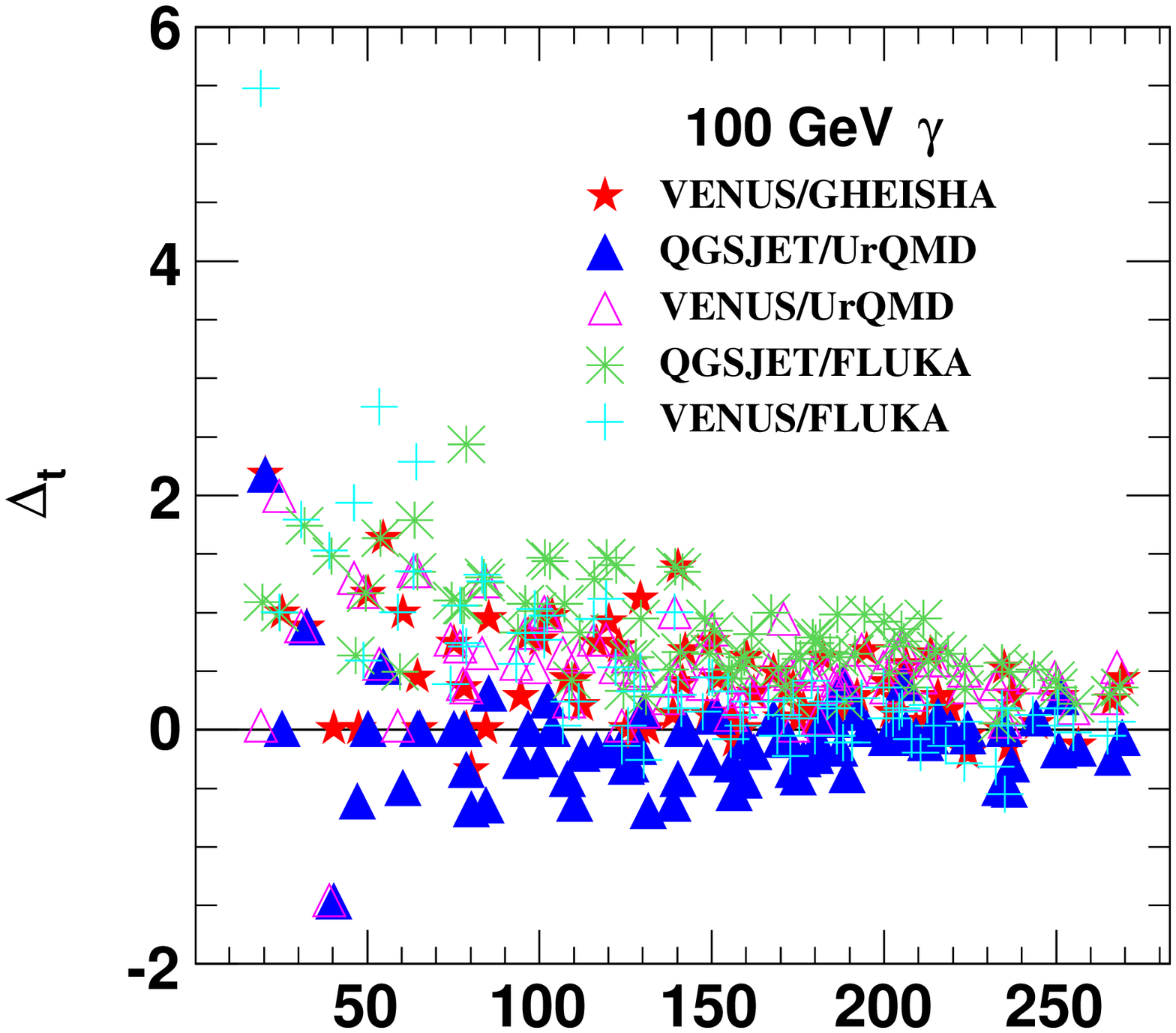}
\includegraphics[width=4.8cm, height=3.9cm]{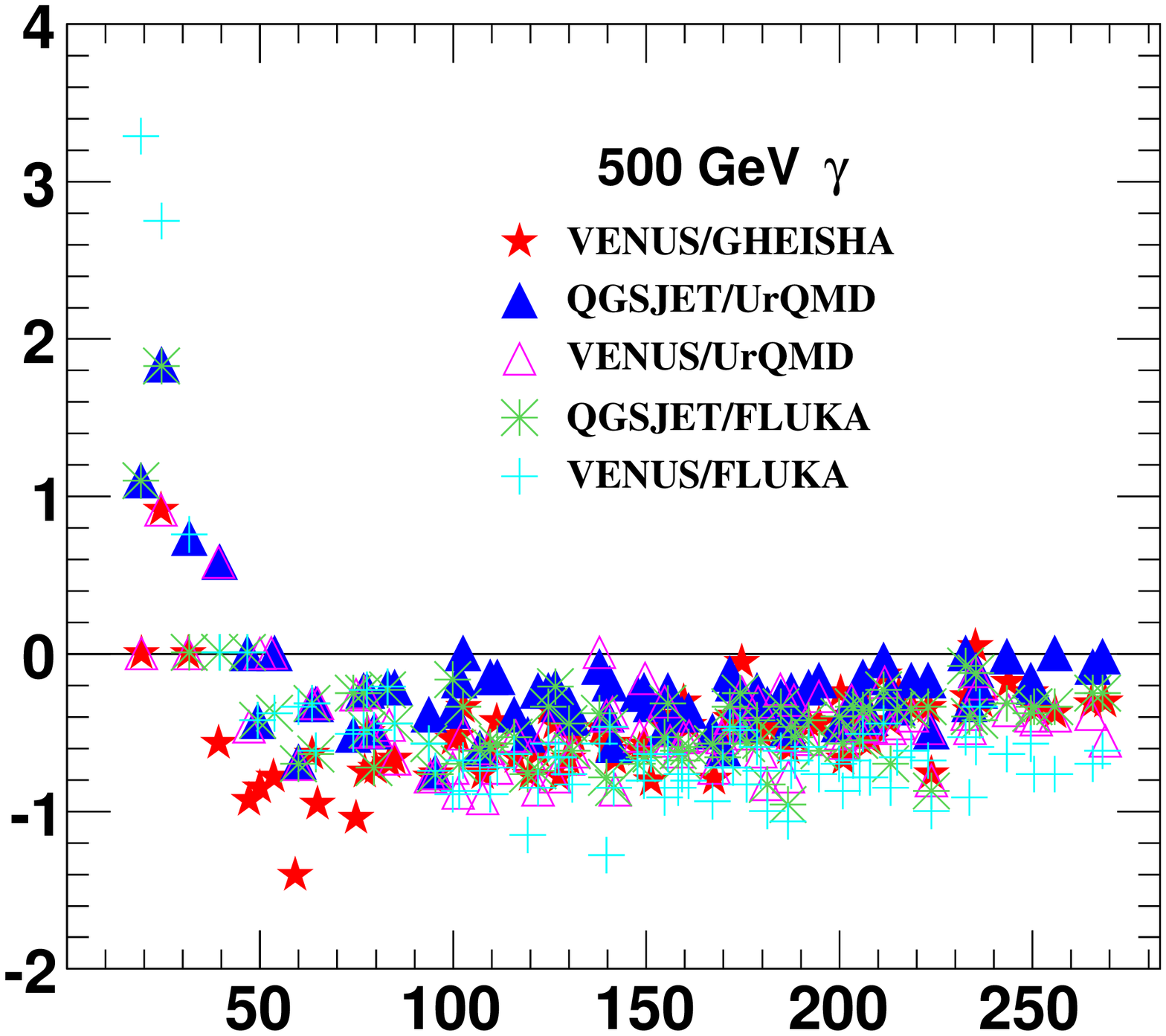}
\includegraphics[width=5cm, height=4.1cm]{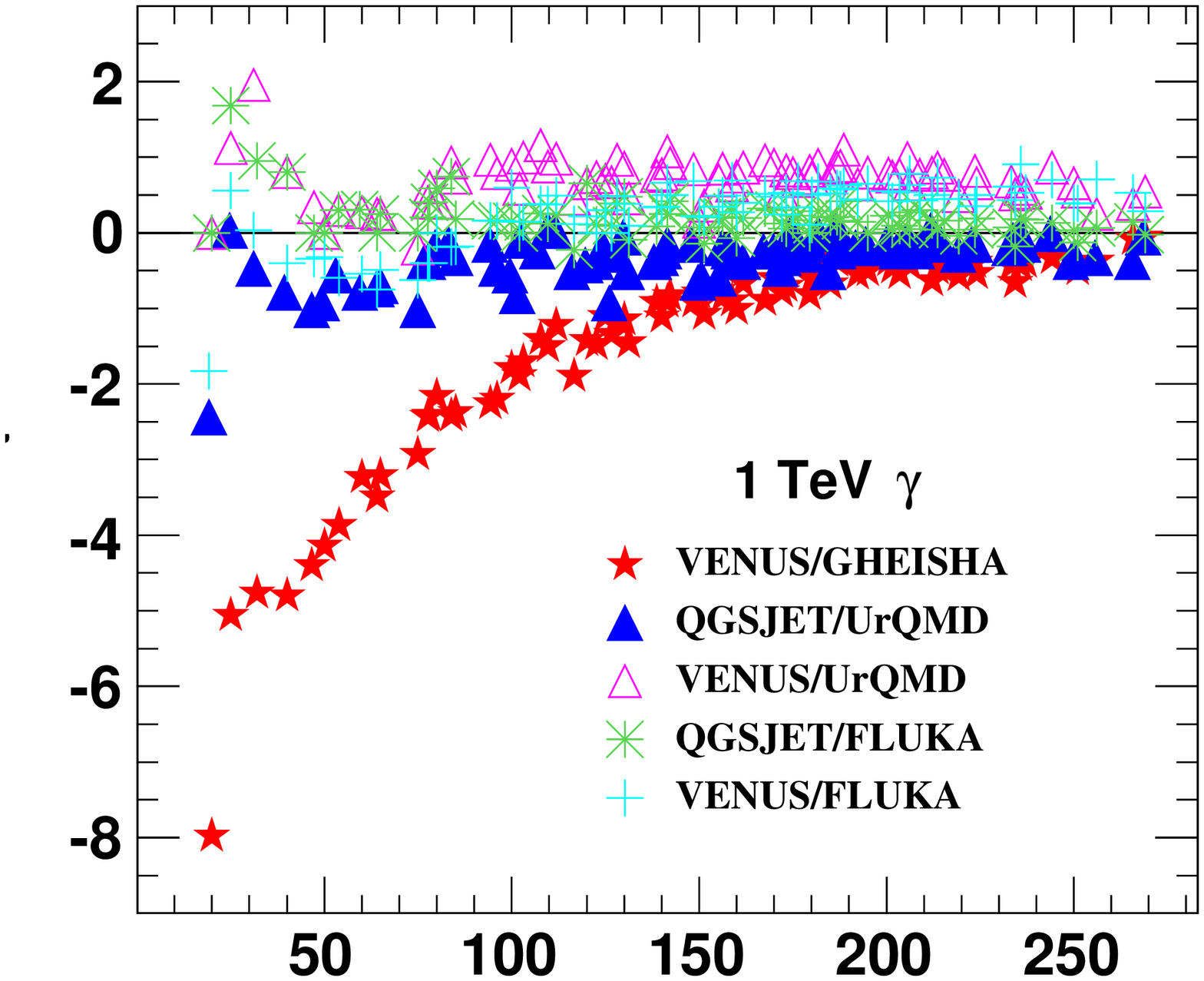}}
\centerline{\includegraphics[width=5.5cm, height=4cm]{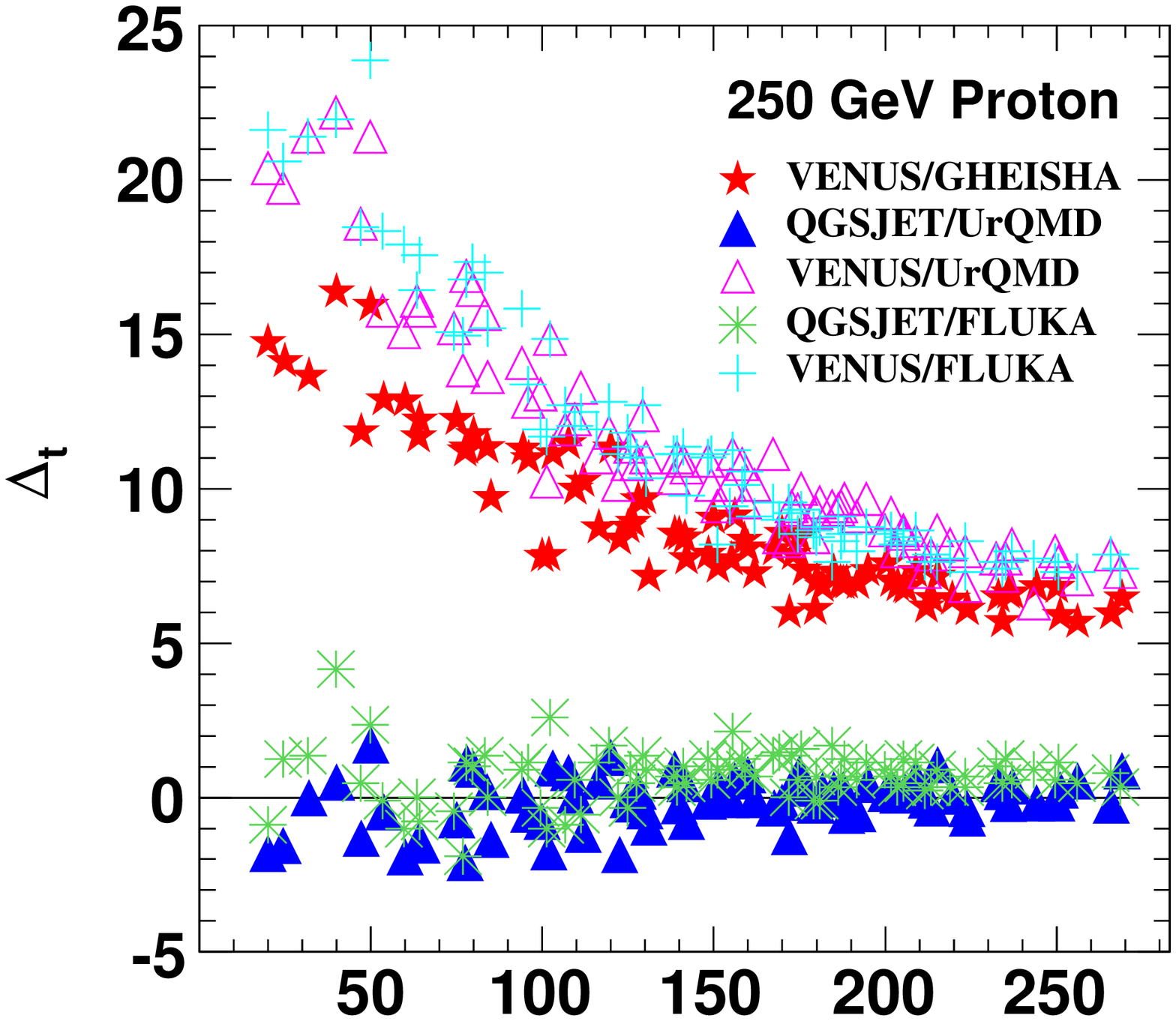}
\includegraphics[width=5cm, height=4cm]{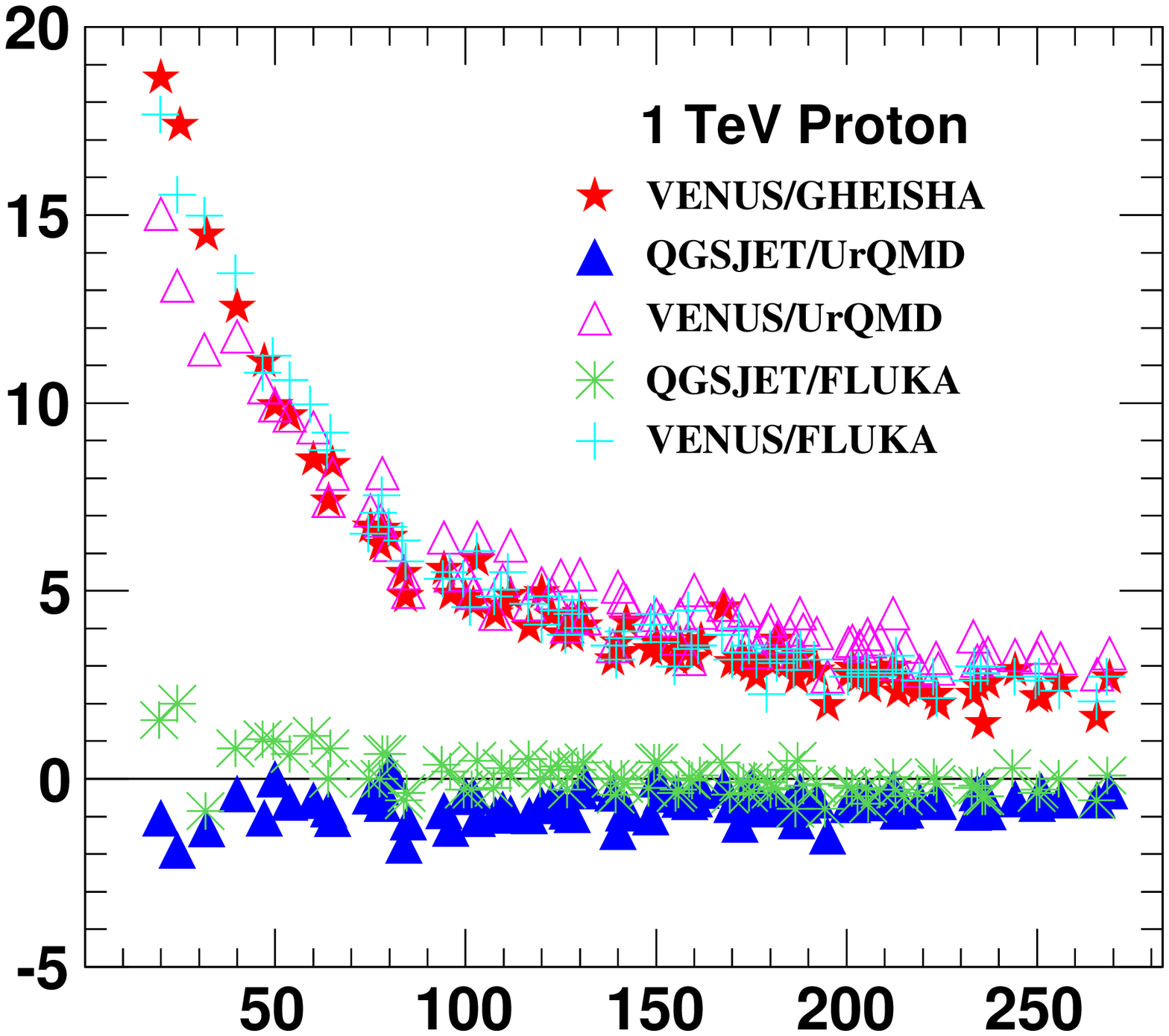}
\includegraphics[width=5cm, height=4cm]{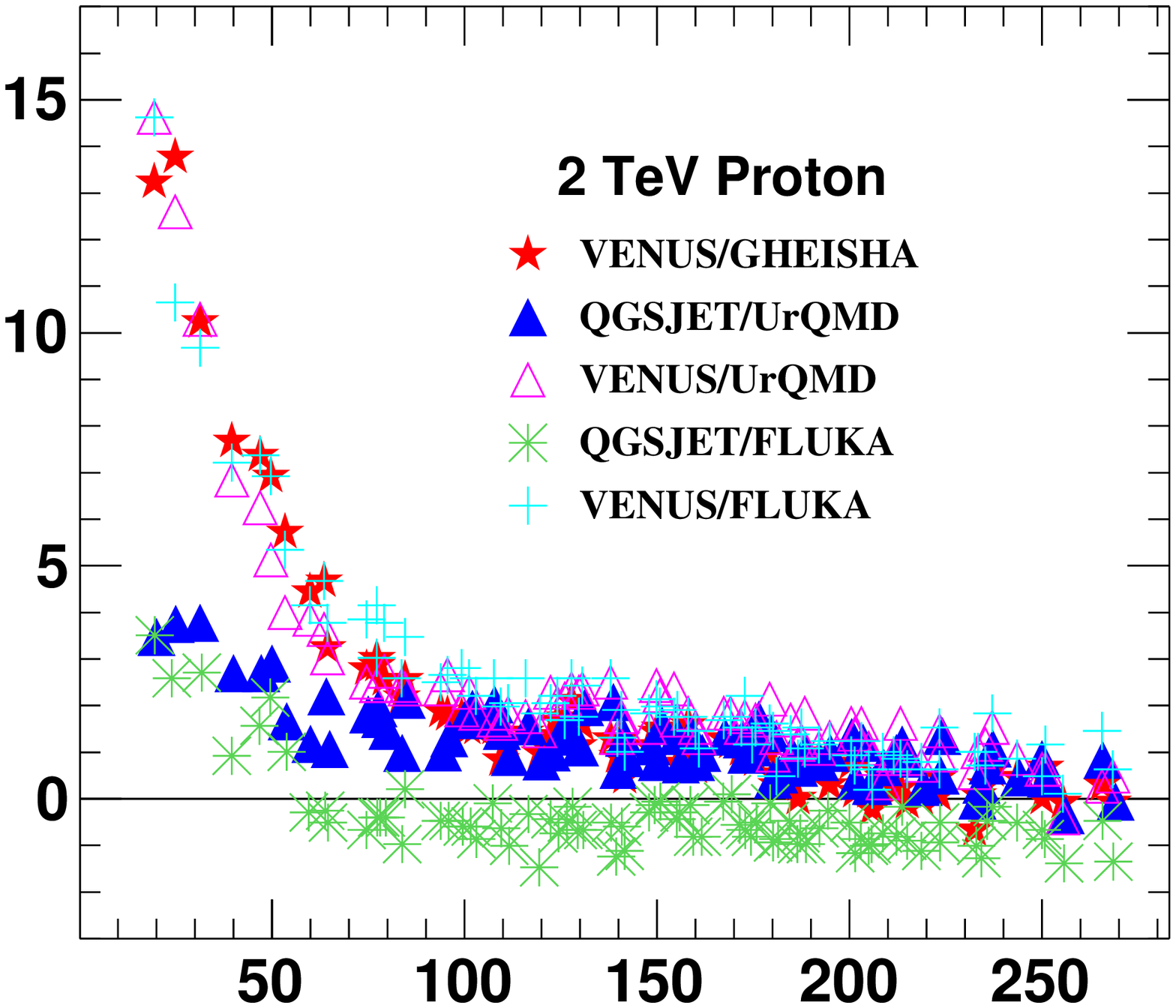}}
\centerline{\includegraphics[width=5.5cm, height=4.5cm]{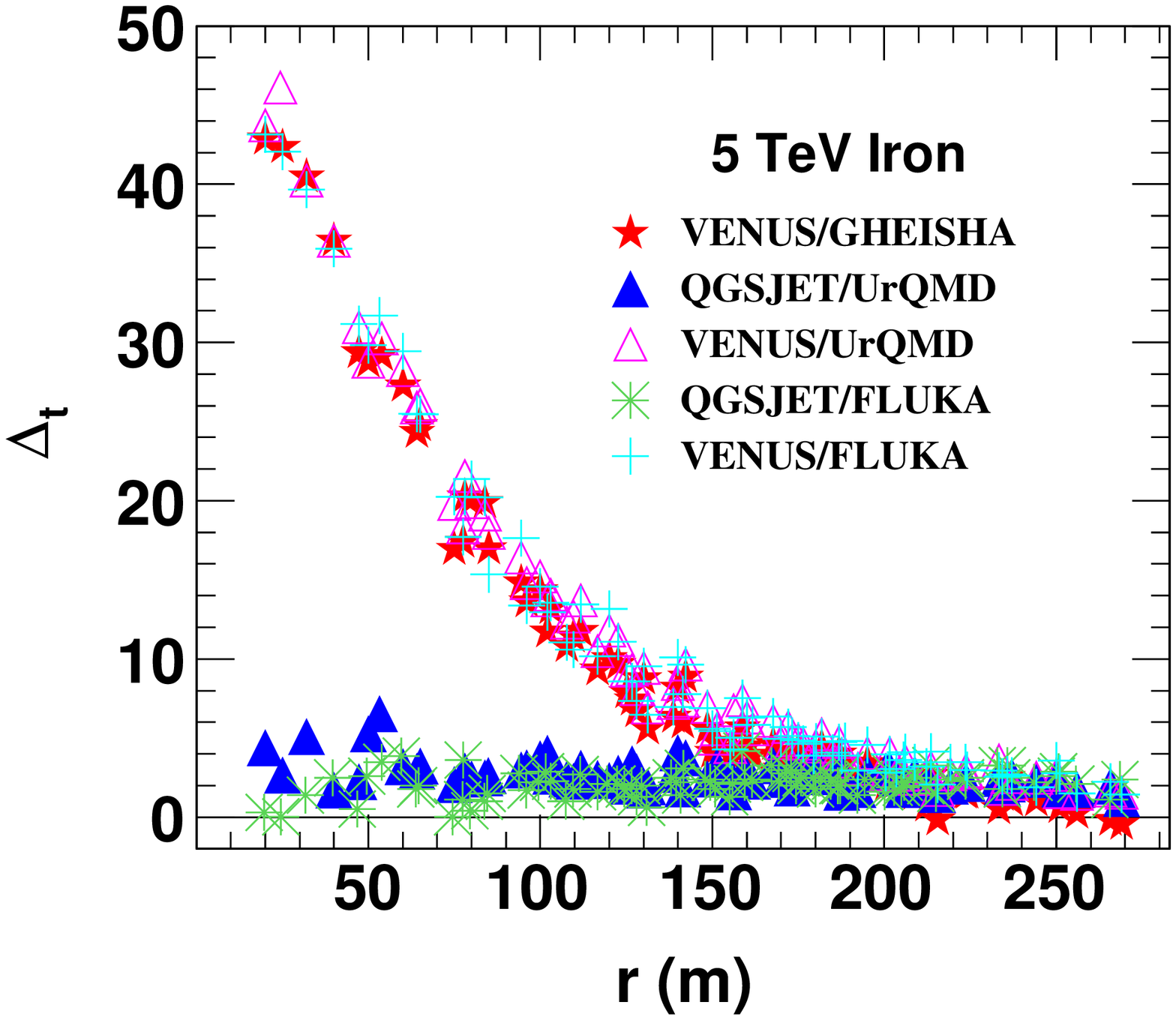}
\includegraphics[width=5cm, height=4.5cm]{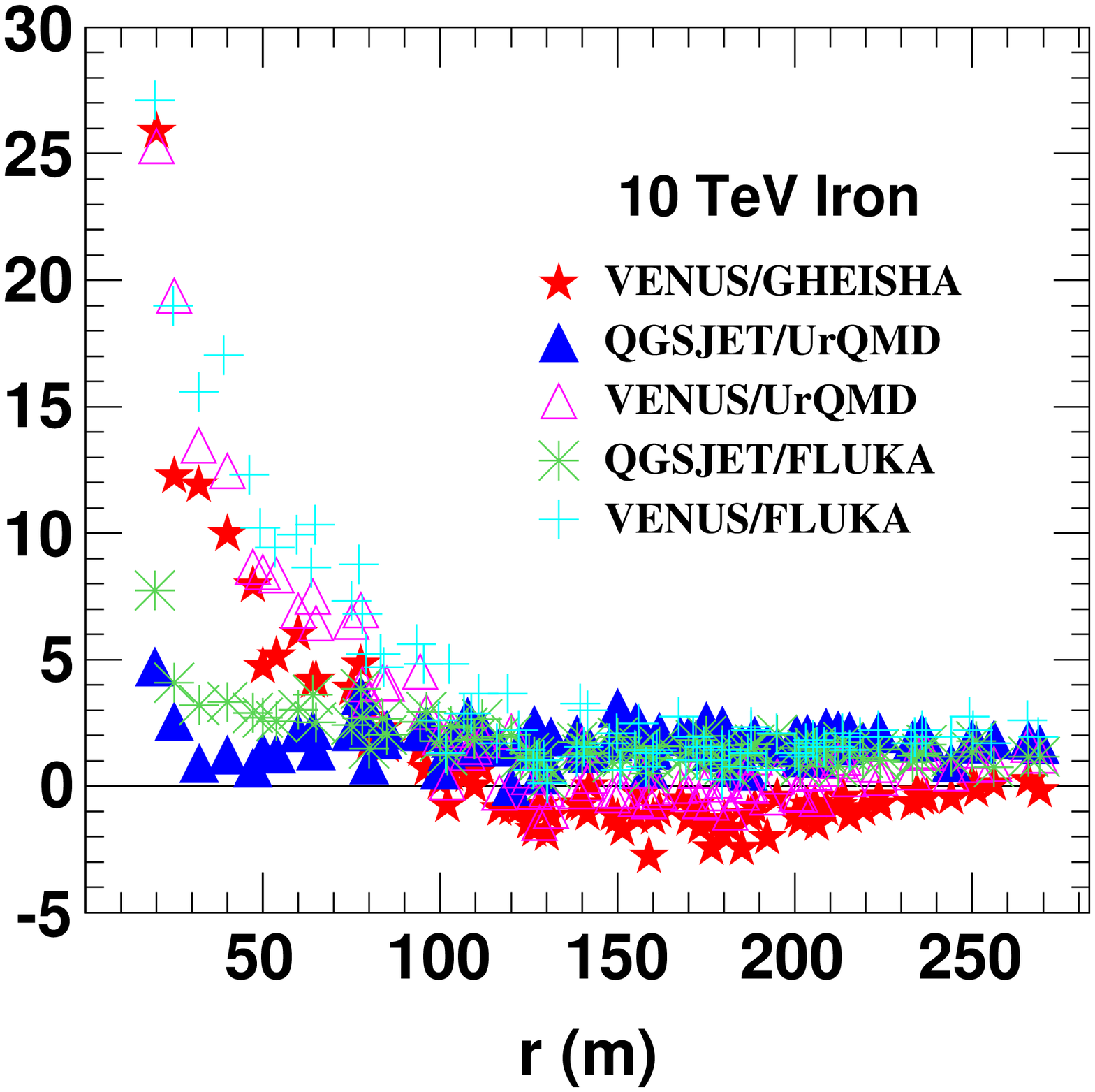}}
\caption{Variations of \% relative arrival time deviations ($\Delta_t$s) of 
Cherenkov photons with respect to distance from the shower core of different 
primaries for different high and low energy hadronic interaction model 
combinations. The solid line in all plots indicates the values of the 
QGSJET-GHEISHA model combination. The QGSJET-GHEISHA model combination is 
considered as the reference for the calculation.}
\label{fig6}
\end{figure*}

As in the case of $\rho_{ch}$, to see differences due to 
combination of models chosen, we study the relative deviations in the 
$t_{ch}$s considering QGSJET-GHEISHA combination as the reference. 
Fig.\ref{fig6} shows the \% relative deviations of the arrival time 
($\Delta_t$) of shower front for various model combinations for various 
monoenergetic primary particles as a function of the core distance.   
From the Fig.\ref{fig6} it is clear that, the differences due to 
different model combinations are small (0 to $\sim$$\pm$8\%) for 
the $\gamma$-ray primaries. For all these primaries, the 
differences above 50 m core distances are negligible ($<$$\pm$ 2\%) 
for all the models combinations 
except for one case, VENUS-GHEISHA combination with 1 TeV $\gamma$-rays. 
For this case, the deviation is the highest, which is upto -8\% near 
the shower core and  negligible only above 100 m core distances. Thus 
there is no
particular trend of deviations of arrival times on the  basis of model 
combination and all model combinations almost agree 
with the QGSJET-GHEISHA combination, at 
all energies of $\gamma$-ray primary.           

For proton and iron primaries the VENUS lead group of model combinations 
differ appreciably from the QGSJET-GHEISHA combination, depending on the 
energy and the type of primary particle. The prominence of these deviations 
increases with decreasing primary energy and with decreasing distance from 
shower core. For 250 GeV proton primary, the VENUS lead group of models 
produced $\sim$5 to 25\% longer arrival times than those produced
by QGSJET-GHEISHA combination, with a gradually decreasing trend with the
increasing core distance.  For 1 TeV and 2 TeV proton primaries this group of 
models produced $\sim$1 to 20\% and $\sim$$\pm$ 1 to 15\% longer arrival 
times respectively than the reference model combination. But in these two 
cases the time differences decrease very fast upto the core distance 
$\sim$100 m. Beyond this distance the arrival time differences become 
gradually negligible with increasing core distance.
On an average, in this group of model combinations, the VENUS-FLUKA 
combination generates highest $\Delta_t$s over all core distances for the 
proton primaries. The trend due to the VENUS lead group of model combinations 
for the iron primaries is almost similar to that of  proton primaries. 
However, here the $\Delta_t$s near the shower core are very large in 
comparison to proton primaries. For instance the $\Delta_t$s nearest to the 
shower cores of 5 TeV and 10 TeV iron primaries are $\sim$48\% and 
$\sim$29\% respectively. In the cases of iron primaries also, on an average 
the VENUS-FLUKA combination generates highest $\Delta_t$s over all core 
distances. For the QGSJET lead group of model combinations, the $\Delta_t$s 
are very small for all energies of proton and iron primary particles,  as in 
the case of the $\gamma$-ray primary. This group of model combinations generate
0 to $\sim$$\pm$5\%  $\Delta_t$s for proton primaries and 0 to 
$\sim$$\pm$8\%  $\Delta_t$s for iron primaries than the reference
model combination.

\subsubsection{Behaviour of fluctuations}
To observe the fluctuations in the $t_{ch}$ distributions, we have calculated 
the $\sigma_{pm}$ and plotted in the Fig.\ref{fig5} for different primaries, 
energies and  model combinations. It is observed that, the fluctuation is 
large near the core for distances below 50 m. This is not clearly 
understandable. The fluctuation decreases with increasing core distance. The 
rate of decrease with the core distance is faster near the core and decreases 
at larger core distances for all primary particles, energies and model 
combinations except for 250 GeV proton primary, for which the rate of decrease 
of the fluctuations is almost same over all core distances. For a given 
primary particle, the fluctuation decreases with increasing energy of the 
particle. There is no particular model dependent trend  seen. It should be 
noted that, for the proton primaries fluctuations produced by different model 
combinations are maximum amongst all primary particles. Moreover, as a whole, 
the differences in fluctuations, produced by different model combinations, are 
maximum near the shower core for all primary particles at all energies.

\subsubsection{Primary energy dependence}
\begin{figure}[hbt]
\centerline
\centerline{\includegraphics[width=6.2cm, height=5.2cm]{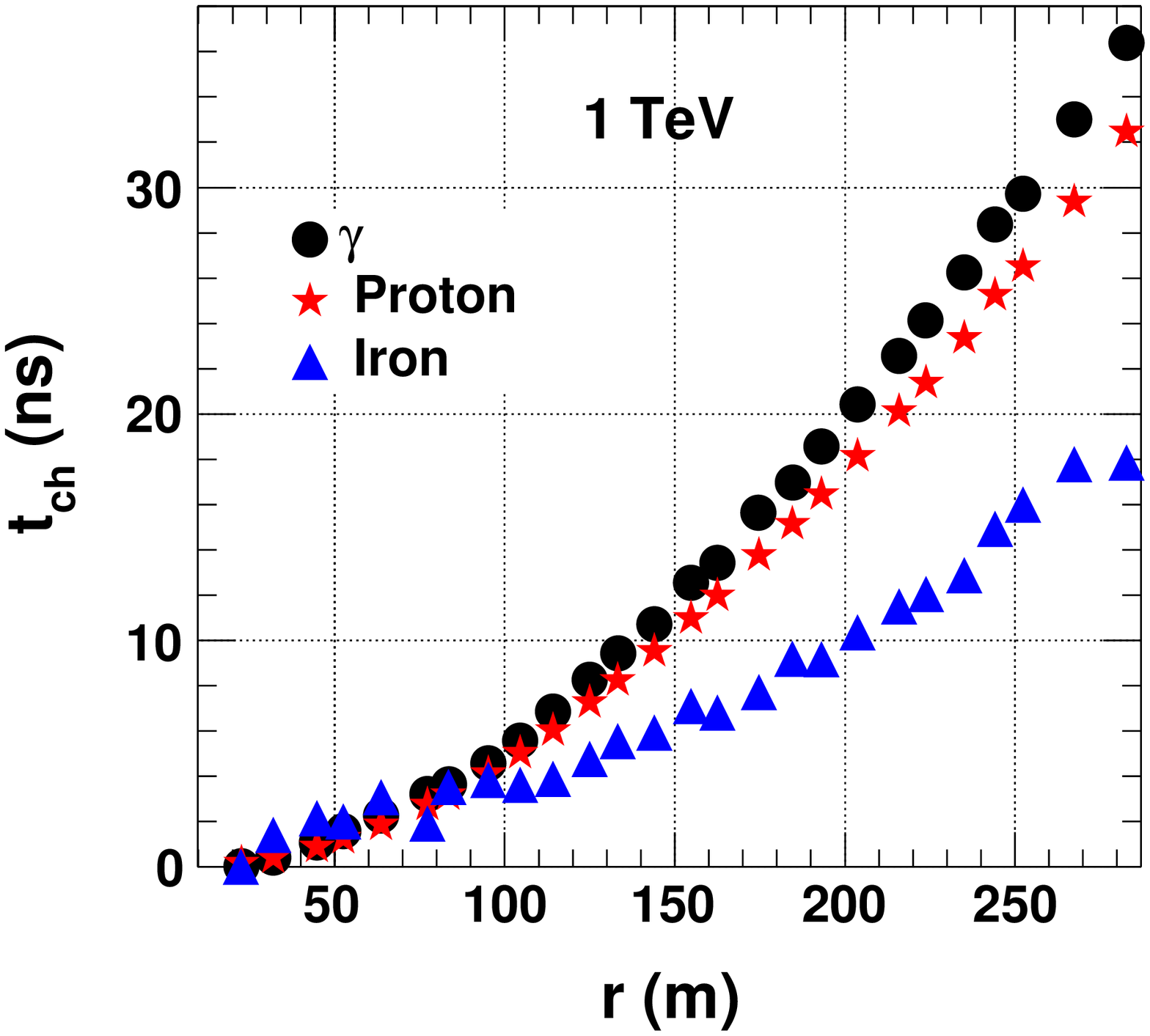}
\vspace{5mm}\\
\includegraphics[width=6cm, height=5cm]{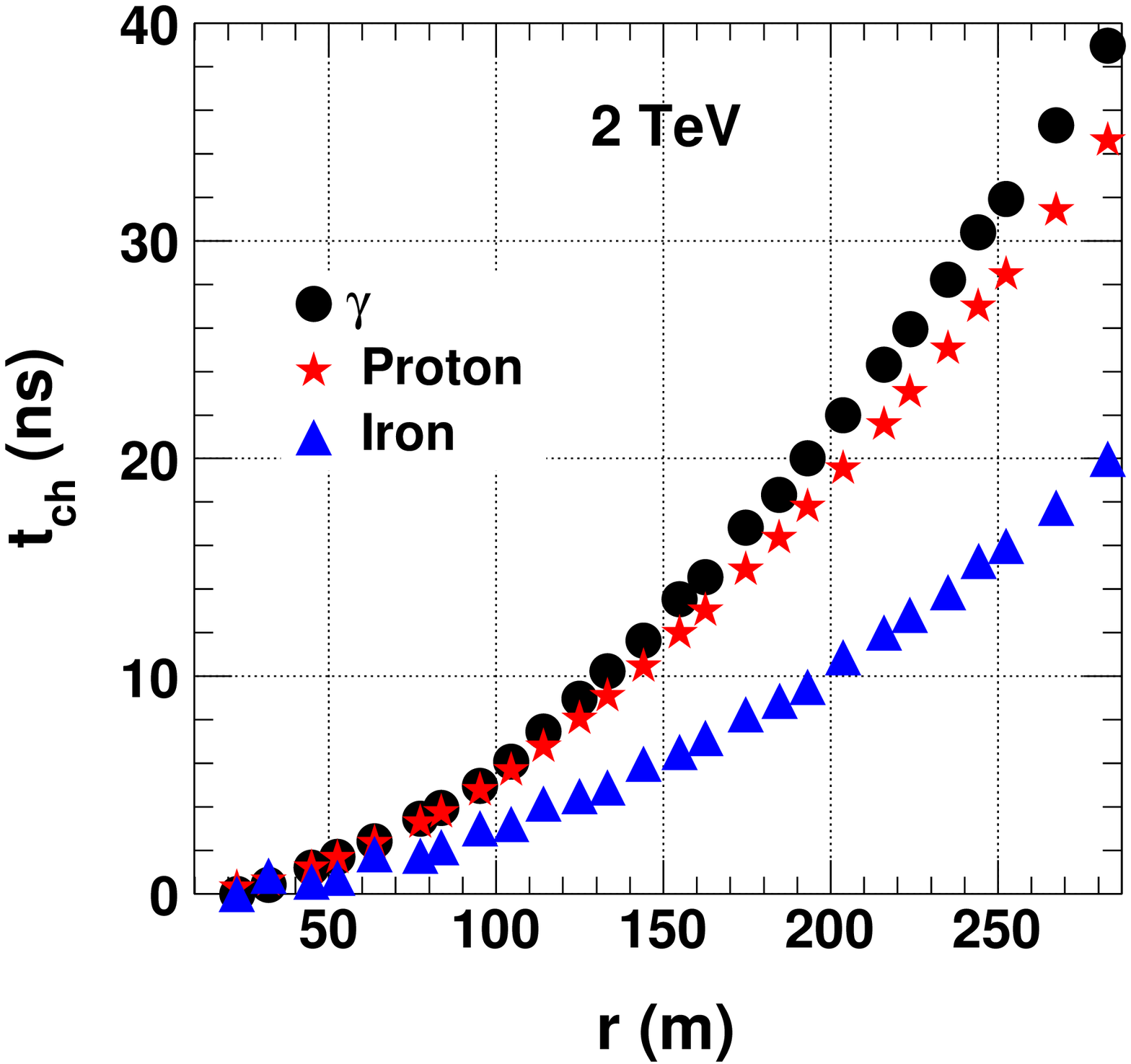}
\vspace{5mm}\\
\includegraphics[width=6.2cm, height=5.2cm]{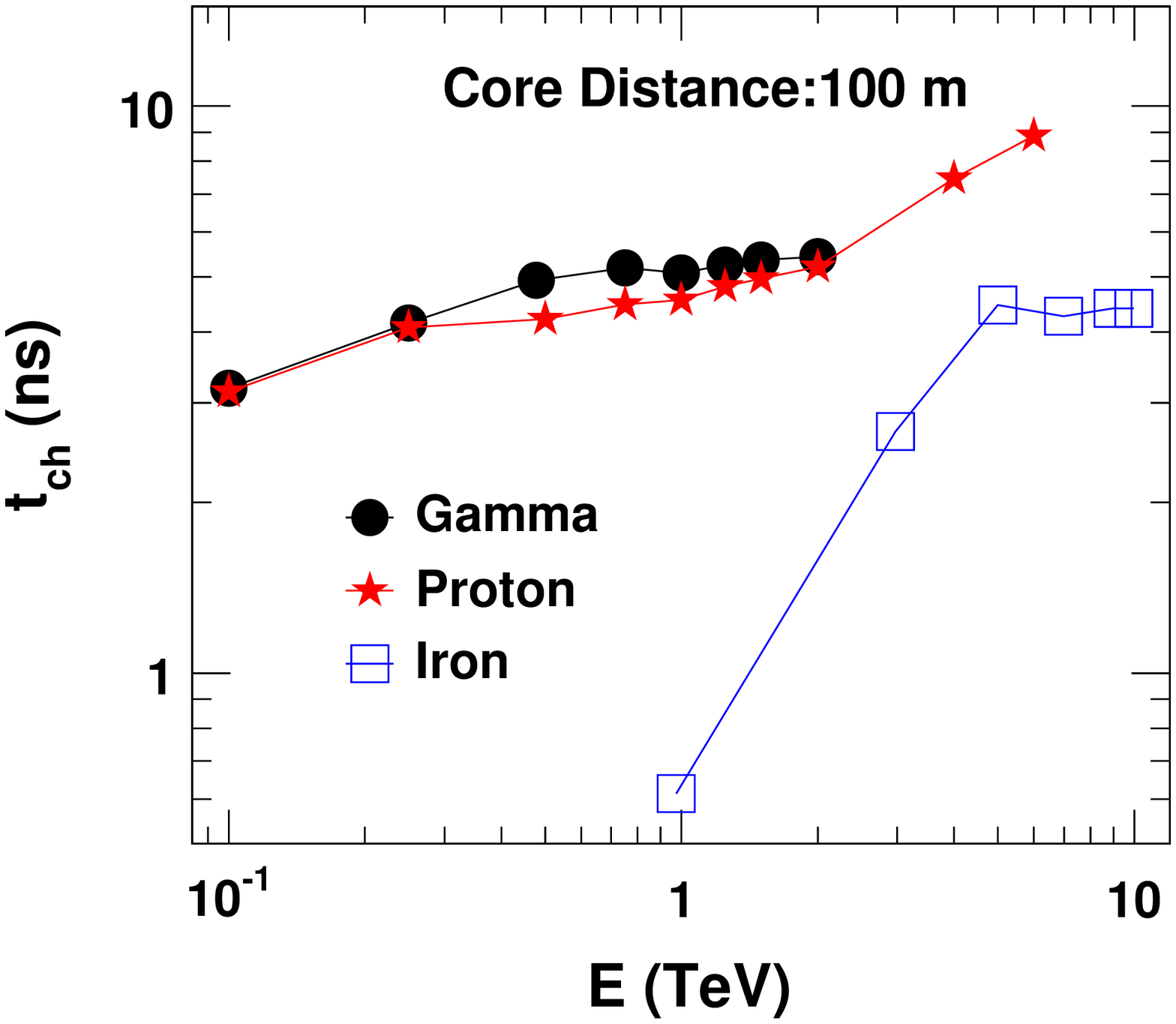}
}
\caption{Top and middle panels: Distributions of offset normalised $t_{ch}$ 
with respect to distance from the shower core of $\gamma$, proton and iron 
primaries at 1 TeV and 2 TeV energies, as given by the VENUS-GHEISHA model 
combination. Bottom panel: Variations of offset normalised $t_{ch}$ with 
respect to primary energy at a distance 100 m from the shower core of 
$\gamma$, proton and iron primaries for the same  model combination as for the 
other two panels.}
\label{fig7}
\end{figure}

For all the three types of primary particles, we have studied the
variation of $t_{ch}$ with core distance for a constant energy. We have also 
studied the variation of $t_{ch}$ at core distance of 100 m as a function of 
primary energy. Results are shown in the Fig.\ref{fig7} for the VENUS-GHEISHA 
combination only.                

\begin{figure}[hbt]
\centerline
\centerline{\includegraphics[width=6cm, height=5cm]{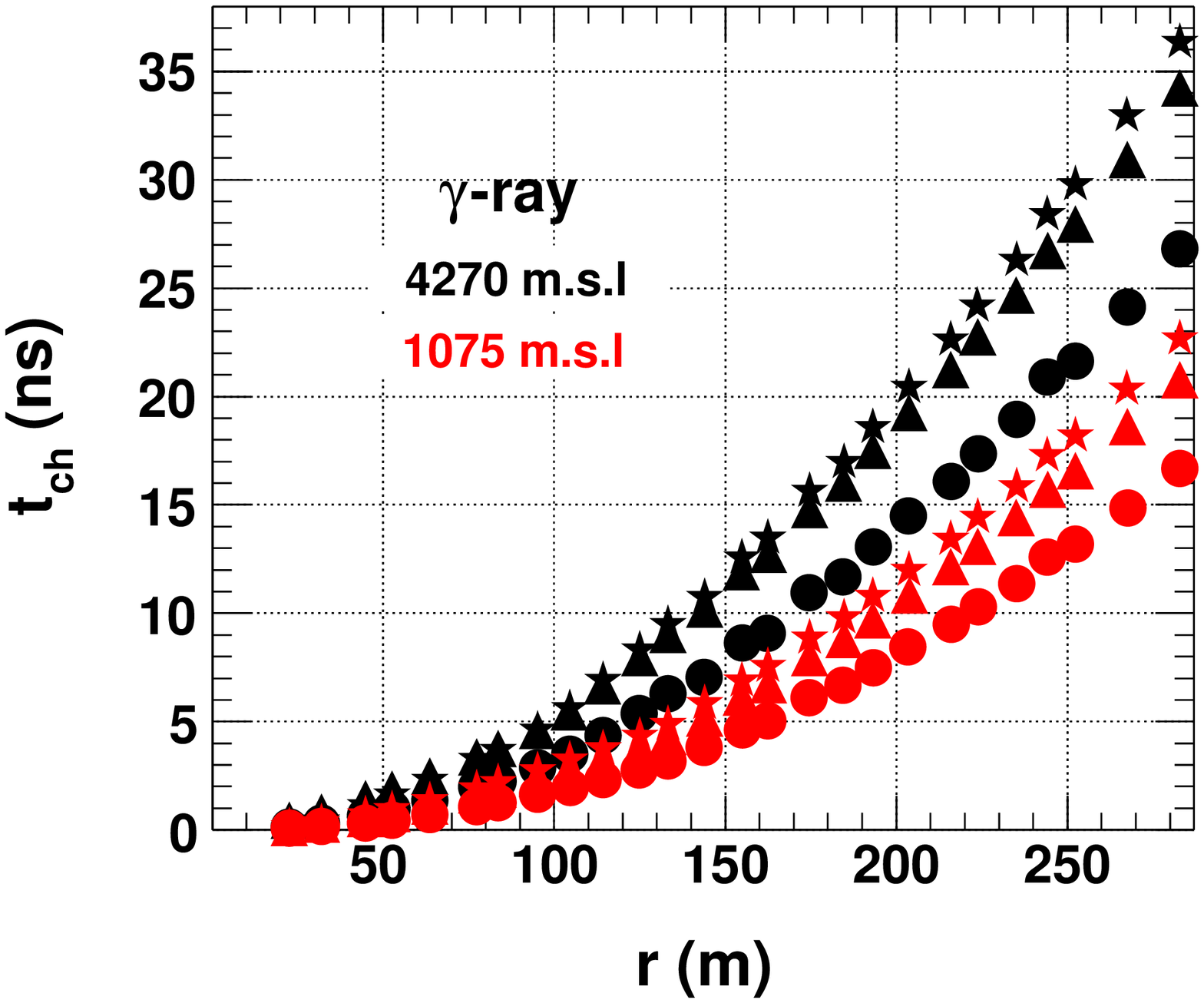}
\vspace{5mm}\\
\includegraphics[width=6cm, height=5cm]{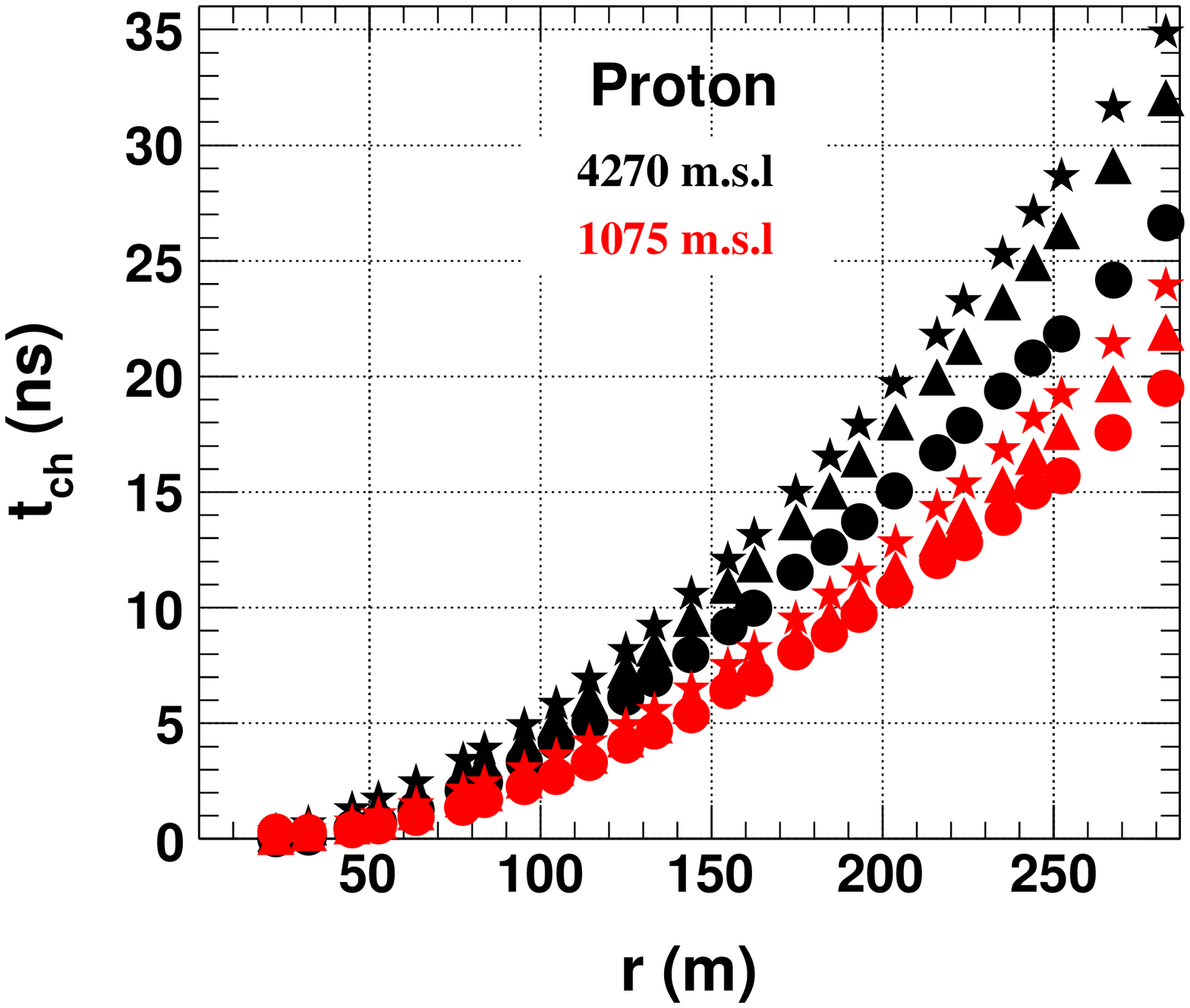}
\vspace{5mm}\\
\includegraphics[width=6cm, height=5cm]{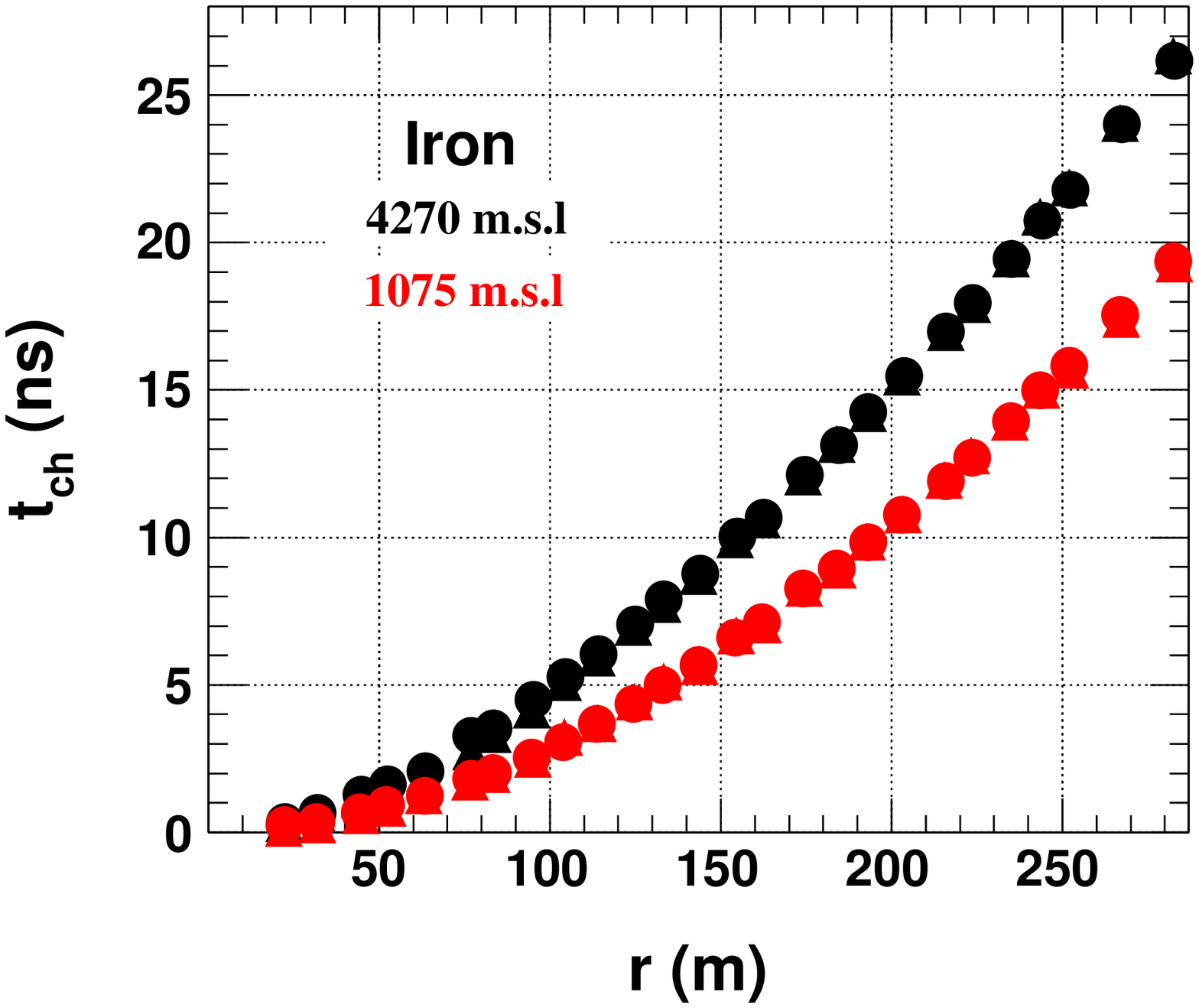}
}
\caption{Distributions of offset normalised $t_{ch}$ with respect to distance
from the shower core of $\gamma$, proton
and iron primaries at different energies given by the QGSJET-GHEISHA model
combination over the observation levels of Hanle (black symbols) and
Pachmarhi (red symbols). In the respective plots,
{\large $\bullet$/\textcolor{r}{$\bullet$}} indicates for
100 GeV $\gamma$, 250 GeV proton and 5 TeV iron primaries;
$\blacktriangle$/\textcolor{r}{$\blacktriangle$} indicates for 500 GeV
$\gamma$, 1 TeV proton and 10 TeV iron primaries; and
$\bigstar$/\textcolor{r}{$\bigstar$} indicates for 1 TeV $\gamma$ and 2 TeV
proton primaries.}
\label{fig8}
\end{figure}

\begin{figure*}[hbt]
\centerline
\centerline{\includegraphics[width=5.5cm, height=4cm]{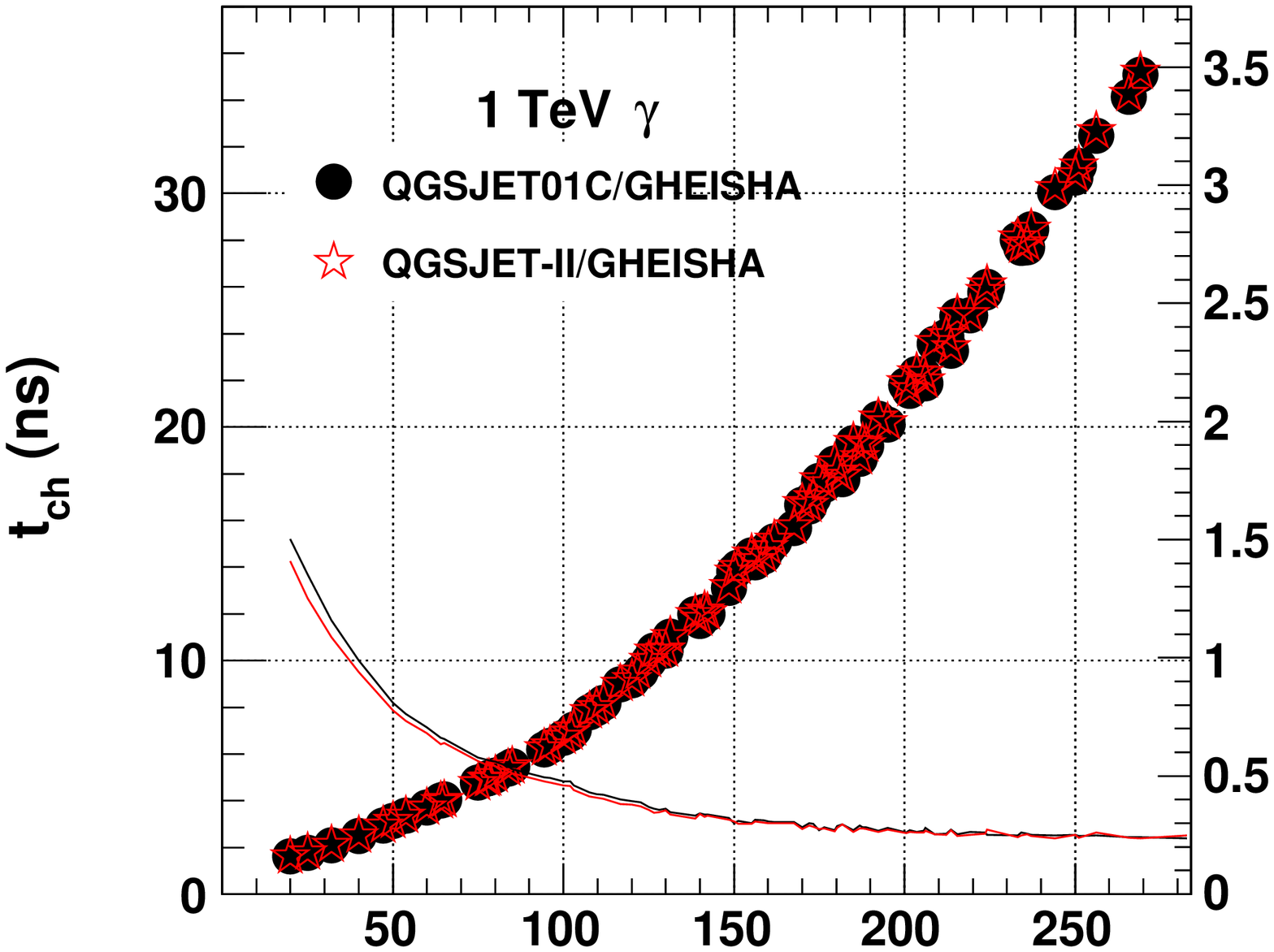}
\includegraphics[width=5cm, height=4cm]{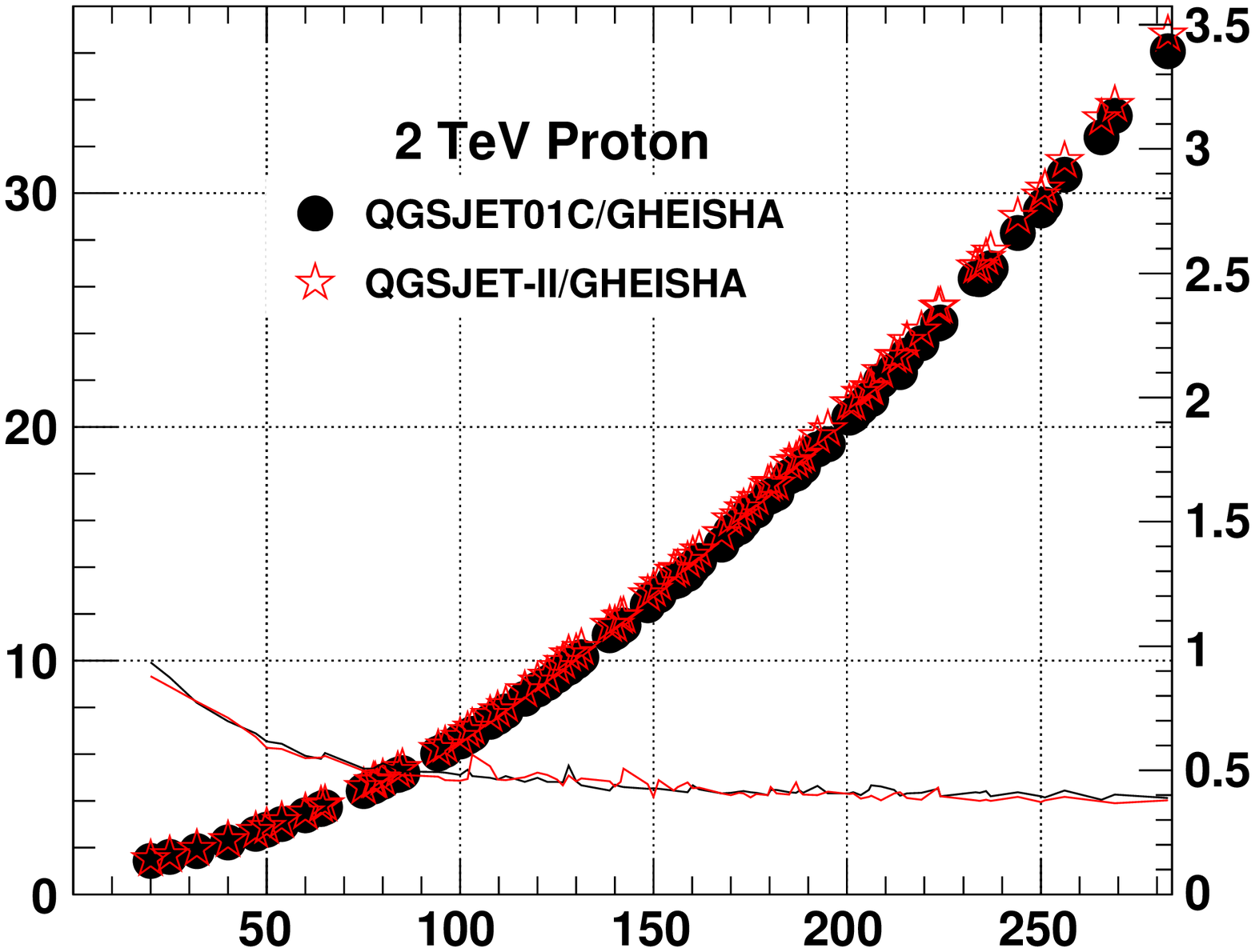}
\includegraphics[width=5.3cm, height=4cm]{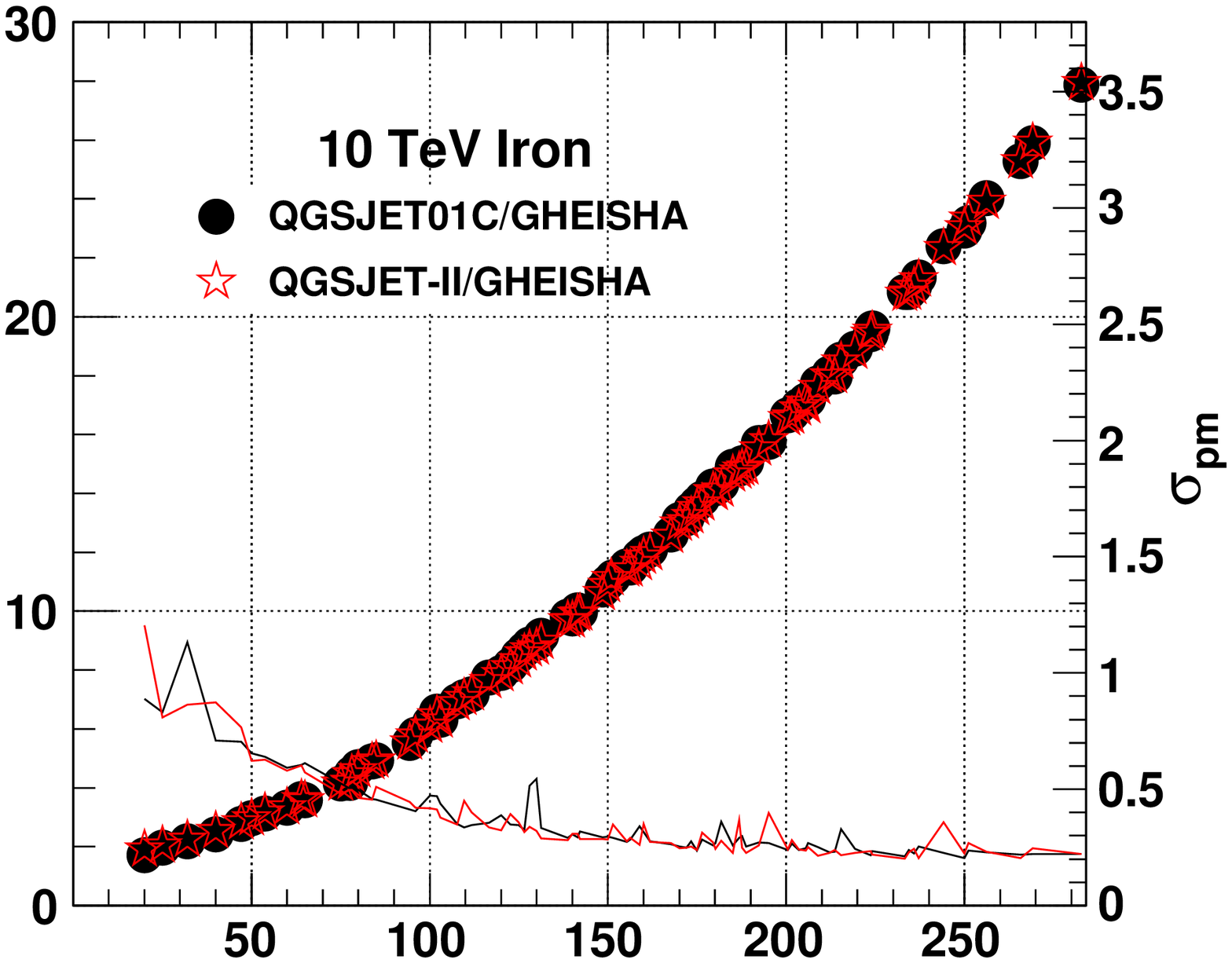}}
\centerline{\hspace{-0.5cm}
\includegraphics[width=5.4cm, height=4.5cm]{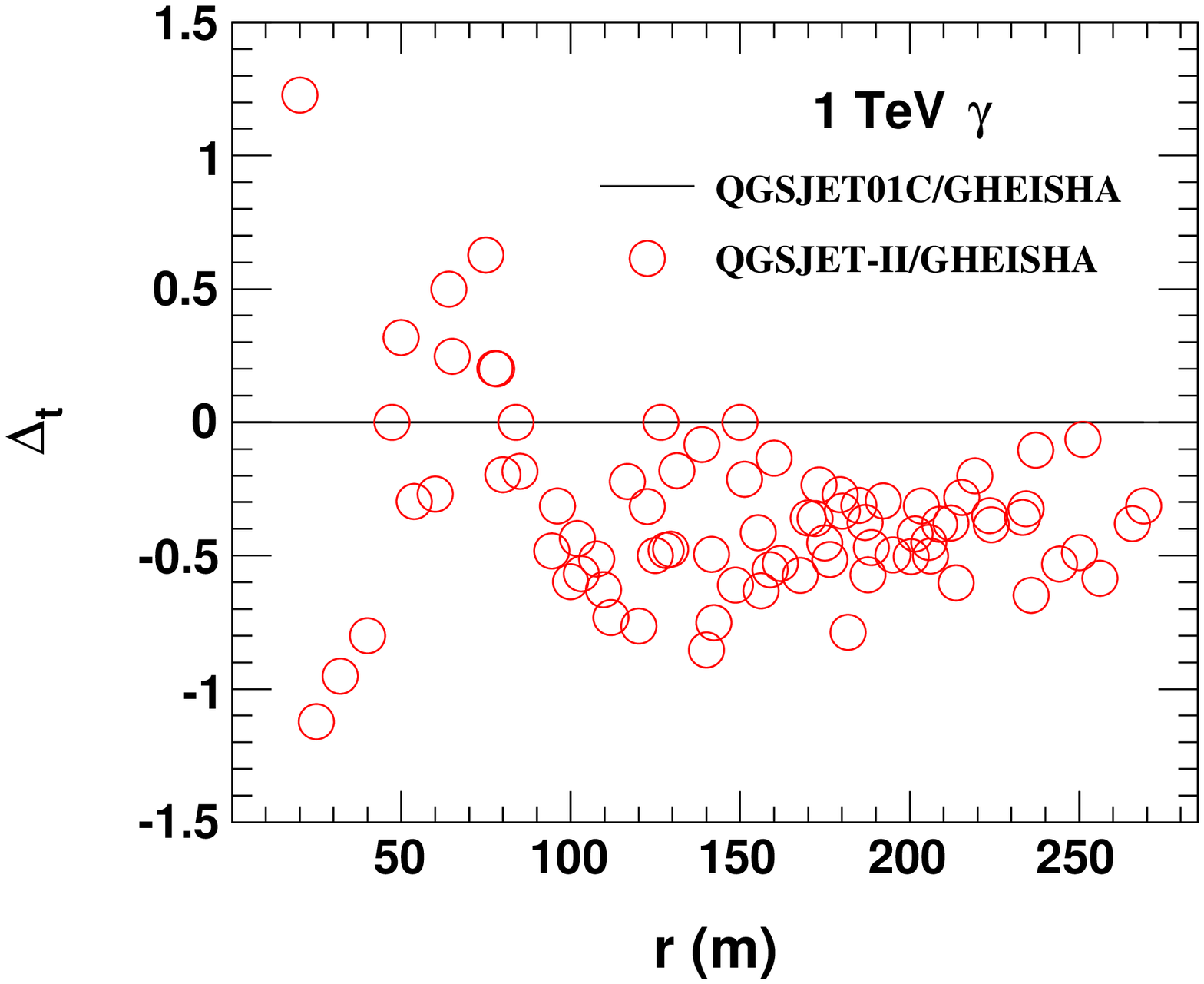}
\hspace{0.2cm}
\includegraphics[width=4.7cm, height=4.5cm]{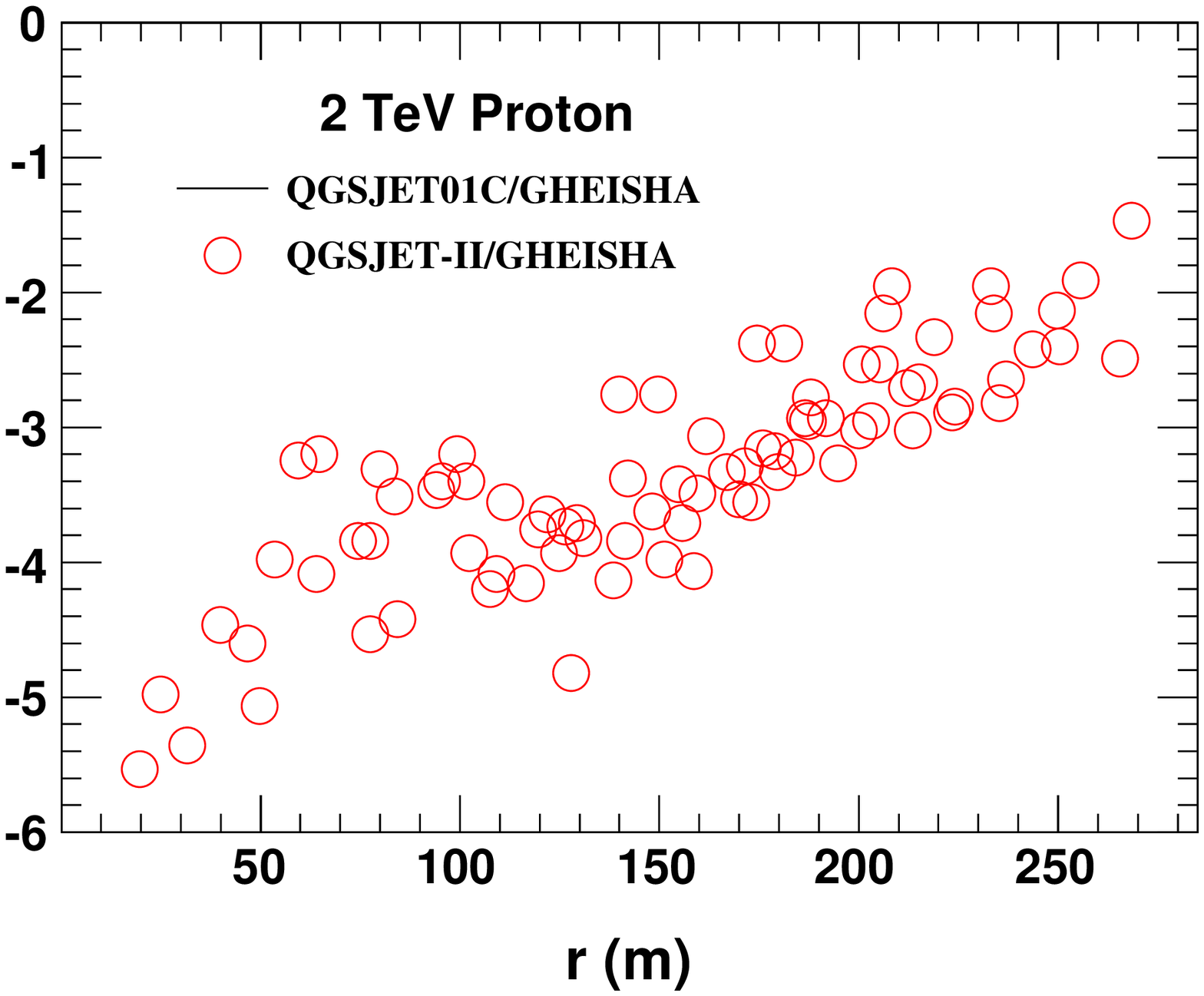}
\hspace{0.2cm}
\includegraphics[width=4.8cm, height=4.5cm]{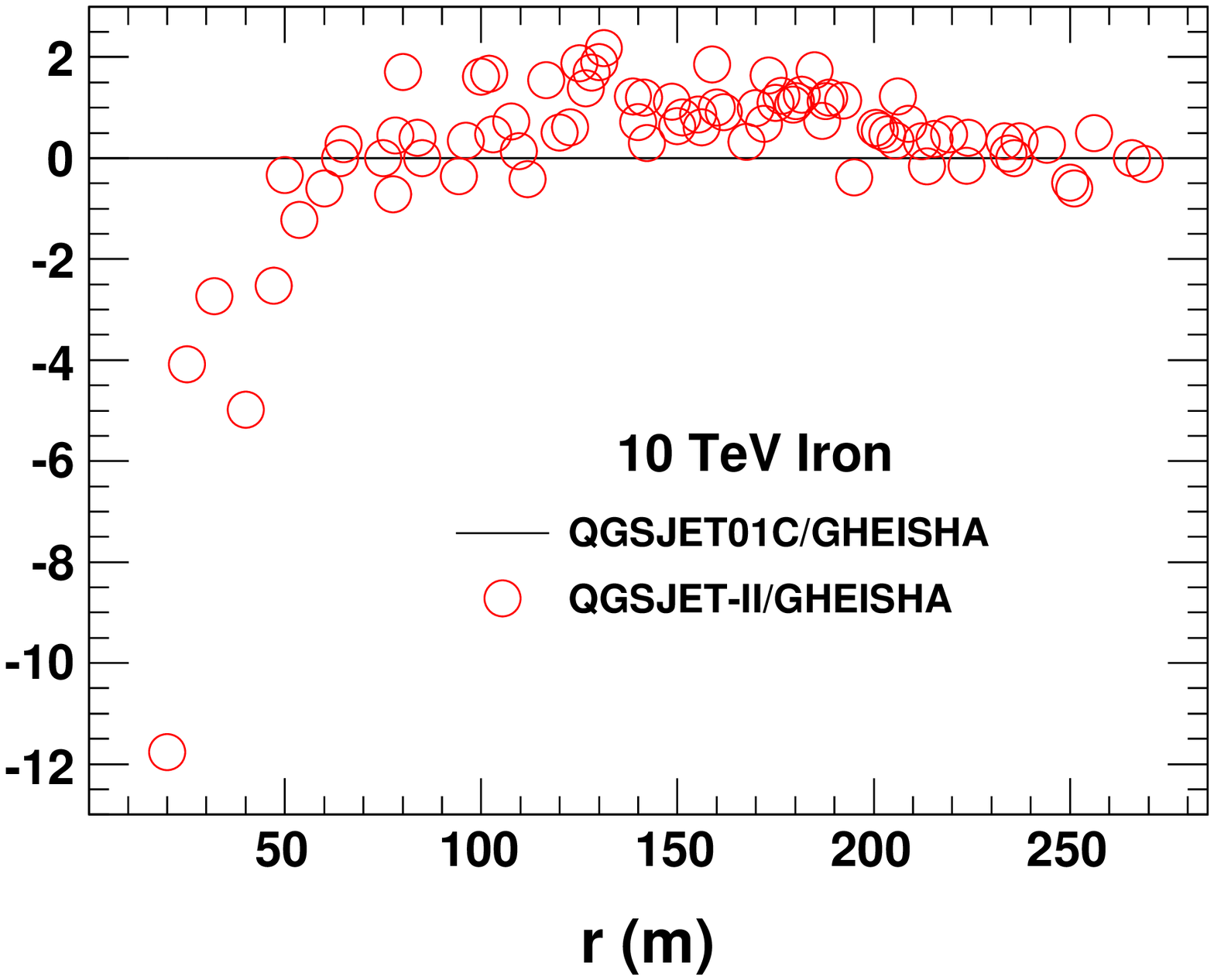}}

\caption{Top panels: Distributions of $t_{ch}$ with
respect to distance from the shower core for 1 TeV $\gamma$, 2 TeV proton and
10 TeV iron primary particles obtained for the QGSJET01-GHEISHA and
QGSJETII-GHEISHA model combinations. Bottom panels: $\Delta_t$ of $t_{ch}$
with respect to distance from the shower core of these primaries
for QGSJETII-GHEISHA model combination from the $t_{ch}$
of QGSJET01-GHEISHA combination.}
\label{fig8a}
\end{figure*}

We see that, at a given energy, on an average, 
Cherenkov photons from a shower initiated by a $\gamma$-ray take the
longest time to reach the observation
level. This time is shortest for the iron primary. The difference in the 
$t_{ch}$ for showers initiated by $\gamma$-ray and proton, at a given core 
distance is very small. While the difference is quite large for the iron 
primary. At a given core distance, the $t_{ch}$ increases with increasing 
energy of the primary particle upto certain energy, depending upon the type of 
primary particle. For the $\gamma$-ray primary this energy is $\sim$2 TeV and 
for the iron primary it is $\sim$5 TeV. But for the proton primary it will
be greater than 6 TeV (see the bottom panel of the Fig.\ref{fig7}).
 
As discussed in the density subsection, at a given primary energy, the 
$\gamma$-ray primary produces highest number and iron primary produces least 
number of Cherenkov photons. While number of Cherenkov photons produced by 
proton primary is slightly less than that produced by the 
$\gamma$-ray primary. So the angular and hence arrival time distributions 
of Cherenkov photons with respect to the first arriving particle over the 
observation level is much wider for
the $\gamma$-ray primary than that for iron and slightly wider that that for
proton primary. This arrival time distribution also 
producing the observed effect on the time distribution.

\subsubsection{Altitude effect}
The effect of altitude of the observational level on the $t_{ch}$
distribution is also studied and is shown in the Fig.\ref{fig8}. It is clear 
from this figure that, for all primary particles
at all energies, the average  
arrival time of Cherenkov 
photons' shower is shorter over the lower observation level than that over
the higher observation level with respect to the same first arriving photon.
Over the same observation level 
it is longer for higher energy primary than that for 
the low energy primary as already mentioned above. The difference between this 
arrival time of two primaries decreases with increasing energy and mass of the
primary particle.                     

When a shower has to a travel long 
distance to arrive at lower observation level, most of the low energetic 
particles get absorbed in the atmosphere, leaving only high energetic 
particles having their energies almost in same order of magnitude. 
Consequently, the relative time differences of different particles, arrive on
the observation level with the first arriving particle are comparatively smaller
than that are observed on the observation level at higher altitude. So the 
$t_{ch}$ at lower observation level are smaller in magnitudes than the 
$t_{ch}$ at higher observational level.

\subsubsection{Comparison of QGSJET01 and QGSJETII} 
We have also studied the difference in $t_{ch}$ 
produced by QGSJET01-GHEISHA and QGSJETII-GHEISHA model combinations. 
The results of this study is shown in the Fig.\ref{fig8a}. It is seen from 
the figure that, for the $\gamma$-ray primary there is no any real difference 
between these two models as the deviations of the $t_{ch}$s generated by these 
two models are only 0 to $\sim$$\pm$1.1\% over all core distances. For the 
proton primary the deviations are $\sim$-2.5 to $\sim$5.5\%, which are  
negligible. On the other hand for the iron primary the effective deviations 
range from 0 to $\sim$$\pm$5\%, which are also negligible in comparison 
deviation of other model combinations discussed above for this primary.       

\section{Summary and conclusion}
In view of the importance in the ACT and lack of sufficient works applicable
to high altitude observation sites, we have made an elaborate study on the 
density and arrival time distributions of Cherenkov photons in EAS using the 
CORSIKA 6.990 simulation package \cite{Heck}. 
Summary of this study and consequent conclusions can be made as follows:  

The lateral density and arrival time distributions of Cherenkov photons follow 
a negative exponential function and a function of the form $t (r) = 
t_{0}e^{\Gamma/r^{\lambda}}$ respectively for all primary particles, energies 
and model combinations. As these functions' parameters are different, the 
geometries of these distributions are obviously different depending upon the 
energy and mass of the primary particle. These parametrisations show that the
analytical descriptions of the lateral density and arrival time distributions 
of Cherenkov photons are possible within some uncertainty. The full scale of
such parametrization as a function of energy and shower angle would be useful 
for analysis of data of $\gamma$-ray telescopes because, it will help to 
disentangle the $\gamma$-ray showers from the hadronic showers over a given 
observation level.           

These distributions of density and arrival time of Cherenkov photons as
a function of core distance for
the  $\gamma$-ray showers are almost independent of hadronic interaction models,
whereas that for the proton and iron showers depend on the hadronic 
interaction models on the basis of the type of models (low and high 
energy), the energy of primary particle and the distance from the shower core. 
In most of the cases the model dependence is significant for the iron showers. 
The systematic effect of the hadronic interaction model dependence has to be 
taken into account when assessing the effectiveness of background rejection 
for a $\gamma$-ray telescope.
Moreover, from the study of shower to shower 
fluctuations ($\Delta_{pm}$) of Cherenkov photons' density and arrival time it 
is clear that they are almost independent of hadronic interaction models, 
but depend on 
the energy and type of the primary particle, number of shower samples used for 
the analysis and the location of detectors. These are very important inference 
to be taken into care on the estimation of systematic uncertainties in a real 
$\gamma$-ray astronomy observation. 

The energy dependent variation of Cherenkov photons' density shows that to get
the equivalent numbers of Cherenkov photons from different primary particles, 
the energy of the particles must be increased to several times with increasing 
mass of them. This explains why we have chosen different specific energies for
different primaries as mentioned in the Sec.II. Similarly, from the study of 
the altitude effect, i.e. the comparison of lateral distributions of Cherenkov 
photons at two observation levels, we can conclude that as we go to the higher
observation levels, it is possible to detect low energy $\gamma$-ray signals 
from a source as well as possible to do the $\gamma$-ray astronomy with a much 
smaller telescope system than at lower observation level. 

As mentioned above the full scale parametrisation of the density and arrival
time distributions of Cherenkov photons for different primary is important for 
the analysis of $\gamma$-ray observation data, so in future we are planning to 
perform such full parametrization for the HAGAR telescope site for the 
effective analysis of the HAGAR telescope \cite{Versha1} data. Again, since 
in this work we have considered the energy only upto 10 TeV, hence we will 
extend our future work upto 100 TeV, the more relevant energy range of 
$\gamma$-ray astronomy. Furthermore, as the pattern of angular distribution of
Cherenkov photons for different primaries is another crucial parameter for
separation of hadron showers from the $\gamma$-ray showers, we will take up
this issue also for our future study.
               
\section*{Acknowledgments}
U.D.G. and P.H. are thankful to Department of Science $\&$ Technology (DST),
Govt of India for financial support through the Project
No. SR/S2/HEP/-12/2010(G). U.D.G. and V.R.C. also thankful to J. Knapp,
Karlsruhe Institute of Technology, Karlsruhe, Germany for his
useful comment on the work during a discussion. Finally we thank the anonymous 
referees for valuable comments which allow us to improve the manuscript.

\end{document}